\documentclass[onecolumn,superscriptaddress,superscriptreference]{revtex4-2}
\usepackage{amsmath}
\usepackage{bbold,bm,amssymb,scalerel,mathtools}
\usepackage{empheq}
\usepackage{graphicx}
\usepackage{color}
\usepackage{enumitem}
\usepackage{algorithm,algpseudocode}
\usepackage{multirow}
\usepackage{colortbl,booktabs}
\usepackage{placeins}
\usepackage[usenames,dvipsnames]{xcolor}
\usepackage{appendix}
\usepackage{tikz}
\usepackage[colorlinks, linkcolor=blue!50!black, urlcolor=blue!50!black, citecolor=blue!50!black]{hyperref}
\usepackage[parfill]{parskip}
\usepackage{stackengine}

\usetikzlibrary{calc}
\newcommand\doverline[1]{%
	\tikz[baseline=(nodeAnchor.base)]{
		\node[inner sep=0] (nodeAnchor) {$#1$}; 
		\draw[line width=0.1ex,line cap=round] 
		($(nodeAnchor.north west)+(0.0em,0.2ex)$) 
		--
		($(nodeAnchor.north east)+(0.0em,0.2ex)$) 
		($(nodeAnchor.north west)+(0.0em,0.5ex)$) 
		--
		($(nodeAnchor.north east)+(0.0em,0.5ex)$) 
		;
}}

\newcommand\equalhat{%
	\let\savearraystretch\arraystretch
	\renewcommand\arraystretch{0.3}
	\begin{array}{c}
		\stretchto{
			\scalerel*[\widthof{=}]{\wedge}
			{\rule{1ex}{3ex}}%
		}{0.5ex}\\ 
		=%
	\end{array}
	\let\arraystretch\savearraystretch
}

\newcommand{\beq}{\begin{equation}}
\newcommand{\eeq}{\end{equation}}
\newcommand{\ben}{\begin{equation*}}
\newcommand{\een}{\end{equation*}}
\newcommand{\bseq}{\begin{subequations}}
\newcommand{\eseq}{\end{subequations}}
\newcommand{\bea}{\begin{eqnarray}}
\newcommand{\eea}{\end{eqnarray}}

\newcommand{\p}{\partial}
\newcommand{\la}{\langle}
\newcommand{\ra}{\rangle}

\newcommand{\Int}{\int\!\!}

\newcommand{\avS}[1]{\la{#1}\ra}
\newcommand{\avM}[1]{\overline{#1}\,}
\newcommand{\avMP}[1]{\doverline{#1}\,}
\newcommand{\avAS}[1]{\la{#1}\ra_\text{AS}}

\newcommand{\br}{\bm{r}}
\newcommand{\bth}{\boldsymbol{\theta}}

\newcommand{\bs}{\bm{s}}

\newcommand{\Dpol}{\mathcal{D}}
\newcommand{\Dmut}{\Dpol_{\policy}}
\newcommand{\meanmem}{\mu_\mem}

\newcommand{\meanmemzero}{\mu_{\mem_0}}
\newcommand{\meanmemone}{\mu_{\mem_1}}
\newcommand{\meanpol}{\mu_\policy}

\newcommand\state{\mathcal{S}}
\newcommand\mem{\mathcal{M}}
\newcommand\reward{\mathcal{R}}
\newcommand\policy{\mathcal{P}}

\newcommand{\Reff}{\avM{\reward}}

\begin{document}
	
	\setlength{\belowdisplayskip}{3pt} \setlength{\belowdisplayshortskip}{3pt}
	\setlength{\abovedisplayskip}{3pt} \setlength{\abovedisplayshortskip}{3pt}
	
	\title{Theory of collective learning in populations of adaptive agents}
	
	\author{Gerhard Jung}	
	\affiliation{Universit\'e Grenoble Alpes, CNRS, LIPhy, 38000 Grenoble, France}
    \affiliation{Institut f\"ur Theoretische Physik, Universit\"at Innsbruck, 6020 Innsbruck, Austria}

    \author{Johann Asnacios}
	\affiliation{Gulliver UMR CNRS 7083, ESPCI Paris, PSL Research University, 10 rue Vauquelin, 75005 Paris, France}
	    
	\author{Misaki Ozawa}
	\affiliation{Universit\'e Grenoble Alpes, CNRS, LIPhy, 38000 Grenoble, France}

    \author{Olivier Dauchot}
	\affiliation{Gulliver UMR CNRS 7083, ESPCI Paris, PSL Research University, 10 rue Vauquelin, 75005 Paris, France}
	
	\author{Eric Bertin}
	\affiliation{Universit\'e Grenoble Alpes, CNRS, LIPhy, 38000 Grenoble, France}

	\date{\today}

    \begin{abstract}
        We investigate homogeneous populations of smart active agents that exchange information with their neighbors to collectively perform a decentralized learning process aimed at achieving a prescribed macroscopic state. Such agents may, for example, represent simple microrobots. The exchanged information comprises tunable parameters governing the agent dynamics, referred to as the individual policy, together with an internal memory encoding previously visited states. This memory is used to evaluate a reward that quantifies the success of a policy to achieve the prescribed state.
         Building on the framework introduced by Jung \textit{et al.} [Phys. Rev. Lett. \textbf{134}, 248302 (2025)], we develop a kinetic-theory description of collective learning in spatially homogeneous systems and derive formal evolution equations for the distribution of policies across the population. A central outcome of our theory is the emergence of an effective reward function that fully determines the evolution of the policy distribution and encapsulates the microscopic details of the agents physical and memory dynamics.  We obtain closed equations for the policy mean and variance which admit explicit time-dependent solutions under the assumption of Gaussian-distributed memories and polices. 
         
        To illustrate the framework, we present a series of minimal microscopic models, considering both perfect and partial separation of physical, memory and policy exchange time scales. Additionally, we study models with one- and two-dimensional policies. The obtained theoretical results compare well with agent-based numerical simulations. Despite the simplicity of the underlying models, the theory captures key aspects of collective learning, including the influence of population diversity and reward fluctuations on learning performance.
        Finally, we discuss potential applications to swarm robotics and machine learning, and highlight connections with classical models of biological evolution, including the Replicator equation and the Moran model. 
	\end{abstract} 
    
	\maketitle

\section{Introduction}

Exchanging information within a population may be a way to enhance adaptability to external conditions or tentatively optimize a collective behavior.
Such adaptation processes have been studied for instance in the context of models of biological evolution \cite{drossel2001biological,sella2005application,houchmandzadeh2011fixation,chia2011statistical}, as well as models of opinion dynamics \cite{lorenz2007continuous} and more general social agent models \cite{castellano2009statistical}.
In many cases, information exchanges occur directly between individuals or agents, rather than through a central unit that would control the adaptation process.
Such decentralized adaptation or optimization processes are also currently attracting interest in the field of robotics \cite{watson2002embodied,bredeche2018embodied,long2018towards,doi:10.1177/1059712320930418,bredeche2022social,ben2023morphological,novkoski2026graspion,fersula2026aggregating}, where (sometimes bio-inspired \cite{verdoucq2022bioinspired}) decentralized processes may be found to perform better than centralized ones \cite{long2018towards,ben2023morphological}.
Beyond such `hard' robotic systems, an alternative `soft' robotics line of research consists in realizing large assemblies of minimalistic microrobots \cite{wang2024robo,ma2024smarticle,ozkan2021collective} made of a soft material, which exhibit responsive or programmable collective properties
\cite{Li2021programming,zhou2022programmable,kotikian2019untethered,zeravzic2017colloquium}.
In both cases, populations of autonomous microrobots with elementary communication and computation skills are found to
lead to adaptive collective behaviors \cite{ben2023morphological,bredeche2022social,kaspar2021rise,levine2023physics,majidi2019soft,mo2023challenges,cazenille2024signaling,fersula2026aggregating},
sometimes called smart or intelligent active matter \cite{PhysRevE.97.042604,pishvar2020foundations,cichos2020machine,kaspar2021rise,levine2023physics,goldman2024robot,nasiri2024smart,te2025artificial,lowen2026towards}.

A collection of a large number of microrobots or communicating agents naturally calls for an analytical description of collective learning in the framework of statistical physics which include additional complexity resulting from memorization and communication processes. Different from usual active matter \cite{ramaswamy2010mechanics,marchetti2013hydrodynamics,ramaswamy2019active} these processes allow the adaptive smart agents to learn prescribed collective properties by adapting their individual dynamical behavior, characterized by a set parameters called `policy'.  From a robotics perspective, turning to a statistical description is also an interesting paradigm shift \cite{ben2023morphological,bredeche2022social,janzen2026active,fersula2026aggregating}.
At the single agent level, statistical physics approaches have recently been used to characterize individual navigation of active particles \cite{liebchen2019optimal,piro2021optimal,piro2022optimal,piro2022efficiency,nasiri2023optimal} or optimal control \cite{floyd2024learning,cocconi2024dissipation,han2025fluctuation}.
Describing collective properties of smart active matter, like dense crowds of pedestrians \cite{echevarria2023body,bonnemain2023pedestrians}
or collective navigation of active units \cite{borra2021optimal,yang2022autonomous}, then requires the development of novel theoretical approaches \cite{ziepke2022multi,jung2025kinetic,ariosto2025replication}, possibly expanding on concepts from stochastic thermodynamics \cite{vansaders2023informational,cocconi2024dissipation} and extending hydrodynamic descriptions \cite{jung2025kinetic,garnier2025hydrodynamics}.

In this paper, we present a general and systematic analytical framework to characterize the dynamics of agents which are able to observe and memorize their environment, communicate information locally and thus adapt their individual policies. The pairwise information exchange of policies between different agents is described using kinetic theory, a technique which has already found useful applications in an active matter context \cite{bertin2006boltzmann,bertin2009hydrodynamic,ihle2011kinetic,bertin2013mesoscopic,ihle2014towards}. Within this framework, we derive a generic evolution equation for the policy distribution to characterize the collective, decentralized learning processes in large assemblies of agents which are homogeneously distributed in space. The present work expands on the first results presented in Ref.~\cite{jung2025kinetic} by providing a more systematic and comprehensive account of the developed framework, emphasizing the separation between the exact derivation steps and the unavoidable approximate closures. The key novel quantity introduced in the present paper is the effective reward which advances previous results both conceptually and also simplifies practical solution of the kinetic theory. We also include here additional ingredients in the general framework which were absent
from Ref.~\cite{jung2025kinetic}, such as the presence of a Langevin noise in the memory dynamics, the possibility of overlapping memory and teaching time scales, as well as the inclusion of more general, multicomponent memory and policy variables.

The paper is organized as follows. The general model of communicating agents with decentralized learning dynamics is introduced in Sec.~\ref{sec:agent:model}.
 Sec.~\ref{sec:kinetic:theory} then presents the kinetic theory framework to describe the statistics of a spatially homogeneous assembly of learning agents. This microscopic theory in then reduced in
Sec.~\ref{sec:evol:policy:dist} to derive a macroscopic statistical description of the decentralized learning dynamics characterized by the time evolution of the policy distribution and the effective reward function. Transforming these expressions into closed equations requires the introduction of closure approximations, which are discussed in Sec.~\ref{sec:policy-closure}.
Two applications to explicit models are presented in Secs.~\ref{sec:AOU} and~\ref{sec:BROWNIAN}, illustrating step by step how the general formalism can be applied to different situations, including overlapping time scales and multidimensional policies.
Finally, the main results, limitations and open issues of the present work are briefly discussed in Sec.~\ref{sec:discussion}, together with the similarities and differences of the learning dynamics with population dynamics in a fitness landscape (as described by simple theoretical models of biological evolution). Potential implications of our work both in robotics and machine learning contexts are also mentioned.

\section{Agent-based model of decentralized learning}
\label{sec:agent:model}

\begin{figure}[b]
    \centering
    \includegraphics[width=0.8\linewidth]{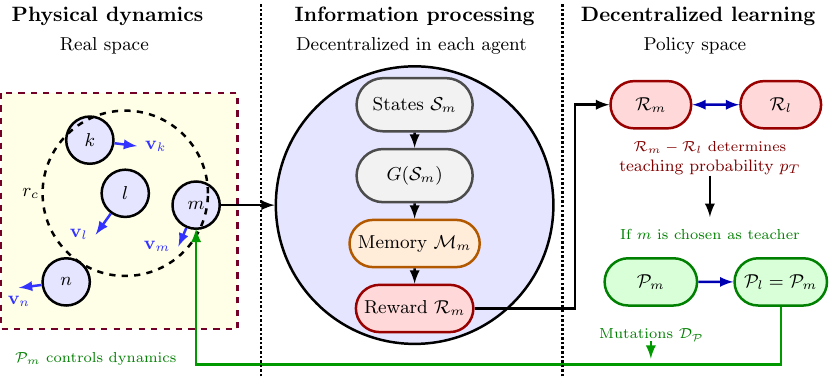}
    \caption{Illustration of the agent-based model. }
    \label{fig:illustration1}
\end{figure}

\subsection{Agents characteristics and physical dynamics}
Our microscopic model consists of $N$ agents located at position $\br$ with velocity $\bm{v}$, in a $d$-dimensional space (often taken as $d=2$). Agents typically represent microrobots which can (i) perceive their environment through sensors measuring for instance external fields, or information related to their neighbors, (ii) process the signal of these sensors and modify their own dynamics accordingly, (iii) exchange information with neighboring agents \cite{bredeche2022social,ben2023morphological}. We assume an overdamped dynamics, characterized for instance by a self-propulsion velocity and/or an active torque, as well as spatial and angular diffusion coefficients, as for usual active particles (although other types of dynamics may be considered too). See Fig.~\ref{fig:illustration1} for an illustration of our microscopic model. 

 A set of parameters characterizing the dynamical rules, called policy $\policy$ (a name suggested by standard robotics nomenclature), can be slowly varied by each agent during the learning process according to some learning rules to be specified below. Each agent has its individual policy, which in general differs from the one of other agents.
 On general ground, the parameters included in the policy $\policy$ may be of two different types: (i) physical parameters themselves (e.g., self-propulsion intensity, active torque, mobility, tumbling rate, etc.),  or (ii) parameters of a function, called controller, which characterizes the dependence of physical parameters on a signal measured by a sensor located on the microrobots. As a simple example of a controller for a microrobot illuminated by a space-dependent light field $I(\br)$, one may consider a function $v_0(I)=v_1(1-I/I_0)$ which relates the agent speed $v_0$ to the locally measured light intensity $I$. In this case, the policy may include the controller parameters $v_1$ and $I_0$ (or possibly only one of them).
 However, in the present work we focus on spatially homogeneous systems and we thus do not consider agents in a spatially varying external field, whose study is deferred to a future work \cite{jung2026inhomogeneous}.

We call $\state$ the dynamical state of agents. In general, the state $\state$ of an agent modeling a microrobot would include all physical degrees of freedom (except position), generically denoted as $\bth$, as well as potential outputs from sensors, denoted as $\bs$, leading to $\state=(\bth,\bs)$.
In practice, since the present paper focuses on spatially homogeneous states, we will not use sensors so that $\state=\bth$ here. Doing so, we assume implicitly that the agent has perfect access to the instantaneous values taken by its physical degree of freedom  $\bth$. However, the sensor notion will need to be kept when dealing with spatially inhomogeneous systems \cite{jung2026inhomogeneous}.

Simple examples of physical dynamics for smart agents modeling small robots may for instance correspond to two-dimensional Active Brownian Particle (ABP), Run-and-Tumble Particle (RTP) or active Ornstein-Uhlenbeck (AOUP) particle dynamics \cite{cates2013when,szamel2014self}. Both ABP and RTP are modeled as point-like particles, at position $\br$, with an overdamped self-propelled dynamics. They are further characterized by the orientation angle $\theta$ of their self-propulsion velocity ${\bm v}=v_0\,{\bm e}(\theta)$, with $v_0$ the self-propulsion speed, and ${\bm e}(\theta)$ the unit vector in direction $\theta$
(here, $\bth$ boils down to $\theta$). The distinction between ABP and RTP lies in the dynamics of the angle $\theta$. For ABP, the angle $\theta$ diffuses with an angular diffusion coefficient $D_R$. For RTP, $\theta$ performs stochastic jumps with rate $\lambda_R$ to a new angle $\theta'$ drawn from a distribution $\psi(\theta'|\theta)$, often taken to be rotationally invariant, i.e., $\psi(\theta'|\theta)=\tilde{\psi}(\theta'-\theta)$.
We generically call $\tau_{\rm phys}$ the characteristic time scale of the physical dynamics. For instance, $\tau_{\rm phys}=D_R^{-1}$ for ABP and $\tau_{\rm phys}=\lambda_R^{-1}$ for RTP. The dynamics of AOUP will be discussed in the first application in Sec.~\ref{sec:AOU}. Beyond these standard examples of active dynamics, it may also be useful to consider, as a toy model, agents with a purely diffusive positional dynamics, see Sec.~\ref{sec:BROWNIAN}. The physical dynamics will, in general, also include physical \emph{interactions} between individual agents.

\subsection{Processed information and memory}
To optimize their behavior agents extract and gather relevant information about the states they attain for a given policy $\policy$. The processed information is denoted as $G(\state)$, and it typically depends on generic properties of the targeted behavior and the specific limitations of the agents.
For instance, if the goal is to reach collective motion along the $x$-direction, $G(\state)$ could be the instantaneous velocity component along $x$.
In general, the function $G$ therefore needs to be chosen such that it allows to differentiate between different policies.

The processed information $G(\state)$ typically fluctuates on the time scale of the physical dynamics. Decision making requires the filtering of those fast fluctuations. Each agent therefore has a memory $\mem$, which averages the processed information $G(\state)$ over a characteristic time scale $\tau_{\mem}=\lambda_\mem^{-1}$.
The dynamics of $\mem$ is given by:
\begin{equation} \label{eq:mem}
\frac{d\mem}{d t} = \lambda_\mem \left(G(\state)-\mem\right) +  \sqrt{2 D_\mem(\policy)}\,\eta(t),
\end{equation}
with $\eta(t)$ a unit Gaussian white noise with zero mean and correlation
$\langle \eta(t)\eta(t') \rangle_{\rm noise} = \delta(t-t')$, 
where $\langle \cdots \rangle_{\rm noise}$ denotes an ensemble average over the noise process, and $\delta(t)$ is the Dirac delta function.
Neglecting the noise term, $\mem(t)$ can be expressed as a convolution of the time-dependent processed information $G(\state)$ with an exponential kernel $\lambda_{\mem}\, e^{-\lambda_{\mem}t}$.
The noise term then adds up fluctuations into the memory, which are smoothed out by the averaging over the time scale $\tau_{\mem}$.
Mathematically speaking, the presence of the noise term occurs if the processed information $G(\state)$ is a function of the degenerate coordinates of the underlying overdamped stochastic process,
like the velocity of a particle with overdamped Langevin dynamics for instance (see Application 2 in Sec.~\ref{sec:BROWNIAN} for an explicit example).
In this case, $G(\state)$ should rather be interpreted as a noise-averaged processed information.
Since fluctuations result from the physical dynamics itself, their amplitude characterized by $D_\mem(\policy)$ generically depends on the policy $\policy$.
In practical applications, fluctuations might also emerge from imperfect memory devices, but we do not explicitly consider such sources of fluctuations in the present manuscript.

In some cases, it may be useful to consider multidimensional observables $G(\state)=\big(G_1(\state),\dots,G_n(\state)\big)$ and memories $\mem=(\mem_1,\dots,\mem_n)$. This situation is described in Appendix~\ref{app:multidimensional}.

\subsection{Policy dynamics}
Each agent quantifies the efficiency of their individual policies through a real-valued reward (or score/fitness) function $\reward(\mem)$, which the agent computes from its own memory $\mem$.
Although the reward function is defined individually for each agent, it may be designed to favor a targeted collective behavior, as discussed below on several examples. 
Agents then tend to maximize their individual reward through information exchange about their policies and respective rewards
(or more precisely, memories allowing them to evaluate their respective rewards).
Upon comparison of their rewards in a pairwise encounter, the agent with the highest reward value is more likely to communicate its policy to the other agent.

The learning dynamics of the model is defined by the following rules. Each agent identifies their neighbors, i.e. all other agents within an interaction radius $r_c$. In this way, a neighbor list of length $\mathcal{N}$ is created containing all possible pairwise interactions for all agents. We then select interaction events by drawing random pairs of agents $i$ and $n$ from the neighbor list with rate $\lambda_T $ $\mathcal{N}$.
 The effective rate of teaching events for an agent $i$ is therefore $\lambda_T$ times the number of its neighbors.  For each selected interaction event, agent $i$ becomes the teacher and agent $n$ the student with a probability $p_T(\reward_i,\reward_n)$ evaluated as
\begin{equation} \label{eq:def:PT}
	p_T(\reward_i,\reward_n) = \frac{1}{2}\left( 1+ \tanh\big[\alpha_T \big(\reward_i - \reward_n \big)\big] \right),
\end{equation}
(where the shorthand notation $\reward_i=\reward(\mem_i)$ and $\reward_n=\reward(\mem_n)$ has been used). Otherwise, the roles are exchanged and agent $n$ becomes the teacher.
Note that only the reward difference matters here.
The policy $\policy_S$ and memory $\mem_S$ of the student are updated by copying the teacher's values $\policy_T$ and memory $\mem_T$ of these variables, namely $\policy^{\text{new}}_S = \policy_T$ and $\mem^{\text{new}}_S = \mem_T$. The parameter $\alpha_T$ quantifies how sharp the probability $p_T$ is as a function of the reward difference. Physically, $\alpha_T$ may also be considered as the uncertainty on reward evaluation and communication.
In these update rules, the student's memory is also updated together with the policy \cite{ben2023morphological}. Updating the memory is not strictly necessary (and might even seem counterintuitive), but it has the advantage that the reward calculated by the student immediately after the teaching interaction is consistent with its updated policy. Otherwise, a transient phase would be required to let the memory, and thus the reward, adapt to the newly taught policy. Directly updating the policy from that of the teacher therefore makes the overall learning dynamics more efficient. We discuss in Appendix~\ref{app:nomemory} how the kinetic theory is modified when memory is not updated during teaching interaction.

Finally, we assume that spontaneous policy changes (here called `mutations' by analogy with biological evolutionary processes) also occur, with a small rate $\Dmut$. The role of these mutations is to maintain at least some amount of variance of policies in the population of agents, since this policy diversity turns out to be key to adaptation (again in analogy to evolution in biology).

The above described adaption cycle is illustrated in Fig.~\ref{fig:illustration1}. The physical dynamics as characterized by the variables $\br$ and $\bth$ is controlled by the policies $\policy$ of the agents and thus determines the emergent memory $\mem$ of each agent. In contrast, the memory $\mem$ does not directly act upon the physical variables and its effect is only indirect in that it determines the reward $\reward$ and thus the policy dynamics.

\section{Kinetic theory of smart agents}
\label{sec:kinetic:theory}
\begin{figure}
    \centering
    \includegraphics[width=0.8\linewidth]{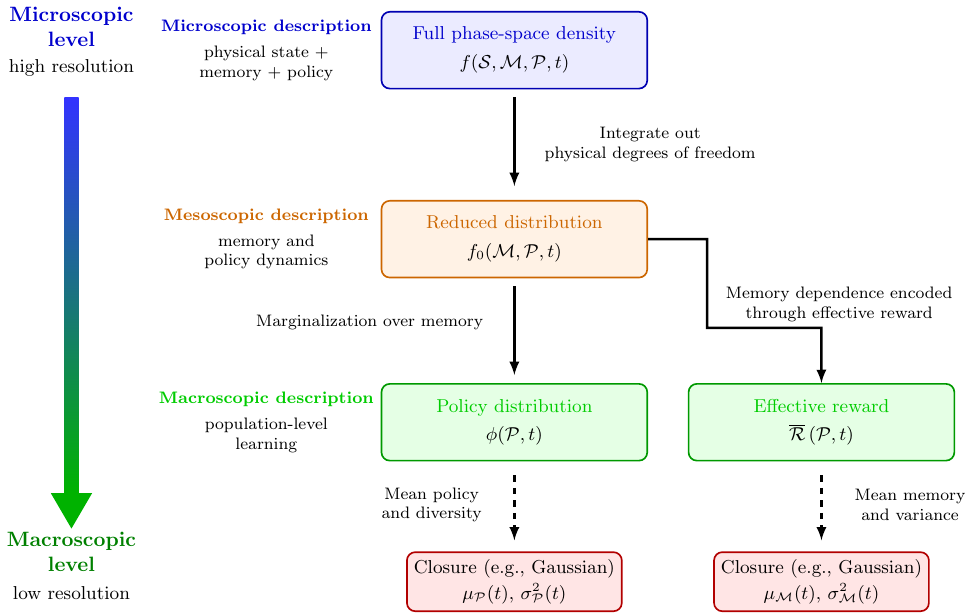}
    \caption{Illustration of the reduction/coarse-graining steps to derive the macroscopic theory of policy dynamics.}
    \label{fig:illustration_KT}
\end{figure}

In this paper, our aim is to systematically derive a continuum description of the learning process taking place in the assembly of agents. In particular, we aim at describing the dynamical evolution of the policy distribution across the population of agents, which is the main statistical characterization of the decentralized learning process described in the above section. The choice of a kinetic theory (Fig.~\ref{fig:illustration_KT}) is natural here  because of the pairwise intermittent nature of the teaching interactions among agents. Yet, we will see that a number of important differences with usual kinetic theories (e.g., of molecular gases, granular gases, or active matter) emerge in the present framework.

\subsection{Evolution equation for the single-agent particle density}
On general grounds, kinetic theories describe the dynamics of systems with pairwise, intermittent interactions in terms of a one-body phase-space density.
Following \cite{jung2025kinetic}, the single-agent phase-space density is in the present model a function $f(\state,\mem,\policy,\br,t)$
of the $\state$, $\mem$, $\policy$ variables, for agents located at position $\br$ at time $t$.
This phase-space density evolves according to the following Boltzmann-type equation:
\begin{equation}
\label{eq:Bolz}
\frac{\p f}{\p t} = I_{\rm phys}[f] + I_{\rm mem}[f] + I_{\rm learn}[f] ,
\end{equation}
where each term on the rhs of Eq.~(\ref{eq:Bolz}) accounts for a specific type of dynamics, classified in physical, memory and learning dynamics.
The term $I_{\rm phys}[f]$ describes the dynamics of the agent's position and of its dynamical state $\state$ (e.g., rotational degrees of freedom, shape for soft robots, etc.).
The physical dynamics is assumed to be overdamped, and may include active forces and torques acting respectively on translational and rotational degrees of freedom.
Rotational dynamics may correspond for instance to rotational diffusion or to run-and-tumble dynamics.
Note that physical interactions like repulsive contact forces may in principle also be present. However, for the sake of simplicity, such physical interactions are neglected
in the present paper, in order to focus on learning interactions. 
Including physical interactions in the form of collisions for instance, would be a natural and probably necessary next step in the development of a kinetic theory of learning.
To briefly give explicit examples of the expression of $I_{\rm phys}[f]$, one has for instance in the case of purely diffusive particles with spatial diffusion coefficient $D$:
\begin{equation} \label{eq:Iphys:Diff}
I_{\rm phys}[f](\br)=D\nabla^2 f(\br),
\end{equation}
while for ABP (see Sec.~\ref{sec:agent:model}):
\begin{equation} \label{eq:Iphys:ABP}
    I_{\rm phys}[f](\br,\theta) = -v_0\,{\bm e}(\theta) \cdot \nabla f(\br,\theta) + D_R \frac{\p^2 f}{\p \theta^2}(\br,\theta).
\end{equation}

The term $I_{\rm mem}[f]$ in Eq.~(\ref{eq:Bolz}) describes the evolution of the single-agent phase-space density $f$ resulting from the dynamics of $\mem$ as defined in Eq.~(\ref{eq:mem}), which can be thought of as a time integrator of the processed information $G(\state)$ with an exponential memory kernel of characteristic time $\tau_{\mem}=\lambda_\mem^{-1}$:
\begin{equation} \label{eq:def:Imem}
I_{\rm mem}[f] = \lambda_\mem \frac{\p }{\p \mem}\left[\left(\mem-G(\state) + \lambda_\mem^{-1} D_\mem(\policy)  \frac{\p }{\p \mem}\right)f\right].
\end{equation}

Finally, the term $I_{\rm learn}[f]$ describes the learning dynamics, which can be decomposed into a teaching contribution $I_{\rm teach}[f]$ describing information exchange (as described below), and a purely random mutation contribution with a (typically small) amplitude $\Dmut$.
If the policy $\policy$ is a one-dimensional real variable, we assume the mutation contribution to be purely diffusive:
\begin{equation} \label{eq:def:Ilearn}
    I_{\rm learn}[f] = I_{\rm teach}[f] + \Dmut \frac{\p^2 f}{\p \policy^2} .
\end{equation}
In some situations, the policy may also take a finite set of discrete values rather than continuous ones. In this case, diffusion in policy space could be represented by stochastic jumps between policy values, with prescribed transition rates $\lambda_{\policy \policy^\prime}$, $ I_{\rm learn}[f] = I_{\rm teach}[f] + \sum_{\policy^\prime} \big(\lambda_{\policy \policy^\prime} f_{\policy'}(t) - \lambda_{\policy^\prime \policy} f_\policy(t)\big).$

In the following, we generically express mutations under the one-dimensional diffusion form as in Eq.~(\ref{eq:def:Ilearn}), keeping in mind possible other forms of the mutation dynamics depending on details of the model. This may include reward-dependent mutation rates \cite{bjedov2003stress,foster2007stress} or multiplicative mutations \cite{komarova2004replicator}. 

The teaching contribution $I_{\rm teach}[f]$ describes information exchange during pairwise interactions between agents, and is thus non-linear in terms of the phase-space density $f$. As stressed above, the degrees of freedom included in the state variable $\state$ are, however, not modified by the learning process and only depend on the physical dynamics. As in other kinetic theories it is thus possible and convenient to express $I_{\rm teach}[f]$ in terms of an effective transition rate $W_{f}(\mem',\policy'|\mem,\policy;\br,t)$ which depends on the phase space density as follows \cite{jung2025kinetic},\\
	\begin{align}
	\hspace{-0.5cm}  
	W_f(\mem',\policy'|\mem,\policy; \br,t) =
    2 \lambda_T \hspace{-0.1cm} \int \hspace{-0.1cm} d\state_2 \hspace{-0.1cm} \int \hspace{-0.1cm} d\mem_2 \hspace{-0.1cm}\int \hspace{-0.1cm} & d \policy_2 \hspace{-0.1cm} \int \hspace{-0.1cm} d\br_2 K(\br_2,\br) p_T\big(\reward(\mem_2),\reward(\mem)\big) f(\state_2,\mem_2,\policy_2,\br_2,t) \delta(\policy^\prime-\policy_{2}) \delta(\mem^\prime-\mem_2).  
	\label{eq:def:WL}
	\end{align}
This transition rate denotes the probability that a tagged agent with policy $\policy$ and $\mem$ learns to adopt the new policy $\policy^\prime$ and $\mem^\prime,$ thus replacing the momentum transfer in typical kinetic theories of active matter \cite{bertin2006boltzmann,bertin2009hydrodynamic,ihle2011kinetic,bertin2013mesoscopic,ihle2014towards}. This teaching transition can occur by exchanging information with another agent $2$ having policy $\policy_2$ and memory $\mem_2.$ The transition depends on the probability of interacting with an agent 2, $\lambda_T K(\br_2,\br) f(\state_2,\mem_2,\policy_2,\br_2,t)$, where the kernel $K(\br_2,\br)=\theta(r_c - |\br_2-\br|)$ is defined by the interaction range $r_c$. 
The fact that the interaction probability is proportional to $f(\state_2,\mem_2,\policy_2,\br_2,t)$ is consistent with the microscopic rule in the agent-based model, where the interaction probability is proportional to the number of neighbors.
The transition rate is weighted by the teaching probability $p_T$, as introduced above. If agent 2 becomes the teacher of the tagged agent, the latter adopts its policy $\policy_2 \rightarrow \policy^\prime $ and memory $\mem_2 \rightarrow \mem^\prime $.

Throughout the remainder of this paper, we focus on spatially homogeneous systems and therefore drop the $\br$ argument from the single-agent phase-space density $f$, assumed to be space independent.
The effective transition rate (\ref{eq:def:WL}) then boils down to,
\begin{equation}
W_{f_0}(\mem',\policy'|\mem,\policy,t) = 2 \lambda_T A_c\, p_T\big( \reward(\mem'),\reward(\mem)\big) \, f_0(\mem',\policy',t),
\label{eq:def:Wf0}
\end{equation}
where the marginal phase-space density $f_0(\mem,\policy,t)$ is defined as
\begin{equation} \label{eq:def:f0}
    f_0(\mem,\policy,t) = \Int d\state f(\state,\mem,\policy,t).
\end{equation}
In Eq.~(\ref{eq:def:Wf0}), $A_c$ is the $d$-dimensional `volume' enclosed by the interaction range $r_c$ (i.e., $A_c=2r_c$ for $d=1$ and $A_c=\pi r_c^2$ for $d=2$).
With the effective transition rate given in Eq.~(\ref{eq:def:Wf0}), $I_{\rm teach}[f]$ formally takes a master equation form:
\begin{align}
I_{\rm teach}[f] =& \Int d\mem'd\policy' \big[ W_{f_0}(\mem,\policy|\mem',\policy',t)f(\state,\mem',\policy',t)-W_{f_0}(\mem',\policy'|\mem,\policy,t)f(\state,\mem,\policy,t) \big] \nonumber\\
=& 2\lambda_T A_c \left[f_0(\mem,\policy)\!\Int d\mem' p_T(\reward,\reward')\!\Int d\policy' f(\state,\mem',\policy') - f(\state,\mem,\policy)\Int d\mem' p_T(\reward',\reward) \Int d\policy' f_0(\mem',\policy') \right], \label{eq:Iteach}
\end{align}
where to lighten notations, the $t$ dependence has been omitted in the last line, and we have used the shorthand notation $\reward=\reward(\mem)$ and $\reward'=\reward(\mem')$.

\subsection{Marginal phase-space densities and averages}

The main goal of the following subsections is to derive coarse-grained equations of motion describing the time-evolution of the statistics of policies $\policy$. This coarse-graining procedure is illustrated in Fig.~\ref{fig:illustration_KT}. We have already introduced the marginal density $f_0$ obtained by integration of $f$ over $\state=\{\bth,\bs\}$. More systematically, we will deal with the following marginal densities:
\begin{align}
\label{eq:marg}
\Phi(\policy,t) =& \Int d\mem \, f_0(\mem,\policy,t) = \Int  d\state d\mem \,f(\state,\mem,\policy,t) = \rho(t) \phi(\policy,t),\\
\rho(t) =& \Int d\policy \, \Phi(\policy,t) = \Int d\state d\mem d\policy \, f(\state,\mem,\policy,t).
\label{eq:marg:rho}
\end{align}
Importantly, due to the assumed spatial homogeneity, the density $\rho$ of agents is space-independent, and is thus constant in time, $\rho=N/V$ (with $V$ the volume of the system) due to the conservation of the number of agents.
Eq.~(\ref{eq:marg:rho}) thus boils down to the normalization condition $\Int d\policy \,\Phi(\policy,t)=\rho$, with $\rho$ a given constant density of agents.

In addition, the average over $\state$ of an arbitrary observable $u(\state,\mem,\policy,t)$ is defined as 
\beq \label{eq:mom0}
\avS{u}(\mem,\policy,t) = f_0(\mem,\policy,t)^{-1} \Int d\state \, f(\state,\mem,\policy,t) \, u(\state,\mem,\policy,t),
\eeq
and averages over $\mem$ and $\policy$ of an observable $U(\mem,\policy,t)$ are defined as
\begin{align}
\label{eq:mom}
\avM{U}(\policy,t) =& \Phi(\policy,t)^{-1} \Int d\mem \, f_0(\mem,\policy,t)\, U(\mem,\policy,t), \\
\avMP{U}(t) =& \rho^{-1} \Int d\mem d\policy\, f_0(\mem,\policy,t)\, U(\mem,\policy,t)
= \rho^{-1} \Int d\policy\, \Phi(\policy,t) \, \avM{U}(\policy,t).
\end{align}
Note that one may in particular choose $U(\mem,\policy,t)=\avS{u}(\mem,\policy,t)$, although other choices may also be relevant.
Additionally, when $U(\mem,\policy,t)$ is independent of $\mem$, i.e., $U(\mem,\policy,t)\equiv U(\policy,t)$, then $\avM{U}(\policy,t)=U(\policy,t)$,
so that $\avMP{U}(t)$ simply denotes in this case an average over $\policy$. To characterize the policy statistics, we will use in the following the policy distribution $\phi(\policy,t)=\Phi(\policy,t)/\rho$,
which is normalized according to $\Int d\policy \,\phi(\policy,t)=1$.

\subsection{Integrating out the state variables $\state$}
The first coarse-graining step is to integrate out the dependence on the state variable $\state=\bth$ 
which is considered as a fast variable. Integrating the Boltzmann equation, Eq.~(\ref{eq:Bolz}), over $\state$, one finds (omitting the explicit $t$ dependence):
\bea
\p_t f_0(\mem,\policy) &=& \int d\state \, I_{\rm phys}[f] + \lambda_\mem \frac{\p}{\p \mem}\left[\left(\mem-\avS{G}(\mem,\policy) + \lambda_\mem^{-1} D_\mem(\policy)  \frac{\p }{\p \mem} \right) f_0(\mem,\policy) \right] \nonumber\\
& & + 2\lambda_T A_c f_0(\mem,\policy) \Int d\mem' \Delta_T(\reward,\reward') \Int d\policy' f_0(\mem',\policy') + \Dmut \frac{\p^2 f_0(\mem,\policy)}{\p \policy^2}
\label{eq:dt:f0:v1}
\eea
with $\Delta_T(\reward,\reward')=p_T(\reward,\reward') - p_T(\reward',\reward)$. We emphasize that following
the general definition in Eq.~(\ref{eq:mom0}) $\avS{G}(\mem,\policy)$ explicitly depends on the memory $\mem$ and the policy $\policy$.

The physical contribution $\int d\state \, I_{\rm phys}[f]$ in general depends on the details of the physical dynamics, e.g., the detailed definition of the physical variables $\bth$, or for different types of active agents, whether the orientational dynamics corresponds for instance to angular diffusion or to run-and-tumble dynamics. Yet, by definition, the physical dynamics only generates fluxes in the physical phase space $(\br,\state)$.
Hence the integral $\int d\br d\state \, I_{\rm phys}[f]$ over the whole physical phase space vanishes.
Under the generic spatial homogeneity assumption made in the present paper, one thus has
$\int d\state \, I_{\rm phys}[f]=V^{-1} \int d\br d\state \, I_{\rm phys}[f]=0$, where $V$ is the system volume.

To make analytical progress, it will be useful in the following to assume $\Delta_T(\reward,\reward')=\alpha_T (\reward-\reward')$, which corresponds to a linearization of the $\tanh$ in Eq.~(\ref{eq:def:PT}) in the small $\alpha_T$ limit. This approximation is analogous in spirit to deep linear neural networks, where nonlinear activation functions are replaced by linear mappings to obtain analytically tractable learning dynamics~\cite{saxe2013exact}.
Under the above assumptions, and introducing the effective teaching rate,
\begin{equation} \label{eq:def:tildelambda}
\tilde\lambda_T = 2\lambda_T \alpha_T A_c,
\end{equation}
Eq.~(\ref{eq:dt:f0:v1}) reads,
\beq
\label{eq:dt:f0:v2}
\p_t f_0(\mem,\policy) = \lambda_\mem \frac{\p}{\p \mem}\left[\left(\mem- \avS G(\mem,\policy)\right) f_0\right] + D_\mem(\policy) \frac{\p^2 f_0}{\p \mem^2} 
 + \tilde\lambda_T \rho f_0 \left(\reward(\mem) - \avMP{\reward} \right) + \Dmut \frac{\p^2 f_0}{\p \policy^2} .
\eeq 
This equation will serve in the following as a starting point to derive an evolution equation for the policy distribution.

In the following, we present the general formalism in the one-dimensional form based on Eq.~(\ref{eq:dt:f0:v2}).
The case of multidimensional policy and memory is described in Appendix~\ref{app:multidimensional}.
This multidimensional case will be useful to deal with the two-dimensional memory and policy model considered in Sec.~\ref{sec:AOU}.


\section{Time evolution of policy distribution and effective reward}
\label{sec:evol:policy:dist}

In this section we will disentangle the complex dynamics of the memory and the policy. Doing so, we pinpoint the role of an effective reward $\Reff(\policy,t)$ emerging from the memory dynamics which then dictates the policy dynamics.

\subsection{Integrating out the memory $\mem$: dynamics of the policy distribution and moments}

\subsubsection{Dynamics of the policy distribution $\phi(\policy,t)$}

Integrating Eq.~(\ref{eq:dt:f0:v2}) over the memory, one finds the following evolution equation for $\phi(\policy,t)$,
\beq
	\p_t \phi(\policy) = \tilde\lambda_T \,\rho\,  \left( \Reff (\policy)- \avMP{\reward}  \right)\phi(\policy) + \Dmut \frac{\p^2 \phi(\policy)}{\p \policy^2} , \label{eq:dt:phi:homog_v1}
\eeq
where
\bea
\avM{\reward}(\policy) &=& [\rho\phi(\policy)]^{-1} \int d\mem\, \reward(\mem) f_0(\mem, \policy) ,\label{eq:def:Reff} \\
\avMP{\reward} &=& \int d\policy\, \avM{\reward}(\policy) \phi(\policy). \label{eq:def:Rdbar}
\eea
Equation~(\ref{eq:dt:phi:homog_v1}) shows that the effective policy-dependent reward $\avM{\reward}(\policy,t)$
plays a key role in setting the dynamics of the policy distribution $\phi(\policy,t)$: policies with effective reward $\Reff(\policy,t)$ above the current average are enhanced by the dynamics, while the others are gradually suppressed. The teaching prefactor $\tilde \lambda_T \rho$ is proportional to $\lambda_T$ and to the mean number of neighbors $\rho A_c$ [see Eq.~(\ref{eq:def:tildelambda})], and is thus consistent with the teaching rate in the microscopic model. The mutation rate $\Dmut$ controls the diffusive exploration of policy space.

Importantly, Eq.~(\ref{eq:dt:phi:homog_v1}) is similar to the Replicator equation with continuous policy space \cite{taylor1978evolutionary} known from evolutionary game theory \cite{hofbauer2003evolutionary} (see Sec.~\ref{sec:evolution} for a more detailed discussion).
However, at variance with macroscopic evolutionary theories in which the effective reward is a predefined input, our coarse-grained theory starts from the microscopic physical and learning dynamics and involves the yet unknown policy-dependent effective reward $\Reff(\policy,t)$ defined in Eq.~(\ref{eq:def:Reff}), from which the mean reward $\avMP{\reward}(t)$ may be deduced. Its determination will be discussed in Sec.~\ref{sec:eff-reward}. At this stage, the effective reward $\Reff(\policy,t)$ still depends on the distribution $f_0(\mem, \policy,t)$ and not only on $\phi(\policy,t)$, so that Eq.~(\ref{eq:dt:phi:homog_v1}) is in general not closed.

\subsubsection{Dynamics of the policy mean and variance}

Instead of the dynamics for the full distribution $\phi(\policy)$, one might only be interested in its first cumulants, the average policy, $\meanpol=\avMP{\policy}$, and the policy variance, $\sigma_\policy^2=\avMP{\delta\policy^2}$, with $\delta\policy=\policy-\avMP{\policy}$ ($\sigma_\policy^2$ is often referred to as the \emph{diversity}). 
Higher order cumulants or moments could also be dealt with in a similar way, but we do not consider them explicitly in this paper.
For definiteness, we assume that the policy $\policy$ is defined over the interval $[\policy_{\min},\policy_{\max}]$, where the bounds $\policy_{\min}$ and $\policy_{\max}$
may be either finite or infinite. 
Importantly, $\partial \phi(\policy)/\partial \policy=0$ at a finite bound $\policy_{\min}$ or $\policy_{\max}$ because the diffusive flux $-\Dmut \partial \phi(\policy)/\partial \policy$ vanishes.
Multiplying Eq.~(\ref{eq:dt:phi:homog_v1}) by either $\policy$ or $(\delta\policy)^2$ and integrating over $\policy$, one finds 
\begin{align}
\frac{d}{dt} \meanpol &= \tilde{\lambda}_T \rho\, \mathrm{Cov}_{\policy}\big(\policy,\Reff(\policy)\big) - \Dmut \left[ \phi(\policy)\right]^{\policy_{\max}}_{\policy_{\min}}, \label{eq:dP:dt} \\
\frac{d}{dt}  \sigma_\policy^2  &= \tilde{\lambda}_T \rho\, \mathrm{Cov}_{\policy}(\delta\policy^2, \Reff\big(\policy)\big) + 2 \Dmut
- 2\Dmut \left[\delta\policy\,\phi(\policy)\right]^{\policy_{\max}}_{\policy_{\min}},
\label{eq:dsigma2P:dt}
\end{align}
where $\mathrm{Cov}_{\policy}(X,Y) = \avMP{XY} - \avMP{X}\avMP{Y}$ is the covariance of $X$ and $Y$ over $\policy$.
In consequence, if the policy distribution is unbounded, mutations enter the time-evolution of the diversity via the additive term $2 \Dmut$ only.
More generally, terms proportional to the mutation rate $\Dmut$ cannot be neglected in the dynamics of $\meanpol$ and $\sigma_\policy^2$ even if $\Dmut$ is small,
since the covariance terms may a priori be arbitrarily small.

In the following sections, we proceed to determine the effective reward $\Reff(\policy,t)$, which appears both in the dynamics of the policy distribution $\phi(\policy)$ and of its mean and variance.

\subsection{Determination of the effective reward $\Reff(\policy,t)$}
\label{sec:eff-reward}

\subsubsection{Generic dynamics of the effective reward $\Reff(\policy,t)$}
\label{sec:eff-reward:dyn}

We start by considering the dynamics of an arbitrary function of the memory,
$\mathcal{A}(\mem)$, averaged over the distribution $f_0(\mem,\policy)$ using Eq.~(\ref{eq:mom}), since such an evolution equation for a generic observable will be useful below on several occasions. The time-evolution of $\avM{\mathcal{A}}(\policy,t)=\avM{\mathcal{A}(\mem)}$ is obtained from the relation,
\begin{equation}
\phi(\policy)\, \p_t \avM{\mathcal{A}}(\policy) = - \avM{\mathcal{A}}(\policy) \p_t \phi(\policy) + \rho^{-1} \int d\mem\, \mathcal{A}(\mem) \p_t f_0(\mem,\policy) ,
\label{eq:dt:R:mem:v0}
\end{equation}
combined with Eqs.~(\ref{eq:dt:f0:v2}, \ref{eq:dt:phi:homog_v1}). Denoting $\mathrm{Cov}_{\mem|\policy}(X,Y)=\avM{XY}-\avM{X}\avM{Y}$ the covariance of generic observables $X$ and $Y$ with respect to $\mem$ (for a given $\policy$), the evolution equation for $\avM{\mathcal{A}}(\policy,t)$ reads,
\begin{align}
\p_t \avM{\mathcal{A}}(\policy) &=  \lambda_\mem \avM{\mathcal{A}'(\mem) \left(\avS{G}(\mem,\policy)-\mem\right)} + D_\mem(\policy) \avM{\mathcal{A}''(\mem)}  
+ \tilde\lambda_T \,\rho\, \mathrm{Cov}_{\mem|\policy}(\mathcal{A},\reward)
\nonumber \\
&\quad + \Dmut \frac{\p^2 \avM{\mathcal{A}}(\policy)}{\p \policy^2} + 2 \Dmut \frac{\p \ln \phi(\policy)}{\p \policy} \, \frac{\p \avM{\mathcal{A}}(\policy)}{\p \policy} ,
\label{eq:dt:R:mem}
\end{align}
where $\mathcal{A}'$ and $\mathcal{A}''$ denote the first and second derivatives of $\mathcal{A}$ with respect to $\mem$, respectively.
Replacing $\mathcal{A}$ by $\reward$ leads to a generic evolution equation for the effective reward $\Reff(\policy,t)$.
However, one faces several difficulties when trying to determine $\Reff(\policy,t)$ using Eq.~(\ref{eq:dt:R:mem}) with $\mathcal{A}(\mem)=\reward(\mem)$.
First of all, Eq.~(\ref{eq:dt:R:mem}) involves $\phi(\policy)$, which is inconvenient as one would rather like to first determine $\Reff(\policy,t)$ and then plug it into the evolution equation
(\ref{eq:dt:R:mem:v0}) for $\phi(\policy)$, rather than jointly solving coupled equations for $\Reff(\policy,t)$ and $\phi(\policy)$.
Although by now we keep the terms proportional to $\Dmut$ in Eq.~(\ref{eq:dt:R:mem}) for the sake of generality, we will eventually get rid of these terms by assuming $\Dmut$ to be very small and thereby get rid of this first difficulty. A second difficulty is that Eq.~(\ref{eq:dt:R:mem}) requires the knowledge of $\avS{G}(\mem,\policy,t)$, which is yet unknown. As illustrated in the applications
discussed in Secs.~\ref{sec:AOU} and~\ref{sec:BROWNIAN}, $\avS{G}(\mem,\policy,t)$ depends on the specific microscopic physical dynamics at stake,
and it can be more easily determined by simultaneously taking the average over $\state$ and $\mem$ when evaluating the covariance terms involving
$\avS{G}(\mem,\policy,t)$ in Eq.~(\ref{eq:dt:R:mem}); see below for a more detailed discussion.
At the present level of generality, one may formally expand $\avS{G}(\mem,\policy,t)$ as polynomials in $\mem$ with $\policy$-dependent coefficients.
Finally, even under these simplifying assumptions, Eq.~(\ref{eq:dt:R:mem}) with $\mathcal{A}=\reward$ remains unclosed due to presence of terms like $\avM{\mathcal{\reward}'(\mem)\, \mem^n}$.
Such a closure issue can be more easily dealt with when considering equations for moments of $\mem$, i.e., $\mathcal{A}(\mem)=\mem^k$, since in this case a term like
$\avM{\mathcal{A}'(\mem)\, \mem^n}$ is proportional to the moment $\avM{\mem^{n+k-1}}$.
In the following, we therefore express $\Reff(\policy,t)$ in terms of low order moments of $\mem$, under some appropriate approximation scheme.

\subsubsection{Expansion of the effective reward $\Reff(\policy,t)$ in terms of moments of $\mem$}

Writing the marginal density $f_0(\mem,\policy)$ as $f_0(\mem,\policy)= \rho\, Q(\mem|\policy) \phi(\policy)$, by introducing the probability distribution $Q(\mem|\policy)$ of $\mem$ conditioned to a given value of $\policy$, one finds from Eq.~(\ref{eq:def:Reff}) that 
\begin{equation}
\Reff(\policy)=\int d\mem \, \reward(\mem) \, Q(\mem|\policy) .
\end{equation}
We now assume that the physical time scale $\tau_{\mathrm{phys}}$ is much shorter than both the memory and teaching time scales $\tau_\mem=\lambda_{\mem}^{-1}$ and $\tau_T=\lambda_T^{-1}$.
In this regime, the memory is integrated over a time interval much longer than the time scale of typical physical fluctuations, and policy changes occur slowly enough so as not to perturb too much the memory statistics.
Following the law of large numbers, one thus expects the conditional memory distribution $Q(\mem|\policy)$ to be narrowly peaked around the average value $\meanmem(\policy)=\avM{\mem}(\policy)$,
with a small variance $\sigma_{\mem}^2(\policy)=\avM{\mem^2}(\policy)-\avM{\mem}^2(\policy)$ of memory fluctuations.
In this small memory fluctuations limit, the effective reward $\Reff(\policy)$ can generically be expanded as
\begin{equation} \label{eq:Reff:approx}
    \Reff(\policy) \approx \reward\big(\meanmem(\policy)\big) + \frac{1}{2} \sigma_{\mem}^2(\policy) \reward''\big(\meanmem(\policy)\big) .
\end{equation}
For a linear or quadratic reward $\reward(\mem)$, the above equation is exact without assuming that $\sigma_{\mem}^2(\policy)$ is small, so that the time scale separation assumption between physical and memory time scales is not required in this case.

The cases of linear or quadratic reward are of particular interest, since they are mathematically simple and are often sufficient to describe basic applications.
A linear reward $\reward(\mem)=\mem$ is useful for instance to maximally harvest a spatially inhomogeneous resource, like a light field in the microrobots experiment of Refs.~\cite{ben2023morphological,fersula2026aggregating}. Since such a set up requires spatial inhomogeneities, we do not consider it in this paper. 
By contrast, encoding a target value $\mem_\odot$ can be conveniently done by choosing a quadratic reward $\reward(\mem)=\reward_\odot-(\mem-\mem_\odot)^2$, and this target case is compatible with the spatial homogeneity assumption. In this case, the effective reward $\Reff(\policy)$ reads,
\begin{equation}
    \Reff(\policy) = \reward\big(\meanmem(\policy)\big) - \sigma_{\mem}^2(\policy) . \label{eq:Reff_target}
\end{equation}
In consequence, the value $\policy^*$ maximizing $\Reff(\policy)$ yields $\meanmem(\policy^*)=\mem_\odot+\delta\mem^*$, with
\begin{equation}
    \delta\mem^* = - \frac{1}{2} \, \frac{\partial \sigma_\mem^2/\partial \policy_{|\policy^*}}{\partial \meanmem/\partial \policy_{|\policy^*}}. \label{eq:delta_mem}
\end{equation}
Hence there is a generic shift of the mean memory reached at the optimal policy with respect to the target $\mem_\odot$ due to the policy dependence of the memory fluctuations.
In other words, the learning procedure generically leads to a slightly biased and suboptimal result with respect to the input reward function $\reward(\mem)$ if the amplitude of memory fluctuations depends on policy.

At this stage, we have thus obtained an expression of the effective reward $\Reff(\policy)$, Eq.~(\ref{eq:Reff:approx}), in terms of the mean $\meanmem$ and variance $\sigma_\mem^2$ of the memory.
We thus need to study the dynamics of $\meanmem$ and $\sigma_\mem^2$ to be able in practice to express $\Reff(\policy)$ from Eq.~(\ref{eq:Reff:approx}). 

\subsubsection{Dynamics of the memory mean and variance}
\label{sec:memory-closure}
To do so, we write down explicitly the evolution equations for the mean $\meanmem(\policy)=\avM{\mem}$ and variance $\sigma_\mem^2(\policy)=\avM{\mem^2}-\avM{\mem}^2$ of the memory $\mem$, replacing $\mathcal{A}(\mem)$ by $\mem$ and $\mem^2$ into the general Eq.~(\ref{eq:dt:R:mem}). We find,
\begin{align}
\p_t \meanmem(\policy) &= \lambda_\mem \left(\avM{\avS{G}}(\policy) - \meanmem(\policy)) \right) + \tilde{\lambda}_T \rho\, \mathrm{Cov}_{\mem|\policy}\big(\mem,\reward(\mem)\big) + \Dmut \frac{\p^2 \meanmem}{\p \policy^2} + 2 \Dmut \frac{\p \ln \phi}{\p \policy} \frac{\p \meanmem}{\p \policy}, \label{eq:dt:meanmem}\\
\p_t \sigma_\mem^2(\policy) &= 2\lambda_\mem \left[\mathrm{Cov}_{\mem|\policy}\big(\mem,\avS{G}(\mem,\policy)\big) - \sigma_\mem^2(\policy) \right] + 2D_\mem(\policy)
+ \tilde{\lambda}_T \rho\, \mathrm{Cov}_{\mem|\policy}\big(\delta\mem^2, \reward(\mem)\big)\nonumber \\
&\quad + \Dmut \frac{\p^2 \sigma_\mem^2}{\p \policy^2} + 2\Dmut \left( \frac{\p \meanmem}{\p \policy} \right)^2
+ 2 \Dmut \frac{\p \ln \phi}{\p \policy} \frac{\p \sigma_\mem^2}{\p \policy},  \label{eq:dt:varmem}
\end{align}
where $\delta\mem = \mem -\mu_\mem$.
As a further simplification, we will assume from now on (as anticipated above) that the mutation rate $\Dmut$ remains much smaller than the memory rate $\lambda_{\mem}$ and teaching rate $\tilde{\lambda}_T \rho$. Hence terms proportional to $\Dmut$ can be neglected in Eqs.~(\ref{eq:dt:meanmem},\ref{eq:dt:varmem}) as compared to other terms, yielding
(we drop explicit arguments to lighten notations),
\begin{align}
\p_t \meanmem  &= \lambda_\mem \left(\avM{\avS{G}} - \meanmem \right) + \tilde{\lambda}_T \rho\, \mathrm{Cov}_{\mem|\policy}(\mem,\reward), \label{eq:dm:dt} \\
\p_t \sigma_\mem^2 &= 2\lambda_\mem \left(\mathrm{Cov}_{\mem|\policy}(\mem,\avS{G}) - \sigma_\mem^2 \right) + 2D_\mem
+ \tilde{\lambda}_T \rho\, \mathrm{Cov}_{\mem|\policy}(\delta\mem^2, \reward). \label{eq:dsigma2:dt}
\end{align}
This simplification applies only to the dynamics of the memory moments, and does not amount to setting $\Dmut=0$ throughout, as $\Dmut$ plays a key role in the dynamics of the policy, as seen from Eq.~(\ref{eq:dt:phi:homog_v1}) where the diffusion term $\Dmut\, \p^2 \phi/\p \policy^2$ is essential to maintain a policy diversity in the population of agents.
In addition, in the limit of well-separated memory and teaching time scales (i.e., $\tilde{\lambda}_T \rho \ll \lambda_\mem$), the teaching terms become negligible in
Eqs.~(\ref{eq:dm:dt}) and (\ref{eq:dsigma2:dt}), which therefore boil down in this limit to
\begin{align}
\p_t \meanmem  &= \lambda_\mem \left(\avM{\avS{G}} - \meanmem \right), \label{eq:dm:dt:noteach} \\
\p_t \sigma_\mem^2 &= 2\lambda_\mem \left(\mathrm{Cov}_{\mem|\policy}(\mem,\avS{G}) - \sigma_\mem^2 \right) + 2D_\mem. \label{eq:dsigma2:dt:noteach}
\end{align}
In the following, we always assume that the effect of $\Dmut$ is negligible on the memory dynamics as in Eqs.~(\ref{eq:dm:dt}) and (\ref{eq:dsigma2:dt}).
Regarding the impact of the teaching dynamics on the memory dynamics, we will consider both the case of perfect time scale separation $\tilde{\lambda}_T \rho \ll \lambda_\mem$
and the case of overlapping time scales, when teaching terms cannot be neglected in the memory dynamics. In the application sections \ref{sec:AOU} and \ref{sec:BROWNIAN} below, we may thus use either Eqs.~(\ref{eq:dm:dt},\ref{eq:dsigma2:dt}) or Eqs.~(\ref{eq:dm:dt:noteach},\ref{eq:dsigma2:dt:noteach}) depending on the assumption made regarding time scale separation.

In practical applications, the quantity $\avS{G}(\mem,\policy)$ may be difficult to evaluate because it involves an average over $\state$ conditioned to a given value of $\mem$ and
$\policy$. While conditioning over $\policy$ is straightforward as $\policy$ is a parameter of the dynamics of $\state$, the conditioning over $\mem$ is difficult to implement in practice because $\mem$ is itself a time integral of $G(\state)$, as seen from the formal solution of Eq.~(\ref{eq:mem}):
\begin{equation}
    \mem(t) = \mem(0) \, e^{-\lambda_{\mem} t} + \int_0^{t} dt'\, G\big(\state(t')\big)\, e^{-\lambda_{\mem} (t-t')}.
\end{equation}
The determination of $\mathrm{Cov}_{\mem|\policy}(\mem,\avS{G})$ may thus be practically easier by rewriting this covariance in terms of joint averages over $\state$ and $\mem$, namely,
\begin{equation} \label{eq:covSM:MG}
    \mathrm{Cov}_{\mem|\policy}(\mem,\avS{G}) = \avM{\avS{\mem G(\state)}} - \avM{\mem} \, \avM{\avS{G(\state)}}.
\end{equation}
The joint averages $\avM{\avS{\mem G(\state)}}$ 
and $\avM{\avS{G(\state)}}$ can then be evaluated explicitly in specific models, as illustrated in Secs.~\ref{sec:AOU} and \ref{sec:BROWNIAN}.
It follows that the closure of Eqs.~(\ref{eq:dm:dt}) and (\ref{eq:dsigma2:dt}) in terms of $\meanmem$ and $\sigma_\mem^2$ is model-dependent, and has to be performed on a
case-by-case basis.

\subsubsection{Gaussian ansatz for the conditional distribution $Q(\mem|\policy)$}
In cases when the conditional average $\avS{G}(\mem,\policy)$ can be evaluated explicitly from the microscopic dynamics, it becomes possible to generically write Eqs.~(\ref{eq:dm:dt}) and (\ref{eq:dsigma2:dt}) in a closed form, that is, to formally evaluate $\avM{\avS{G}}$ and the covariance terms appearing in Eqs.~(\ref{eq:dm:dt}) and (\ref{eq:dsigma2:dt}) as a function of $\meanmem$ and $\sigma_\mem^2$.
This requires an additional assumption on the conditional distribution $Q(\mem|\policy)$ of the memory $\mem$ for a given policy $\policy$.
Consistently with the above assumption that $Q(\mem|\policy)$ is narrowly peaked when $\tau_{\rm phys}\ll \tau_{\mem}$, we assume based on the Central Limit Theorem that $Q(\mem|\policy)$ takes a Gaussian form:
\begin{equation}
    Q(\mem|\policy) = \frac{1}{\sqrt{2\pi \sigma_{\mem}^2(\policy)}} \, \exp\left( -\frac{\big(\mem-\meanmem(\policy)\big)^2}{2\sigma_{\mem}^2(\policy)} \right).
    \label{eq:gaussian:mem}
\end{equation}
This Gaussian ansatz is convenient, since all average quantities (e.g., all moments $\avM{\mem^n}(\policy)$ with $n>2$) can be expressed in terms of the first two cumulants $\meanmem(\policy)$ and $\sigma_{\mem}^2(\policy)$.
The covariances appearing in Eqs.~(\ref{eq:dm:dt}) and (\ref{eq:dsigma2:dt}) may then be computed explicitly using the Gaussian ansatz (\ref{eq:gaussian:mem}).
Alternatively, one may transform the covariances using the Stein identities for Gaussian variables: $\mathrm{Cov}_{\mem|\policy}(\mem, \reward) = \sigma_\mem^2 \avM{\reward'(\mem)}$, $\mathrm{Cov}_{\mem|\policy}(\delta\mem^2, \reward) = \sigma_\mem^4 \avM{\reward''(\mem)}$ and $\mathrm{Cov}_{\mem|\policy}(\mem, \langle G\rangle ) = \sigma_\mem^2 \avM{\partial_\mem \langle G\rangle}$.
The equations for the two first cumulants of $\mem$, Eqs.~(\ref{eq:dm:dt}) and (\ref{eq:dsigma2:dt}), are now closed and read, within the Gaussian approximation, as: 
\begin{align}
\p_t \meanmem &= \lambda_\mem \left( \avM{\avS{G}} - \meanmem \right) + \tilde{\lambda}_T \rho \, \avM{\reward'(\mem)} \sigma_\mem^2, \label{eq:dm:dt:gauss} \\
\p_t \sigma_\mem^2 &= 2\lambda_\mem \left( \avM{\partial_\mem\avS{G}} - 1\right) \sigma_\mem^2 + 2D_\mem(\policy) + \tilde{\lambda}_T \rho \, \avM{\reward''(\mem)} \sigma_\mem^4  .  \label{eq:dsigma2:dt:gauss}
\end{align}
These closed equations, together with Eq.~(\ref{eq:Reff:approx}) determine the dynamics of the effective reward $\Reff(\policy)$, and thereby that of the policy distribution $\phi(\policy)$ through Eq.~(\ref{eq:dt:phi:homog_v1}). In the absence of time scale separation between the physical and memory time scales $\tau_{\rm phys}$ and $\tau_{\mem}$, one may still use the Gaussian ansatz (\ref{eq:gaussian:mem}), but it becomes in this case an uncontrolled approximation.
In addition, $\sigma_\mem^2$ would no longer be small in this case.

\subsubsection{Summary of the set of equations used in the application section}
Altogether, the dynamics for the policy distribution, namely Eq.~(\ref{eq:dt:phi:homog_v1}), with definitions~(\ref{eq:def:Reff}),~(\ref{eq:def:Rdbar}), together with the approximate
expression~(\ref{eq:Reff:approx}) of the effective reward and the simplified dynamics for the first two moments of the memory $\mem$ as given in Eqs.~(\ref{eq:dm:dt}) and (\ref{eq:dsigma2:dt}), represent the basis of our kinetic theory analysis. We recapitulate these equations below for the sake of clarity, following the style of Fig.~\ref{fig:illustration_KT}.

\begin{tikzpicture}[
	>=latex,
	font=\small,
	node distance=1.8cm,	
	level3/.style={
		rectangle,
		rounded corners,
		draw=green!60!black,
		thick,
		fill=green!10,
		minimum width=4.5cm,
		minimum height=1.3cm,
		align=center
	},
	statbox/.style={
		rectangle,
		rounded corners,
		draw=red!70!black,
		thick,
		fill=red!10,
		minimum width=3.2cm,
		minimum height=1cm,
		align=center
	},
	arrow/.style={->, thick},
	note/.style={align=center,font=\footnotesize}
	]
	\node[level3] (reff) at (8.2,-2.5)
	{
		\textcolor{green!80!black}{Effective reward}\\[1mm]
		Eq.~(\ref{eq:Reff:approx}): $
		\Reff(\policy) = \reward\big(\meanmem(\policy)\big) + \frac{1}{2} \sigma_{\mem}^2(\policy) \reward''\big(\meanmem(\policy)\big)
		$
	};
	\draw[arrow, dashed, very thick]
	(7.7,-3.2) -- (7.7,-4.4);
	\node[statbox] (memorymom) at (7.7,-5.2)
	{
		Closure\\[0.5mm]
\hspace{-4.0cm}	Eq.~(\ref{eq:dm:dt}):	$
\p_t \meanmem =  \lambda_\mem \left(\avM{\avS{G}} - \meanmem \right) + \tilde{\lambda}_T \rho\, \mathrm{Cov}_{\mem|\policy}(\mem,\reward) $ \\
Eq.~(\ref{eq:dsigma2:dt}):	$\p_t \sigma_\mem^2 = 2\lambda_\mem \left(\mathrm{Cov}_{\mem|\policy}(\mem,\avS{G}) - \sigma_\mem^2 \right) + 2D_\mem(\policy) + \tilde{\lambda}_T \rho\, \mathrm{Cov}_{\mem|\policy}(\delta\mem^2, \reward)
		$
	};
	\node[note] at (9.2,-3.9)
	{
		Mean memory $\meanmem$\\
		and variance $\sigma_\mem^2$
	};
	\node[level3] (policydist) at (-0.5,-2.5)
	{
		\textcolor{green!80!black}{Policy distribution}\\[1mm]
		Eq.~(\ref{eq:dt:phi:homog_v1}): $
		\p_t \phi(\policy) = \tilde\lambda_T \,\rho\,  \left( \Reff (\policy)- \stackon[1pt]{\stackon[1pt]{$\reward$}{\rule{8.0pt}{0.08ex}}}
		{\rule{8.0pt}{0.08ex}}  \right)\phi(\policy) + \Dmut \frac{\p^2 \phi(\policy)}{\p \policy^2}
		$ 
	};
\end{tikzpicture}

We have selected Eqs.~(\ref{eq:dm:dt}) and (\ref{eq:dsigma2:dt}) rather than Eqs.~(\ref{eq:dm:dt:gauss}) and (\ref{eq:dsigma2:dt:gauss})  to describe the dynamics of the memory because, as noted above, the quantity $\avS{G}(\mem, \policy)$ may be more difficult to evaluate in practice than the covariance $\mathrm{Cov}_{\mem|\policy}(\mem,\avS{G})$ written in terms of joint averages over $\state$ and $\mem$.

\subsection{Closure of evolution equations for the policy mean and variance}
\label{sec:policy-closure}

Having determined $\Reff(\policy,t)$, we can in principle determine $\phi(\policy,t)$ by integrating Eq.~(\ref{eq:dt:phi:homog_v1}), which in most cases can only be done numerically.
This requires the joint integration of Eqs.~(\ref{eq:dm:dt}) and (\ref{eq:dsigma2:dt}) for $\meanmem$ and $\sigma_\mem^2$, through which $\Reff(\policy,t)$ can be evaluated using Eq.~(\ref{eq:Reff:approx}). 
As already mentioned above, we consider that $\avM{\avS{G}}$ and $\mathrm{Cov}_{\mem|\policy}(\mem,\avS{G})$ can be determined explicitly in specific models, once the physical dynamics is specified (see the application sections below for explicit examples).
Instead of integrating in time the whole function $\phi(\policy,t)$, a simpler alternative route may be to reduce the statistical information to the mean and variance of the policy, using Eqs.~(\ref{eq:dP:dt}) and (\ref{eq:dsigma2P:dt}), which thus need to be closed.
Such a simplified approach may more easily allow for analytically tractable results, but requires additional approximations to close the equations.

\subsubsection{Ansatz for the policy distribution $\phi(\policy)$}
\label{sec:ansatz:policy:dist}

In order to close Eqs.~(\ref{eq:dP:dt}) and (\ref{eq:dsigma2P:dt}) one needs an ansatz for the distribution $\phi(\policy)$ such that higher-order moments can be expressed in terms of a finite set of lower-order moments. 
Depending on the applications, different distributions might be required and this is the reason why we have kept the boundary terms in Eqs.~(\ref{eq:dP:dt}) and (\ref{eq:dsigma2P:dt}).
In the case of an unbounded policy defined over the whole real axis, a Gaussian distribution may be a natural and convenient ansatz (albeit not backed by the Central Limit Theorem in this case, at variance with the memory distribution):
\begin{equation}
    \phi(\policy) = \frac{1}{\sqrt{2\pi \sigma_{\policy}^2}} \, \exp\left( -\frac{\big(\policy-\meanpol\big)^2}{2\sigma_{\policy}^2} \right).
\end{equation}
For bounded or semi-bounded variables, other types of distributions may be useful (see for instance Sec.~\ref{sec:AOU:degenerate} for an example where a distribution bounded from below is used).
Having made an ansatz on the shape of the distribution, we then determine the time-dependent parameters $\meanpol$ and/or $\sigma_{\policy}^2$ which fully characterize the distribution $\phi(\policy)$ given its fixed shape. In the case of the Gaussian distribution, the two parameters $\meanpol$ and $\sigma_{\policy}^2$ are required because the shape of the distribution is invariant by translation and scale transformation. In contrast, when the policy is bounded from below (or above), the policy dynamics may be dominated by the bound, and the translation degree of freedom is lost so that a fixed-shape distribution has a single free parameter, which can be chosen either as $\meanpol$ or $\sigma_{\policy}^2$.

In the case of a Gaussian distribution, boundary terms vanish in Eqs.~(\ref{eq:dP:dt}) and (\ref{eq:dsigma2P:dt}), and the covariance terms can be reformulated using again the Stein identity, leading to the simplified equations:
\begin{align}
\frac{d}{d t}  \meanpol =& \tilde{\lambda}_T \rho \, \avMP{\Reff'(\policy)} \sigma_\policy^2,
\label{eq:hatP_unif:gauss}\\
\frac{d}{d t}  \sigma^2_{\policy} =& \tilde{\lambda}_T \rho \, \avMP{\Reff''(\policy)} \sigma_\policy^4 + 2 \Dmut,
\label{eq:sigmahatP_unif:gauss}
\end{align}
where $\avMP{\Reff'(\policy)}$ and $\avMP{\Reff''(\policy)}$ are functions of $\meanpol$ and $\sigma^2_{\policy}$.
The expressions $\Reff'(\policy)$ and $\Reff''(\policy)$ denote derivatives of $\Reff$ with respect to the policy $\policy$ (and not derivatives of $\reward$ with respect to the memory $\mem$).
If the effective reward $\Reff(\policy)$ is concave and quadratic,
\begin{equation}
    \Reff(\policy)= \Reff_{\max}-\frac{1}{2}\reward_2 (\policy -\policy^*)^2,
\end{equation}
where $\Reff_{\max}$, $\reward_2>0$ and $\policy^*$ are model-dependent constants,
one has $\avMP{\Reff'(\policy)} = -\reward_2 (\meanpol - \policy^*)$ and $\avMP{\Reff''(\policy)} = -\reward_2$.
In this case, the diversity $\sigma^2_{\policy}$ converges to a nonzero value resulting from the presence of mutations, $\Dmut>0$. And, as long as the diversity $\sigma_\policy^2>0$, the mean policy converges toward the local maxima of $\Reff(\policy)$ because it follows the gradient of $\Reff$.
If $\Reff(\policy)$ includes higher order contributions around $\policy=\policy^*$, e.g., a cubic term $\reward_3 (\policy-\policy^*)^3/3$, then $\meanpol$ converges to a slightly shifted value
$\tilde{\policy}^*=\policy^* + (\reward_3/\reward_2) \, \sigma^2_{\policy}$, assuming $\sigma^2_{\policy}$ to be small. In consequence, the long-time policy will not converge towards the maximum of the effective reward due to the non-zero diversity caused by the presence of mutations $\Dmut$

We emphasize that even if $\reward(\mem)$ is (at most) quadratic in $\mem$, $\Reff(\policy)$ may acquire higher than quadratic contributions in $\policy$ due to the policy dependence of the
variance $\sigma^2_{\mem}(\policy)$, as seen from Eq.~(\ref{eq:Reff_target}).

\subsubsection{Explicit time solutions for the policy mean and variance}

We again assume here that the policy variance is small, and that the effective reward is concave and at most quadratic, 
so that $\avMP{\Reff'(\policy)} = -\reward_2 (\meanpol - \policy^*)$ and $\avMP{\Reff''(\policy)} \equiv -\reward_2$ is a negative constant.
In this simplified setting Eqs.~(\ref{eq:hatP_unif:gauss}) and (\ref{eq:sigmahatP_unif:gauss}) boil down to

\begin{align}
\frac{d}{d t}  \meanpol =& - \tilde{\lambda}_T \rho \, \reward_2 (\meanpol - \policy^*) \sigma_\policy^2,
\label{eq:hatP_unif:gauss:simple}\\
\frac{d}{d t}  \sigma^2_{\policy} =& - \tilde{\lambda}_T \rho \, \reward_2 \sigma_\policy^4 + 2 \Dmut.
\label{eq:sigmahatP_unif:gauss:simple}
\end{align}
In particular, we find that the time-evolution of the mean policy $\meanpol$ is directly proportional to the diversity $\sigma_\policy^2,$ which is known as Fisher's fundamental theorem of natural selection \cite{ewens1989interpretation}.
Eq.~(\ref{eq:sigmahatP_unif:gauss:simple}) is a Ricatti equation, which can be integrated exactly and one obtains:
\begin{align}
\meanpol(t) &= \policy^* + \left(\meanpol(0) - \policy^*\right) \frac{{\sigma_\policy^2}^\infty}{{\sigma_\policy^2}^\infty \cosh \gamma t + \sigma_\policy^2(0) \sinh \gamma t } , \label{eq:ricatti:mup} \\
\sigma_\policy^2(t) &= {\sigma_\policy^2}^\infty \frac{{\sigma_\policy^2}^\infty \tanh{\gamma t} + \sigma_\policy^2(0)}{{\sigma_\policy^2}^\infty + \sigma_\policy^2(0) \tanh{\gamma t}}, \label{eq:ricatti:Dp}
\end{align}
with 
\begin{equation}
\lambda_0 = \tilde{\lambda}_T \rho \,\reward_2,\qquad
\gamma = \sqrt{2 \Dmut \lambda_0}, \qquad
{\sigma_\policy^2}^\infty = \sqrt{2 \Dmut /\lambda_0}, 
\end{equation}
where $\lambda_0$ is an effective learning rate.
The evolution of $\meanpol(t)$ and $\sigma_\policy^2(t)$ in Eqs.~(\ref{eq:ricatti:mup}) and (\ref{eq:ricatti:Dp}), respectively,  involve a relaxation over two distinct time scales
\begin{equation} \label{eq:def:taui}
    \tau_1 = \big( \sigma_\policy^2(0)\lambda_0 \big)^{-1}, \qquad \tau_2=\gamma^{-1}= (2 \Dmut \lambda_0)^{-1/2}.
\end{equation}
Here, $\tau_1$ is connected to learning due to the initial diversity $\sigma_\policy^2(0)$ and $\tau_2$ requires the presence of mutations.
Typically the initial policy is widely distributed and $\sigma_\policy^2(t)$ is a decreasing function of time, so that $\tau_1/\tau_2={\sigma_\policy^2}^\infty/\sigma_\policy^2(0) \ll 1$. 
The expressions (\ref{eq:ricatti:mup}) and (\ref{eq:ricatti:Dp}) of $\meanpol(t)$ and $\sigma_\policy^2(t)$ can be rewritten as (using the approximation $\tau_1 \ll \tau_2$ to simplify the expression of $\sigma_\policy^2$)
\begin{equation}
    \meanpol(t) = \policy^* + \Delta \meanpol(0)
    \left[ \cosh \frac{t}{\tau_2} \left( 1 + \frac{\tau_2}{\tau_1} \tanh \frac{t}{\tau_2} \right) \right]^{-1} \!\!\!,
    \qquad
    \sigma_\policy^2(t) = \sigma_\policy^2(0) \left( 1 + \frac{\tau_2}{\tau_1} \tanh \frac{t}{\tau_2} \right)^{-1} \!\!\!,
\end{equation}
with $\Delta \meanpol(0) = \meanpol(0) - \policy^*$.
Assuming again $\tau_1 \ll \tau_2$, these expressions simplify in the different time regimes:
\begin{align}
&\text{for} \quad t \sim \tau_1, \qquad  \qquad \;\;\,\meanpol(t) = \policy^* + \frac{\Delta \meanpol(0)}{1+t/\tau_1}, \qquad\qquad\qquad\quad\;
\sigma_\policy^2(t) = \frac{{\sigma_\policy^2}(0)}{1+t/\tau_1},\\ 
&\text{for} \quad \tau_1 \ll t \ll \tau_2, \qquad \meanpol(t) = \policy^* + \Delta \meanpol(0) \, \frac{\tau_1}{t}, \qquad\qquad\qquad\;\,
\sigma_\policy^2(t) = {\sigma_\policy^2}(0)\, \frac{\tau_1}{t},\\
&\text{for}\quad  t \sim \tau_2, \qquad \qquad  \;\;\,\meanpol(t) = \policy^* + \Delta \meanpol(0) \frac{\tau_1}{\tau_2 \, \sinh(t/\tau_2)}, \qquad
\sigma_\policy^2(t) = \frac{{\sigma_\policy^2}^\infty}{\tanh(t/\tau_2)},
\end{align}
where we recall that ${\sigma_\policy^2}^\infty = \sigma_\policy^2(0)\, \tau_1/\tau_2$.
When $\Dmut = 0$, $\tau_2 \rightarrow \infty$ according to Eq.~(\ref{eq:def:taui}), and $\sigma_\policy^2$ decays algebraically up to arbitrarily large times.
When $\Dmut >0$, $\lambda_0$ combines with $\Dmut$ to set both the learning times $\tau_1$ and $\tau_2$ and the asymptotic diversity ${\sigma_\policy^2}^\infty$. As already noticed in Ref.~\cite{jung2025kinetic}, decreasing $\Dmut$ in order to decrease the uncertainty on the final policy also decreases the learning rate; conversely increasing $\lambda_0$ both decreases the final uncertainty and increases the learning rate. One recovers here the `uncertainty relation' introduced in Ref.~\cite{jung2025kinetic},
\begin{equation}
   {\sigma_\policy^2}^\infty \gamma^{-1} = 1/\lambda_0 ,
   \label{eq:uncertainty}
\end{equation}
relating the asymptotic diversity and the effective learning rate.

In the following, we illustrate two different applications of the present kinetic theory formalism in simple, yet illuminating examples, in order to guide the reader through a practical implementation of the general framework.


\section{Application 1: Active Ornstein-Uhlenbeck (AOU) Agents with Adaptive mobility and activity}
\label{sec:AOU}

Robots or microorganisms usually have the ability to actively move into a certain direction (a property called motility), rather than being slaved by thermal noise. As a first application we thus introduce a simple active matter model and study active Ornstein-Uhlenbeck agents in two-dimensions which are defined by the physical dynamics \cite{szamel2014self,bonilla2019active},
\begin{align}
\frac{d \br_i(t)}{d t} &\equiv \bm V_i(t) = b_i \bm F + \bm{v}_i(t), \label{eq:AOUA_ri}\\
\tau \frac{d \bm{v}_i(t)}{d t} &= -\bm v_i(t) + \sqrt{2 D_i} \bm W_i(t), \label{eq:AOUA_va}
\end{align}
with external force $\bm F = (F_x,0)$, active propulsion velocity $\bm{v}_i$, persistence time $\tau$ (which corresponds to the physical time scale) and propulsion strength $D_i$;
$\bm{W}_i(t)$ is a vectorial unit Gaussian white noise, with correlation $\langle W_{i,\alpha}(t) W_{j,\beta}(t') \rangle_{\rm noise} = \delta_{ij} \, \delta_{\alpha\beta} \delta(t-t')$,
where Greek indices stand for Cartesian coordinates.
Following the general definitions introduced in Sec.~\ref{sec:agent:model}, the state $\state_i$ of agent $i$ is given by $\state_i=\bm{v}_i$.
We assume that agents can generally control both their mobility $b_i$ and their non-thermal activity $D_i$, so that the policy may be chosen as $\policy_i=(b_i,D_i)$ or a subset of these parameters.
The policy $b_i$ may take any real value (including negative ones), while we assume $D_i>0$. Agent-based simulations are performed for $N=16000$ agents with box size $L=40$ and $r_c = \pi^{-1/2}$. 
We study 16 independent learning procedures with different microscopic initial conditions. Results show the average policy and diversity. Error bars correspond to the standard deviation across different initial conditions. 

In the following, we will study various learning scenarios. We will start with a one-dimensional policy $\policy_i=b_i$, by fixing the activity to $D_i=D$, a constant and identical activity for all agents.
This allows us to investigate the fundamental time scales emerging in the learning dynamics.
Subsequently, we investigate learning when both mobility $b_i$ and activity $D_i$ are adaptable, in which case one has the two-dimensional policy $\policy_i=(b_i,D_i)$.

\subsection{One-dimensional policy, with perfect separation of the memory and teaching time scales ($\tilde \lambda_T \rho, \Dmut \ll \lambda_\mem$)}
\label{sec:AOU:1d:ts}

In this model, the goal of collective learning is to obtain a macroscopic flow along the $x$-direction, with mean velocity $\avAS{V_x} = V_\odot$, where $\avAS{\dots}$ is the steady-state ensemble- and time-average in agent-based simulations, and $V_\odot$ is the target velocity.
This goal can be achieved by agents adapting their mobility $b_i$. Omitting the subscript $i$, we can thus identify the policy $\policy_0=b$ (the subindex `$0$' refers to this first protocol). In the present example, the optimization problem has a `trivial' solution  $b_\odot=V_\odot/F_x$, obtained by averaging Eq.~(\ref{eq:AOUA_ri}) and imposing
$\avAS{V}=V_\odot$. We can thus compare the solution found by the collective learning with $b_\odot$ and assess whether the optimization was successful. In more complex cases, $b_\odot$ is generally not analytically accessible. In the present example, the relevant observable $G(\state)\equiv G_0(\bm{v})$ defining the memory dynamics is chosen as the instantaneous velocity of the agents in the $x-$direction,
\begin{equation} \label{eq:def:GS:AOU:v0}
    G_0(\bm{v})=V_{x}=b F_x + v_{x}.
\end{equation}
A running average of the velocity $V_{x}$ is stored into the memory $\mem_{0}$, whose dynamics is given by
\begin{align} \label{eq:def:mem:AOU:v0}
\frac{d \mem_{0}(t)}{d t} &= \lambda_\mem \big(b F_x + v_{x}(t) - \mem_{0}(t)\big).
\end{align}
Here, $D_\mem(\policy)=0$ since there is no white noise acting on the memory dynamics.
The reward function is chosen as 
\begin{equation} \label{eq:reward:def:AOUP:1d}
\reward(\mem_0)= \reward_\odot - (\mem_0 - V_\odot)^2,
\end{equation}
to target the velocity $V_\odot$ along the $x$-axis.
The reward function $\reward(\mem_0)$ is in principle dimensionless, so that the term $(\mem_0 - V_\odot)^2$ should be normalized by a characteristic squared velocity $v_c^2$. We choose length and time units so that $v_c=1$.

In this first subsection, we assume well-separated memory and teaching time scales (i.e., $\Dmut, \tilde{\lambda}_T \rho \ll \lambda_\mem$), so that the dynamics of $\meanmemzero(b)$ and $\sigma_{\mem_0}^{2}(b)$ are given by Eqs.~(\ref{eq:dm:dt:noteach}) and (\ref{eq:dsigma2:dt:noteach}), which now read 
\begin{align}
\p_t \meanmemzero &= \lambda_\mem (b F_x - \meanmemzero),  \label{eq:meanM0:AOUP:1d} \\
\p_t \sigma_{\mem_0}^{2} &= 2 \lambda_\mem [  \mathrm{Cov}_{\mem_0|b}(\mem_0,\avS{V_x}) - \sigma_{\mem_0}^{2} ]
\label{eq:sigma2M0:AOUP:1d},
\end{align}
where we have used $\avM{\avS {G_0}}(\bm{v})=b F_x$. Under the time scale separation assumption, the policy dynamics is determined by the effective reward $\Reff(b)$,
which depends only on the stationary values of $\meanmemzero$ and $\sigma_{\mem_0}^{2}$ for a fixed policy $b$. 
These stationary values are obtained from Eqs.~(\ref{eq:meanM0:AOUP:1d}) and (\ref{eq:sigma2M0:AOUP:1d}),
\begin{align} 
\meanmemzero &= b F_x, \label{eq:muM:bF}  \\
\sigma_{\mem_0}^{2} &= \mathrm{Cov}_{\mem_0|b}(\mem_0,\avS{V_x}). \label{eq:sigma2:covMv}
\end{align}
The stationary value of $\mathrm{Cov}_{\mem_0|\policy_0}(\mem_0,\avS{G_0})$ can be evaluated using Eq.~(\ref{eq:covSM:MG}).
For AOU particles, we have $\avM{\langle v_{x,y} \rangle} = 0$, $\avM{\langle v_{x,y}^2 \rangle} = D / \tau$ and $\avM{\langle v_{x,y}^4 \rangle} =  3 (D / \tau)^2$, due to the Gaussian statistics of $v_{x,y}$.
We thus have for the stationary covariance,
\begin{equation}
\mathrm{Cov}_{\mem_0|b}(\mem_0,\avS{G_0}) = \avM{\avS{\mem_0 V_x}} - \avM{\avS{\mem_0}}\,\avM{\avS{V_x}} = \avM{\avS{\mem_0 v_x}}.
\end{equation}
The time evolution of $\avM{\avS{\mem_0 v_x}}$ can be derived from Eqs.~(\ref{eq:AOUA_va}), (\ref{eq:def:GS:AOU:v0}) and (\ref{eq:def:mem:AOU:v0}),
and we get in the stationary regime:
\begin{equation} \label{eq:cov:M0vx}
    \avM{\avS{\mem_0 v_x}} =  \frac{D \lambda_\mem}{1 + \lambda_\mem \tau} .
\end{equation}
In this model we do not assume a time scale separation between the physical time scale $\tau$ and the memory time scale $\lambda_\mem^{-1}$: the product $\lambda_\mem \tau$ in Eq.~(\ref{eq:cov:M0vx}), which quantifies the time scale separation, is not assumed to be small.
It follows from Eqs.~(\ref{eq:sigma2:covMv}) to (\ref{eq:cov:M0vx}) that in the stationary state, for a fixed policy $b$,
\begin{equation} \label{eq:mean:var:AOU:1d:ts}
\meanmemzero = b F_x, \qquad
\sigma_{\mem_0}^{2} = \frac{D \lambda_\mem}{1 + \lambda_\mem \tau},
\end{equation}
from which we can calculate the effective reward $\Reff(b)$, using the expression $\Reff(b)=\reward(\meanmemzero(b))- \sigma^2_{\mem_0}(b)$ given in Eq.~(\ref{eq:Reff_target}), which is exact for a quadratic reward $\reward(\mem)$.
We then find
\begin{equation} \label{eq:Reff:AOUP:v0}
\Reff(b) = \reward_\odot - (b F_x - V_\odot)^2 - \frac{D \lambda_\mem}{1 + \lambda_\mem \tau}.
\end{equation}

We can now investigate the policy dynamics on the learning time scale.
Assuming a Gaussian statistics for $b$, the dynamics of the policy mean $\mu_b=\avMP{b}$ and variance $\sigma_b^2=\avMP{b^2}-\avMP{b}^2$
is obtained from Eq.~(\ref{eq:hatP_unif:gauss}) and (\ref{eq:sigmahatP_unif:gauss}), which read
\begin{align}
\frac{d}{d t} \mu_b =& -2\tilde{\lambda}_T \,\rho \, F_x^2 \left(\mu_b - \frac{V_\odot}{F_x}\right) \sigma_b^2,
\label{eq:mub:AOUP:1d:v0}\\
\frac{d}{d t} \sigma^2_b =& -2\tilde{\lambda}_T \rho \, F_x^2 \sigma_b^4 + 2 \Dpol_b,
\label{eq:sigmab:AOUP1d:v0}
\end{align}
where we have used $\avMP{\p_b \Reff(b)}=-2F_x(\mu_b F_x-V_\odot)$ and $\avMP{\p^2_b \Reff(b)}=-2F_x^2$, with $\p^n_b$ the $n^{\rm th}$-derivative with respect to $b$.
Due to the simplicity of the model, $\Reff''(b)=-2F_x^2$ is a constant so that Eqs.~(\ref{eq:mub:AOUP:1d:v0}) and (\ref{eq:sigmab:AOUP1d:v0})
take the form of Eqs.~(\ref{eq:hatP_unif:gauss:simple}) and (\ref{eq:sigmahatP_unif:gauss:simple}), with $\reward_2=2F_x^2$ and $\policy_0^*=b^*=V_\odot/F_x$.
We can thus use the time-dependent analytical solutions given in Eqs.~(\ref{eq:ricatti:mup}) and (\ref{eq:ricatti:Dp}).
Quite importantly, we observe that in the long-time limit $\mu_b$ converges to the value $b^*$ 
which maximizes the effective reward $\Reff(b)$. The reward-maximizing policy $b^*$ therefore coincides in this case with the `trivial' solution $b_\odot$ of the optimization problem.

The dynamics of $\mu_b$ and $\sigma^2_b$ can also be integrated numerically, to take into account the memory dynamics and lift the assumption of stationary memory statistics. This is useful even in the present time scale separation limit, to see the effect of the initial policy variance $\sigma^2_{b}(t=0)=\sigma_0^2$. For this purpose we use the explicit expression Eq.~(\ref{eq:Reff_target}) of the effective reward $\Reff(b)$ combined with the time-dependent memory dynamics [Eqs.~(\ref{eq:meanM0:AOUP:1d},\ref{eq:sigma2M0:AOUP:1d})] to get the time-dependent effective reward $\Reff(b,t)$,
\begin{equation} \label{eq:Reff:AOUP:v1}
\Reff(b,t) = \reward_\odot - (\meanmemzero(b,t) - V_\odot)^2 - \sigma^2_{\mem_0}(b,t).
\end{equation}
Still assuming a Gaussian statistics for $b$, this allows us to integrate numerically Eqs.~(\ref{eq:hatP_unif:gauss}) and (\ref{eq:sigmahatP_unif:gauss}) for the dynamics of the policy mean $\mu_b$ 
and variance $\sigma^2_b$ 
which we rewrite explicitly here for the sake of clarity,
\begin{align}
\frac{d}{d t}  \mu_b =& \tilde{\lambda}_T \rho \, \avMP{\p_b \Reff(b,t)} \sigma_b^2,
\label{eq:hatP_unif:gauss:AOU1d}\\
\frac{d}{d t}  \sigma^2_b =& \tilde{\lambda}_T \rho \, \avMP{\p^2_b \Reff(b,t)} \sigma_b^4 + 2 \Dpol_b.
\label{eq:sigmahatP_unif:gauss:AOU1d}
\end{align}
The derivatives of $\Reff$ with respect to the policy $b$ and the policy integrals can be easily evaluated numerically by reconstructing the underlying distribution $\phi(\policy,t)$ using the moments $\mu_b(t)$ and $\sigma^2_b(t)$.
We will pursue a similar route in all of the following applications. Importantly, the introduction of the effective reward is therefore not only conceptually interesting, but also allows us to overcome the expansion of $\mu_\mem(\policy)$ and $\sigma^2_\mem(\policy)$ used in Ref.~\cite{jung2025kinetic}. If the policy has non-Gaussian statistics, we have to resort to Eqs.~(\ref{eq:dP:dt},\ref{eq:dsigma2P:dt}) instead.

\begin{figure}
    \centering
    \includegraphics[width=0.95\linewidth]{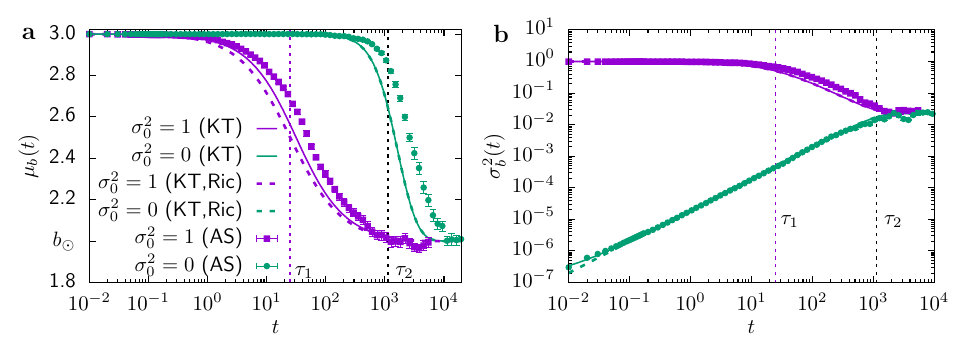}
    \caption{ Learning dynamics for the one-dimensional policy $\policy_0 = b$ with time scale separation. Results predicted analytically in Eqs.~(\ref{eq:ricatti:mup}) and (\ref{eq:ricatti:Dp}) (KT, Ric) and numerically in Eqs.~(\ref{eq:hatP_unif:gauss}) and (\ref{eq:sigmahatP_unif:gauss}) using kinetic theory (KT), compared to results from agent-based simulations (AS).  We study two scenarios with a broad initial distribution of policies ($\sigma^2_{b}(0)=\sigma^2_0=1$) and a delta-distribution ($\sigma^2_0=0$) with $\mu_{b}(0)=3$. Vertical lines feature the two time scales derived in Sec.~\ref{sec:policy-closure}, where $\tau_1(\sigma_0^2=0) \rightarrow \infty$ and $\tau_2$ independent of $\sigma^2_0$. \textbf{(a)} Average policy $\mu_{b}(t)$. \textbf{(b)} Diversities $\sigma^2_{b}(t)$. The parameters are $\rho \tilde \lambda_T=0.02$, $\lambda_\mem=1.0$, $\Dpol_{b} = 10^{-5}, \tau=0.1, F_x=1$,  and $V_\odot=2.0.$ }
    \label{fig:AOU1d:dynamics_v1}
\end{figure}

The learning dynamics are plotted in Fig.~\ref{fig:AOU1d:dynamics_v1} for two different scenarios. In both cases, $F_x=1, V_\odot = 2$, so that the target policy is $b_\odot = V_\odot/F_x = 2$.
In the first scenario, the initial policy distribution is a relatively broad Gaussian distribution centered on $b=3$, with variance $\sigma^2_{b}(0)=1.$  The target policy is therefore already in the set of existing initial policies and the learning process is just a matter of teaching it to all the other agents, which occurs on a time scale $\tau_1$ (see Fig.~\ref{fig:AOU1d:dynamics_v1}a). During the learning the diversity decreases, because specific policies around $b = b_\odot$ are selected, until it reaches the long-time steady-state value $\sigma^{2\,\infty}_{b}$ (see Fig.~\ref{fig:AOU1d:dynamics_v1}b).
The second scenario starts from a delta-distributed policy 
with vanishing diversity $\sigma^2_{b}(0)=0.$ In this case, the best policy first needs to be found by mutations which obviously increases the required time scale for learning to $\tau_2.$ For $t < \tau_2 $ the diversity increases linearly, as expected by simple diffusive mutations, as shown in Fig.~\ref{fig:AOU1d:dynamics_v1}b. It is only for $t>\tau_2$, that the actual teaching process takes place, as seen in Fig.~\ref{fig:AOU1d:dynamics_v1}a. In both cases, the results from kinetic theory provide a very good description of the dynamics observed in the agent-based simulations and allow for a clear understanding of the role of the initial distribution of policy, as well as that of the mutation.

\subsection{One-dimensional policy with overlapping memory and teaching time scales ($\Dmut \ll \tilde \lambda_T \rho, \lambda_\mem$)}
In this subsection, we briefly explore the consequences of a lack of time scale separation between memory and teaching time scales.
We still consider that $\Dmut \ll \tilde \lambda_T \rho, \lambda_\mem$, and that the physical dynamics is much faster than the memory and learning dynamics. 
As an approximation, we assume that despite overlapping teaching and memory time scales, memory-dependent quantities involved in the policy dynamics may still be evaluated from the stationary statistics of the memory $\mem$, as done in Sec.~\ref{sec:AOU:1d:ts}, since convergence to the stationary statistics is expected to be sufficient on the teaching time scale $\tilde \lambda_T \rho$. 
However, Eqs.~(\ref{eq:dm:dt:noteach}) and (\ref{eq:dsigma2:dt:noteach}) for the dynamics of the first memory cumulants need to be replaced by
Eqs.~(\ref{eq:dm:dt}) and (\ref{eq:dsigma2:dt}) to take into account the teaching contributions to the memory dynamics.
Since $\mem_0$ has a Gaussian statistics, the teaching terms proportional to $\tilde{\lambda}_T$
in Eqs.~(\ref{eq:dm:dt}) and (\ref{eq:dsigma2:dt}) can be conveniently evaluated using their expression given in
Eqs.~(\ref{eq:dm:dt:gauss}) and (\ref{eq:dsigma2:dt:gauss}). 
We thus need to evaluate $\avM{\reward'(\mem_0)}$ and $\avM{\reward''(\mem_0)}$, leading from Eq.~(\ref{eq:reward:def:AOUP:1d}) to $\avM{\reward'(\mem_0)} = 2 (V_\odot - \mu_{\mem_0})$ and $\avM{\reward''(\mem_0)} = -2 $.
The evolution equations for $\meanmemzero(b,t)$ and $\sigma_{\mem_0}^{2}(b,t)$ then read
\begin{align}
\p_t \meanmemzero &= \lambda_\mem (b F_x - \meanmemzero) + 2 \tilde \lambda_T \rho (V_\odot - \mu_{\mem_0}) \sigma_{\mem_0}^2 , \label{eq:app1:overlapping:mu}\\
\p_t \sigma_{\mem_0}^{2} &= 2 \lambda_\mem \left [\frac{\lambda_\mem D}{1 + \lambda_\mem \tau} - \sigma_{\mem_0}^{2} \right ]  - 2 \tilde \lambda_T \rho \sigma_{\mem_0}^4. \label{eq:app1:overlapping:sigma}
\end{align}
Inserting the time-dependent memory cumulants $\meanmemzero(b,t)$ and $\sigma_{\mem_0}^{2}(b,t)$, which are evaluated numerically, 
into Eq.~(\ref{eq:Reff:AOUP:v1}), we obtain the effective reward $\Reff(b,t)$.
Again assuming a Gaussian policy distribution $\phi(b)$, the dynamics of the policy mean $\mu_b$ and variance $\sigma_b^2$ are then given by Eqs.~(\ref{eq:hatP_unif:gauss:AOU1d}) and (\ref{eq:sigmahatP_unif:gauss:AOU1d}), which are integrated numerically in time, simultaneously to the memory equations (\ref{eq:app1:overlapping:mu}) and (\ref{eq:app1:overlapping:sigma}).

\begin{figure}
    \centering
    \includegraphics[width=0.95\linewidth]{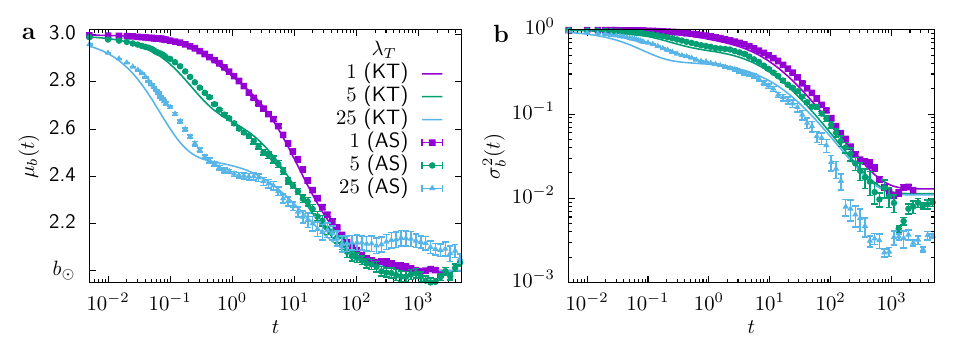}
    \caption{Learning dynamics for the one-dimensional policy $\policy_0=b$ with overlapping memory and teaching time scales. Results predicted by kinetic theory in Eqs.~(\ref{eq:hatP_unif:gauss}) and (\ref{eq:sigmahatP_unif:gauss}) are compared to results from agent-based simulations (AS).   \textbf{(a)} Average policy $\mu_{b}(t)$. \textbf{(b)} Diversities $\sigma^2_{b}(t)$. The parameters are $\rho \tilde \lambda_T=0.2, 1, 5$ (for $\lambda_T=1, 5, 25$, respectively), $\lambda_\mem=1.0$, $\Dpol_{b} = 10^{-5},$ $V_\odot=2.0,$ $F_x=1$ and $\tau=0.01$.  }
    \label{fig:AOU1d:dynamics_v2}
\end{figure}

Figure~\ref{fig:AOU1d:dynamics_v2} illustrates the result of a learning procedure, starting from initial conditions for the policy $\mu_{b}(t=0) = 3$, $\sigma^2_{b}(t=0) = 1 $, and the memory $\mu_{\mem_0}(b,t=0) = bF_x$ and $\sigma^2_{\mem_0}(b,t=0) = 0.$ The memory is therefore initialized such that the effective reward favors $b F_x = V_\odot$, from the very beginning of the procedure. Yet, when investigating the learning we observe clearly the emergence of two different time scales for both the dynamics of the mean $\mu_{b}$ and diversity $\sigma^2_{b}$. The shorter time scale corresponds to $\tau_1 = \lambda_0^{-1}$ and describes efficient learning based on the initial memory. However, the second term in
Eqs.~(\ref{eq:app1:overlapping:mu}) and (\ref{eq:app1:overlapping:sigma}) distorts the memory of the agents, such that it converges towards $\mu_{\mem_0}(b,t) = V_\odot$ independent of their policy. This transient dynamics disrupts the optimization procedure and it emerges because agents with $\mem_{0,i}=V_\odot$ most likely become teacher and thus transfer their policy and memory to other agents. Only on a time scale $\tau_\mem = \lambda_\mem^{-1}$ the agents gain significant contributions from the first term in Eqs.~(\ref{eq:app1:overlapping:mu}) and (\ref{eq:app1:overlapping:sigma})
and converge towards the correct long-time policy, $b_\odot$, which also in this case corresponds to the maximum $b^*$ of the effective reward $\Reff(b).$ In consequence, a faster teaching rate does not imply faster convergence towards the long-time goal. In fact, we observe in agent-based simulations that for $\lambda_T=25$ the learning dynamics shows very large sample-to-sample fluctuations as indicated by the error bars in Fig.~\ref{fig:AOU1d:dynamics_v2}a, and on average did not converge properly towards the goal $b_\odot,$ even at $t>5\cdot 10^3.$ Additionally, highly non-Gaussian, bimodal policy distributions emerge which differ from sample to sample. These bimodal distributions, combined with the fact that for $\lambda_T=25$ the assumption of stationary memory statistics needed to derive Eqs.~(\ref{eq:app1:overlapping:mu}) and (\ref{eq:app1:overlapping:sigma}) becomes questionable, explain the differences between kinetic theory and agent-based simulations observed in Fig.~\ref{fig:AOU1d:dynamics_v2}. We conclude that this example highlights the importance of limiting the teaching rate to achieve a more stable learning procedure.

\subsection{Two-dimensional policy with a degenerate component}
\label{sec:AOU:degenerate}

To illustrate the flexibility of the approach, we will now consider agents which have the possibility of adjusting multiple policy parameters to achieve their goal. The activity $D$ is therefore now an adaptable parameter and the policy is two-dimensional, $\policy = (\policy_0, \policy_1)= (b,D)$. The extension of the general formalism to a multidimensional policy is presented in Appendix~\ref{app:multidimensional}.
We again assume a perfect time scale separation between the memory and learning dynamics as in Sec.~\ref{sec:AOU:1d:ts}, namely $\tilde \lambda_T \rho, \Dmut \ll \lambda_\mem$.

In the scenario considered here, the definitions of other variables and parameters remain unchanged. One still has $G(\state)=G_0=V_x$ as in Eq.~(\ref{eq:def:GS:AOU:v0}), leading to the dynamics of the memory $\mem_0$ through Eq.~(\ref{eq:def:mem:AOU:v0}), and to the reward $\reward(\mem_0)$ defined in Eq.~(\ref{eq:reward:def:AOUP:1d}).
Let us emphasize that the model is now based on a single memory variable $\mem_0$ to control the adaptation of a two-dimensional policy $(b,D)$.
We call this scenario a `policy-degenerate' case since $D$ has no direct influence in the reward evaluation, hence a priori on the learning process. One could therefore expect that many different policies $\policy=(b,D)$ should maximize the reward. As we shall see below, this is actually not the case. 

Due to the time scale separation assumption, one can still use the stationary values of the mean $\meanmemzero$ and variance $\sigma_{\mem_0}^2$ of the memory given in Eq.~(\ref{eq:mean:var:AOU:1d:ts}). The effective reward $\Reff(b,D)$ reads
\begin{equation} \label{eq:Reff:AOUP:v1_two_dim}
\Reff(b,D) = \reward_\odot - (b F_x - V_\odot)^2 - \frac{D \lambda_\mem}{1 + \lambda_\mem \tau}.
\end{equation}

The expression of $\Reff$ is formally identical to Eq.~(\ref{eq:Reff:AOUP:v0}), but now considered as a function of both policy variables $(b,D)$.
We observe that $\Reff(b,D)$ is maximized for $(b,D)=(V_\odot/F_x,0)$, hence for a vanishing activity $D$.

\begin{figure}[t!]
    \centering
    \includegraphics[width=0.95\linewidth]{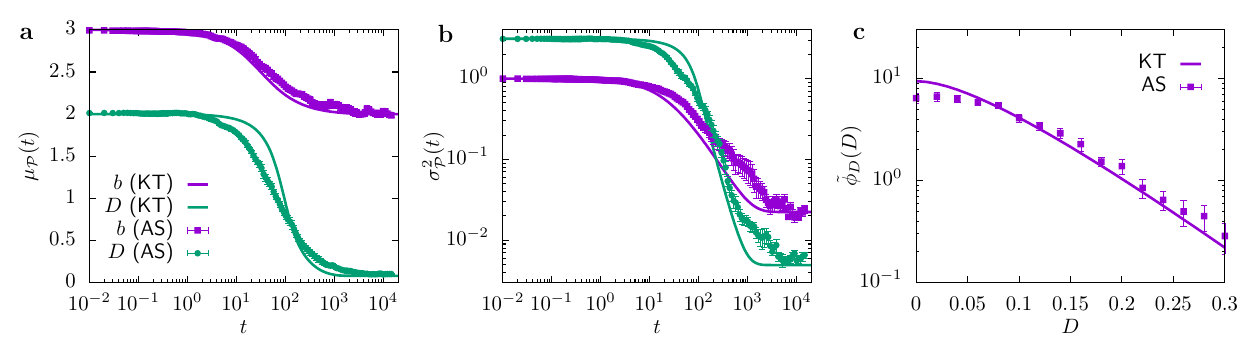}
    \caption{ Learning dynamics as predicted by kinetic theory (KT) in Eqs.~(\ref{eq:AOU:deg:b_dynamics})-(\ref{eq:AOU:deg:D_dynamics}) for the degenerate scenario compared to results from agent-based simulations (AS).  \textbf{(a)} Average policies $\mu_b(t)$ and $\mu_D(t)$. \textbf{(b)} Diversities $\sigma^2_{b}(t)$ and $\sigma^2_{D}(t) = 7\mu_D(t)^2/9$.  \textbf{(c)} Long-time normalized steady-state distribution of activities $\tilde \phi_D(D)$ as introduced in Eq.~(\ref{eq:fact:hyp:phibD}).  The parameters are $\rho \tilde \lambda_T=0.02$, $\lambda_\mem=1.0$, $\Dpol_b = \Dpol_D = 10^{-5}, V_\odot=2.0$, $F_x=1$ and $\tau=0.1$.  }
    \label{fig:AOU:dynamics}
\end{figure}

Following the approach presented in Appendix~\ref{app:multidimensional} for multidimensional policies,
we assume as an approximation that the policy density $\phi(b,D)$ is factorized,
\begin{equation} \label{eq:fact:hyp:phibD}
    \phi(b,D) = \tilde{\phi}_b(b) \tilde{\phi}_D(D).
\end{equation}
This assumption is useful here to deal with boundary terms, as explained below.
Importantly, the factorization assumption (\ref{eq:fact:hyp:phibD}) does not necessarily imply that $b$ and $D$ do not influence each other. Interactions are, for example, possible due the dependence of $\sigma^2_{\mem_0}(D)$ on $D$, which itself could influence the dynamics of $\mu_b$.
The marginal distribution $\tilde{\phi}_b(b)$ is then assumed to be a Gaussian, with average $\mu_b>0$ and variance $\sigma_b^2$ (we recall that negative values of $b$ are allowed).
In contrast, since $D>0$ the distribution of $D$ may not be described as a Gaussian distribution, especially in the present case when the lower bound $D=0$ has a relevant role in the dynamics of $D$ (see below). Empirical observations in agent-based simulations (see Fig.~\ref{fig:AOU:dynamics}(c)) shows that the marginal distribution $\tilde{\phi}_D(D)$ roughly behaves as an exponential,
but with a zero slope at $D=0$ to ensure that the diffusive flux vanishes at the boundary.
As a simple ansatz satisfying these conditions, we choose the following form
\begin{equation}\label{eq:app1:deg:dist}
    \tilde{\phi}_D(D) = \frac{3}{4\mu_D} \left( 1 + \frac{3D}{2\mu_D}\right) e^{-3D/2\mu_D}
\end{equation}
of mean $\mu_D$. The distribution $\tilde{\phi}_D(D)$ is then described by the single parameter $\mu_D$ (see the discussion of this point in Sec.~\ref{sec:ansatz:policy:dist}), and the variance is fixed to $\sigma^2_D=7\mu_D^2/9$.
Under these assumptions, we can describe the dynamics of $\mu_b$, $\sigma^2_b$ and $\mu_D$ on the learning time scale by using Eqs.~(\ref{eq:dP:dt:multidim:fact}) and (\ref{eq:dsigma2P:dt:multidim:fact}):
\begin{align}
\frac{d\mu_b}{dt} &= \tilde\lambda_T \rho \, \mathrm{Cov}_{b,D}\big(b,\Reff(b,D)\big) = 2 \tilde\lambda_T \rho F_x (V_\odot - \mu_b F_x)\sigma_{b}^2 
\label{eq:AOU:deg:b_dynamics} , \\
\frac{d\sigma_{b}^2}{dt} &= \tilde\lambda_T \rho \, \mathrm{Cov}_{b,D}\big((\delta b)^2,\Reff(b,D)\big) = -2 \tilde \lambda_T \rho F_x^2 \sigma_{b}^4 + 2 \Dpol_b 
\label{eq:AOU:deg:b_variance} ,\\ 
\frac{d\mu_D}{dt} &= \tilde\lambda_T \rho \, \mathrm{Cov}_{b,D}\big(D,\Reff(b,D)\big) + \Dpol_D \tilde{\phi}_D(0)
= -\frac{7}{9} \tilde\lambda_T \rho  \frac{ \lambda_\mem}{1 + \lambda_\mem \tau} \mu_D^2  + \frac{3\Dpol_D}{4\mu_D}.
\label{eq:AOU:deg:D_dynamics}
\end{align}
Thanks to the Gaussian statistics of $b$, we have used Eqs.~(\ref{eq:hatP_unif:gauss}) and~(\ref{eq:sigmahatP_unif:gauss}) to evaluate the covariance terms involved in the evolution equations for $\mu_b$ and $\sigma_{b}^2$,
\begin{align}
    \mathrm{Cov}_{b,D}\big(b,\Reff(b,D)\big) &= \avMP{\frac{\partial \Reff}{\partial b}} \sigma_{b}^2 = 2F_x (V_\odot - \mu_b F_x) \sigma_{b}^2,\\
    \mathrm{Cov}_{b,D}\big((\delta b)^2,\Reff(b,D)\big) &= \avMP{\frac{\partial^2 \Reff}{\partial b^2}} \sigma_{b}^4 = -2F_x^2  \sigma_{b}^4.
\end{align}
The covariance term $\mathrm{Cov}_{b,D}\big(D,\Reff(b,D)\big)$ is evaluated directly from its expression since the statistics of $D$ is non-Gaussian.

We thus find that $\mu_b$ converges to the value $b^*$ maximizing $\Reff(b,D)$ with respect to $b$, but $D$ fails to reach the optimum value $D=0$ due to the presence of mutations which enter the dynamical equations via the boundary term in Eq.~(\ref{eq:AOU:deg:D_dynamics}). Instead, it reaches the mutation-dependent value,
\begin{equation}\label{eq:AOU:deg:Dlong}
    \mu_D(t\rightarrow\infty) = \left( \frac{{27} \Dpol_D (1+\lambda_\mem \tau)}{{28} \tilde \lambda_T \rho \lambda_\mem} \right)^{1/3}.
\end{equation}

While the dynamics of $\mu_b$ and $\sigma^2_{b}$ simply obey the Ricatti equation given by Eq.~(\ref{eq:ricatti:mup}) and~(\ref{eq:ricatti:Dp}), the evolution of $\mu_D(t)$ has two interesting new terms. First, there is an inherently negative term pushing $\mu_D(t)$ towards zero. This term emerges because it is desirable for the agents to reduce fluctuations around the target state $V_\odot$. This can be explained by the fact that any activity-induced fluctuations lead to departure from the `ideal' memory $\mem_{0,i} = V_\odot$ and thus to a reduced reward. This effect can already be seen by the negative contribution of the activity-dependent variance $\sigma^2_{\mem_0}(D)$ to the effective reward. Agents therefore aim to minimize $D$. In consequence, learning does change $D$ albeit it does not occur directly inside the reward function. The assumed degeneracy of policies is thus resolved. Second, $\mu_D$ converges towards a non-zero value in the limit $t\rightarrow \infty$, because of the mutations term, which now enters the dynamics of $\mu_D$. This term arises because of the presence of a lower bound, $D\geq 0$, in the exponential distribution ansatz for $\tilde{\phi}_D(D)$, with a non-zero value $\tilde{\phi}_D(0)$ at the lower bound.

As shown in Fig.~\ref{fig:AOU:dynamics}, the solution of the above equations yields very good agreement with agent-based simulations. The mobility converges in a very similar way as observed for the one-dimensional case (see Fig.~\ref{fig:AOU1d:dynamics_v1}). Importantly, the final system-averaged policy $\mu_b(t) F_x \rightarrow V_\odot$ is independent of the strength of the activity (see Fig.~\ref{fig:AOU:dynamics}a). In contrast, $D$ is minimized but converges towards a small, non-zero value due to mutations, consistent with Eq.~(\ref {eq:AOU:deg:Dlong}). Due to the learning, the variance decays strongly and eventually converges towards a plateau for long times which is defined by the mutation rate. The final distribution of activities is shown in Fig.~\ref{fig:AOU:dynamics}c.  The agreement between kinetic theory (KT) and simulations is good. For small $D$ there are some deviations, which are likely caused by the assumed empirical distribution $\tilde \phi_D$ postulated for KT in Eq.~(\ref{eq:app1:deg:dist}).

\subsection{Two-dimensional policy and memory}
\label{sec:2D:policy:memory}

In this second scenario, we keep the two-dimensional policy $\policy=(b,D)$, and the goal is now to obtain a macroscopic flow with both a mean velocity in the $x$-direction $\avAS{V_x } = V_\odot$ and transverse velocity fluctuations $\avAS{v_y^2} = \sigma^2_\odot$ in the $y$-direction. 
We therefore define the observable $G(\state)\equiv \big(G_0(\bm{v}),G_1(\bm{v})\big)$ as
\begin{equation}
    G(\state) = (V_x,V_y^2)=(bF_x+v_x,v_y^2),
\end{equation}
using Eq.~(\ref{eq:AOUA_ri}). The dynamics of the two-dimensional memory $\mem=(\mem_0,\mem_1)$ is then given for $\mem_0$ by Eq.~(\ref{eq:def:mem:AOU:v0}), and for $\mem_1$ by
\begin{align}
\frac{d \mem_1(t)}{d t} &= \lambda_\mem \big(v_y(t)^2 - \mem_1(t)\big), \label{eq:mem_1}
\end{align}
where $\mem_1$ allows agents to memorize the velocity fluctuations in $y-$direction (note that $D_{\mem_1}(\policy)=0$).
The reward function is then chosen as 
\begin{equation} \label{eq:Reff:AOU:2D}
    \reward(\mem_0,\mem_1)=\reward_\odot - v_c^{-2}(\mem_0-V_\odot)^2- \sigma_c^{-4}(\mem_1-\sigma^2_\odot)^2.
\end{equation}
The dimensionless form of the reward has been made explicit by introducing normalization prefactors $v_c^{-2}$ and $\sigma_c^{-4}$,
where $v_c$ and $\sigma_c$ both have the dimension of velocity.
The reward $\reward(\mem_0,\mem_1)$ is a combination of the reward $\reward(\mem_0)$ studied in the previous subsections and of an additional contribution favoring an explicit target $\sigma^2_\odot$ for the velocity fluctuations in $y$-direction.
Doing so, the effective reward becomes
\begin{equation} \label{eq:Reff:AOU2d:v0}
    \Reff(b,D) = \reward_\odot - v_c^{-2}\Big(\meanmemzero(b,D) - V_\odot\Big)^2 - \sigma_c^{-4}\Big(\meanmemone(b,D) - \sigma^2_\odot\Big)^2 - v_c^{-2}\sigma^2_{\mem_0}(b,D) - \sigma_c^{-4} \sigma^2_{\mem_1}(b,D).
\end{equation}
We thus need to determine $\meanmemzero$, $\meanmemone$, $\sigma^2_{\mem_0}$ and $\sigma^2_{\mem_1}$.
Since the memory components $\mem_0$ and $\mem_1$ are uncoupled, their determination can be treated independently. We still assume a time scale separation $\tilde \lambda_T \rho, \Dmut \ll \lambda_\mem$, so that the stationary values of $\meanmemzero$ and $\sigma^2_{\mem_0}$ are given by Eq.~(\ref{eq:mean:var:AOU:1d:ts}).
The dynamics of $\meanmemone$ and $\sigma^2_{\mem_1}$ follows Eqs.~(\ref{eq:dm:dt:noteach}) and (\ref{eq:dsigma2:dt:noteach}) respectively, which read
\begin{align}
\p_t \meanmemone  &= \lambda_\mem \left( \avM{\avS{v_y^2}} - \meanmemone \right), \label{eq:mumemone:AOU2d}\\
\p_t \sigma_{\mem_1}^{2} &= 2 \lambda_\mem  \left[ \mathrm{Cov}_{\mem_1|b,D}(\mem_1,\avS{v_y^2}) - \sigma_{\mem_1}^{2}\right], \label{eq:sigma2memone:AOU2d}
\end{align}
where $\mathrm{Cov}_{\mem_1|b,D}(\mem_1,\avS{v_y^2})=\avM{\mem_1 \avS{v_y^2}}-\avM{\mem_1} \avM{\avS{v_y^2}} = \avM{\avS{\mem_1 v_y^2}}-\meanmemone \avM{\avS{v_y^2}}$,
using the fact that $\avS{\cdots}$ is an average over $v_y$ at fixed $\mem_1$.
We thus need to evaluate $\avM{\avS{v_y^2}}$ and $\avM{\avS{\mem_1 v_y^2}}$. In the stationary state of the memory dynamics, at fixed policy $(b,D)$, we get using It\=o calculus,
\begin{equation} \label{eq:AOU2d:mem:mean:correl}
    \avM{\avS{v_y^2}} = \frac{D}{\tau}, \qquad
    \avM{\langle \mem_1 v_y^2 \rangle} =  \frac{ D^2 (2+3\lambda_\mem \tau)}{\tau^2(2 + \lambda_\mem \tau)}.
\end{equation}
Combining Eqs.~(\ref{eq:mumemone:AOU2d}), (\ref{eq:sigma2memone:AOU2d}) and (\ref{eq:AOU2d:mem:mean:correl}), we obtain in the long-time limit
\begin{align}
\meanmemone &= \avM{\avS{v_y^2}} = D/\tau,\label{eq:app2:muD}\\
\sigma_{\mem_1}^{2} &= \avM{\langle \mem_1 v_y^2 \rangle} - \meanmemone \avM{\avS{v_y^2}}
= \frac{2 D^2 \lambda_\mem}{ \tau (2+\lambda_\mem \tau)} .
\end{align}
The effective reward function $\Reff(b,D)$ given in Eq.~(\ref{eq:Reff:AOU2d:v0}) then takes the explicit form
\begin{equation}
    \Reff(b,D) = \reward_\odot - v_c^{-2} F_x^2 \left(b-\frac{V_\odot}{F_x}\right)^2 - \sigma_c^{-4} \tau^{-2}\left( D - \tau\sigma_\odot^2 \right)^2
    - v_c^{-2} \frac{D\lambda_\mem}{1 + \lambda_\mem \tau}
    - \sigma_c^{-4} \frac{2 D^2\lambda_\mem}{\tau (2 + \lambda_\mem \tau)}.
    \label{eq:effective_reward_b_and_D}
\end{equation}
We wish to select the characteristic scales $v_c$ and $\sigma_c$ such that the learning dynamics of $b$ and $D$ occurs on a comparable time scale.
Since the policy dynamics is governed by the derivatives of $\Reff(b,D)$, we choose as a simple criterion that the coefficient in front of the target terms $(b-V_\odot/F_x)^2$ and $( D - \tau\sigma_\odot^2)^2$ should be equal, yielding $\sigma_c^{-4}=v_c^2 F_x^2 \tau^2$.
In the following, we choose units such that $v_c=1$ (consistently with the one-dimensional policy case studied above), yielding for the effective reward the final expression:
\begin{equation}
    \Reff(b,D) = \reward_\odot -(bF_x-V_\odot)^2 - F_x^2 \left( D - \tau\sigma_\odot^2 \right)^2 - \frac{D\lambda_\mem}{1 + \lambda_\mem \tau}
    - \frac{{2} F_x^2 \tau D^2\lambda_\mem}{2 + \lambda_\mem \tau}.
    \label{eq:effective_reward_b_and_D}
\end{equation}

\begin{figure}
    \centering
    \includegraphics[width=0.95\linewidth]{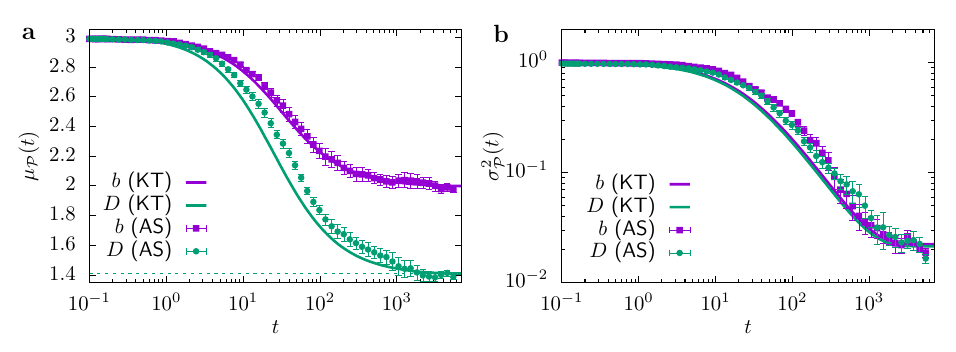}
    \caption{ Learning dynamics as predicted by kinetic theory (KT) in Eqs.~(\ref{eq:AOU:multi:bmean})-(\ref{eq:AOU:multi:Dvar}) for the two-dimensional memory scenario compared to results from agent-based simulations (AS).  \textbf{(a)} Average policies $\mu_b(t)$ and $\mu_D(t)$. The green dotted line corresponds to Eq.~(\ref{eq:AOU:multi:muDtinf}). \textbf{(b)} Diversities $\sigma^2_{b}(t)$ and $\sigma^2_{D}(t)$. The parameters are $\rho \tilde \lambda_T=0.02$, $\lambda_\mem=1.0$, $\Dpol_b=\Dpol_D=10^{-5}$, $V_\odot=2.0$, $\sigma^2_\odot=20.0,$ $F_x=1$ and $\tau=0.1.$}
    \label{fig:AOU:2d_dynamics}
\end{figure}
Turning to the policy dynamics occurring on the learning time scale, we again assume that the factorization property of $\phi(b,D)$ holds as in Eq.~(\ref{eq:fact:hyp:phibD}), thereby neglecting correlations between $b$ and $D$.
As we will see below, $D$ is now driven to a non-zero effective target value by the learning process, so that the dynamics of $D$ is no longer dominated by the lower bound $D=0$ as in Sec.~\ref{sec:AOU:degenerate}. As a result, we now assume the marginal distribution $\tilde{\phi}_D(D)$ to take a Gaussian form, such that the lower bound $D=0$ is located far in the left tail and the value $\tilde{\phi}_D(0)$ can be neglected in the boundary terms.
The dynamics of $\mu_b$, $\sigma^2_b$, $\mu_D$ and $\sigma^2_b$ obtained from Eqs.~(\ref{eq:dP:dt:multidim:fact}) and (\ref{eq:dsigma2P:dt:multidim:fact}) then reads
\begin{align}
\frac{d\mu_b}{dt} &= \tilde\lambda_T \rho \avMP{\frac{\partial \Reff}{\partial b}}\sigma_{b}^2 
= 2 \tilde\lambda_T \rho F_x (V_\odot - \mu_b F_x)\sigma_{b}^2 \label{eq:AOU:multi:bmean} , \\
\frac{d \sigma_{b}^2}{dt} &= \tilde \lambda_T \rho \avMP{\frac{\partial^2 \Reff}{\partial b^2}} \sigma_{b}^4 + 2 \Dpol_b  = - 2 \tilde \lambda_T \rho F_x^2 \sigma_{b}^4 + 2 \Dpol_b ,
\label{eq:AOU:multi:bvar}\\
\frac{d\mu_D}{dt} &=\tilde\lambda_T \rho \avMP{\frac{\partial \Reff}{\partial D}}\sigma_{D}^2 = -\tilde\lambda_T \rho \left[ 2 F_x^2 \left(\mu_D - \tau\sigma_\odot^2 \right)
+ \frac{\lambda_\mem}{1 + \lambda_\mem \tau} + \frac{{4} F_x^2 \lambda_\mem \tau \mu_D }{2+\lambda_\mem \tau} \right] \sigma_{D}^2 \label{eq:AOU:multi:Dmean} , \\
\frac{d \sigma_{D}^2}{dt} &= \tilde \lambda_T \rho \avMP{\frac{\partial^2 \Reff}{ \partial D^2}} \sigma_{D}^4 + 2 \Dpol_D
= - \tilde\lambda_T \rho F_x^2 \, \frac{{4+ 6} \lambda_\mem \tau}{2+\lambda_\mem \tau} \,\sigma_{D}^4 + 2 \Dpol_D . \label{eq:AOU:multi:Dvar}
\end{align}
We also provide in Appendix~\ref{app:Gauss:correl:AOU2d} a more detailed analysis of the policy dynamics, taking into account correlations by using a bivariate Gaussian ansatz for $\phi(b,D)$.
This analysis shows that correlations eventually disappear in the long time limit, which a posteriori justifies the approximation made here to neglect correlations. For long times, we find $\mu_b(t\rightarrow \infty)  = b_\odot  $, as in the previous example. In contrast, $\mu_D$ now also has a `trivial' target $D_\odot = \tau \sigma_\odot^2 > 0$ in addition to the negative terms pushing $\mu_D \rightarrow 0$ as discussed in the previous section. Combined we therefore find the long-time steady-state value,
\begin{equation}\label{eq:AOU:multi:muDtinf}
    \mu_D(t\rightarrow \infty) = D_\odot \frac{(2 + \lambda_\mem \tau) }{({2+3}  \lambda_\mem\tau)} - \frac{\lambda_\mem  (2+\lambda_\mem \tau )}{2F_x^2(1 + \lambda_\mem \tau){(2 + 3\lambda_\mem \tau)}} .
\end{equation}
In consequence, we observe that the desire of the collective learning to minimize fluctuations leads to a long-time policy $\mu_D(t\rightarrow \infty) < D_\odot$ which actually depends on the learning process itself.

Comparing the prediction from kinetic theory to agent-based simulations we find that the theory is indeed able to describe this two-dimensional adaptation scenario, in particular the fact that $\mu_D(t\rightarrow \infty)<D_\odot$ (see Fig.~\ref{fig:AOU:2d_dynamics}). We also observe that the adaptation of the mobility $b$ and activity $D$ happens on the same time scale,
consistently with the choice made for the normalizing scales $v_c$ and $\sigma_c$, which introduces the weight factor $\tau^2$ into the reward function.
Generally, this effect emerges because teaching efficiency does not only depend on which reward is larger but also on how much larger it is. Since the memory $\mu_{\mem_1}$ scales as $\tau^{-1}$ as shown in Eq.~(\ref{eq:app2:muD}), without the weight factor emerging reward differences would be larger for the policy $D$ as compared to the policy $b$ thus agents would react more sensitively to fluctuations in $D$. This observation underlines an important feature of the reward function, which should be specifically designed to tune the relative importance and thus sensitivity of the different components of a multi-dimensional policy.

More generally, this application to two-dimensional policies and memories shows that our kinetic theory can be applied to multi-dimensional problems without conceptual difficulties, which is essential for future applications to more complex optimization problems.

\section{Application 2: Brownian Agents with adaptive mobility and coupled diffusivity}
\label{sec:BROWNIAN}

In this second application, we would like to illustrate in a minimal model the impact of policy-dependent fluctuations on the learning process, and how such fluctuations may bias the optimal policy reached with respect to the initial target, as generically seen in Eqs.~(\ref{eq:Reff_target}) and (\ref{eq:delta_mem}). In addition, we also aim to illustrate the impact of the presence of a noise term in the evolution equation of the memory,
since such a noise term was absent from previous applications considered in Sec.~\ref{sec:AOU}.
With these goals in mind, we introduce a one-dimensional model of Brownian agents driven by an external force. The equation of motion for each agent $i$ reads
\begin{equation}\label{eq:brownian:eom}
V_i=\frac{d r_i}{d t} = b_i F_x + \sqrt{2 k_B T b_i} \eta_i(t),
\end{equation}
with position $r_i$, external force $F_x$, mobility $b_i$, temperature $k_B T$ and unit Gaussian white noise $\eta_i(t)$
satisfying $\langle \eta_i(t)\eta_j(t') \rangle_{\rm noise} = \delta_{ij} \delta(t-t')$.
At first glance, this equation looks simpler than the active Ornstein-Uhlenbeck agents introduced in the first application. However, there is a non-trivial coupling between the mobility and fluctuation strength for the Brownian agents which introduces interesting new features into the learning dynamics. 
In this simplistic setting, the agent has a single degree of freedom which is its position $r$. Hence
the agent does not have an internal state $\state$, so that there is no need to average over $\state$
(we recall that $\state$ gather the agent's degrees of freedom apart from position).

As in the previous section, the goal of the collective learning is to adapt the mobility $\policy = b$, taken as the policy, and to try to obtain a macroscopic flow with a target mean velocity $\avAS{ V } = V_\odot$. As before, the `trivial' solution is therefore $b_\odot = V_\odot / F_x $ . 
We choose the observable $V$ to define the memory (i.e., agents memorize the average drift), and set a reward function, $\reward(\mem) = \reward_\odot - (\mem - V_\odot)^2$.
The policy, the target observable and the reward function are thus a priori similar to the ones considered in Sec.~\ref{sec:AOU:1d:ts}, but the dynamics of the model differs in important ways.
The equation for the dynamics of the memory, Eq.~(\ref{eq:mem}) reads,
\begin{equation} \label{eq:mem:app1}
\frac{d\mem}{dt} = \lambda_\mem \left(V-\mem\right)
          = \lambda_\mem \left(b F_x - \mem\right) + \sqrt{2\lambda_\mem^2 b\,k_B T} \,\eta(t).
\end{equation}
Since the velocity $V$ is a degenerate coordinate, we can identify the noise-averaged processed information $G = b F_x$ and the noise on the memory is made explicit, with an amplitude $D_\mem = \lambda_\mem^2 k_B T b$.
The noise-averaged processed information $G = b F_x$ is a constant since there is no state $\state$ in the model.
As a result, one has in this simple model $\avS{G}\equiv G$ (which thus does not depend on $\mem$).
Accordingly $\avM{\avS{G}} = bF_x$ and $\partial_\mem \avS{G} = 0$. 
No time scale separation between physical and memory dynamics is needed in this model, due to its simplicity.

We now consider the dynamics of the memory mean $\meanmem$ and variance $\sigma_\mem^{2}$.
One simply has $\avM{\reward'(\mem)}= 2(V_\odot - \mu_\mem )$ and $\avM{\reward''(\mem)}=-2$, so that Eqs.~(\ref{eq:dm:dt:gauss}) and (\ref{eq:dsigma2:dt:gauss}) read:\\
\begin{align}
\p_t \meanmem &= \lambda_\mem (b F_x - \meanmem) + 2 \tilde \lambda_T \rho \,(V_\odot - \meanmem)\, \sigma_\mem^2\label{eq:mu_mem:app1},\\
\p_t  \sigma_\mem^{2} &= 2  \lambda_\mem \left [\lambda_\mem k_B T b  - \sigma_\mem^{2}\right ] - 2 \tilde \lambda_T \rho \, \sigma_\mem^4 , \label{eq:sigma_mem:app1}
\end{align}
For the following KT results, we solve these equations numerically, insert them into Eq.~(\ref{eq:Reff:approx}) and thus solve the dynamics of the mean policy and diversity using Eqs.~(\ref{eq:hatP_unif:gauss}) and (\ref{eq:sigmahatP_unif:gauss}).

\subsection{Perfect separation of the teaching time scale: $\tilde \lambda_T \rho, \Dpol_b \ll \lambda_\mem$}

\begin{figure}
    \centering
    \includegraphics[width=0.95\linewidth]{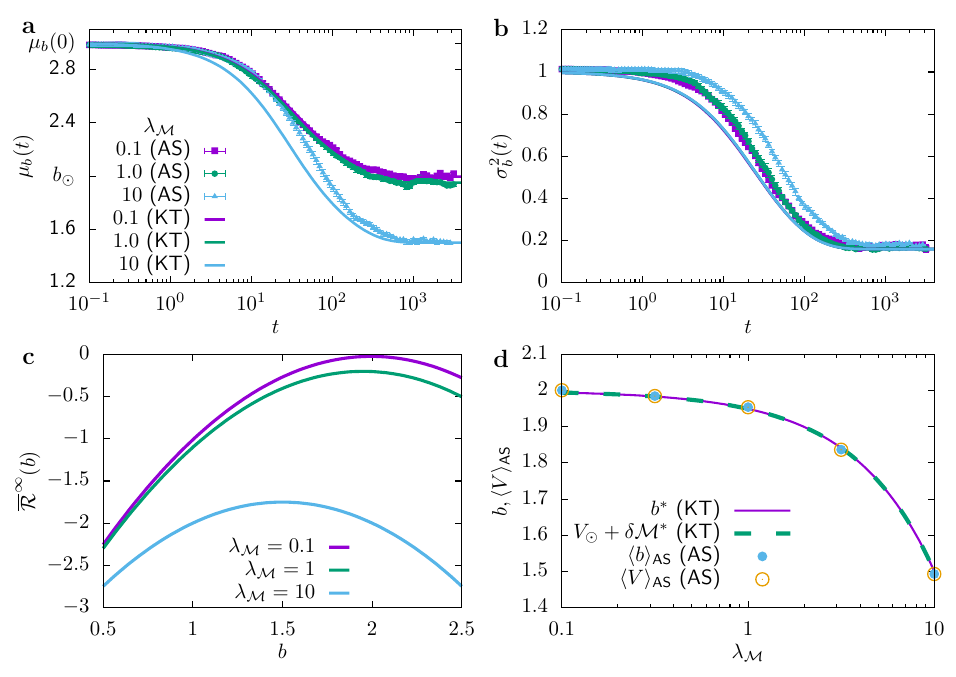}
    \caption{{\bf Learning the mobility $\policy=b$ with large time scale separation between memory and teaching, $\tilde \lambda_T \rho, \Dmut \ll \lambda_\mem$.} \textbf{(a),(b)} Average policy $\mu_b(t)$ and diversity $\sigma^2_bt)$, for the same set of memory rates $\lambda_\mem$, as obtained from the kinetic theory (KT) and from the agent based simulations (AS). \textbf{(c)} Effective long-time reward $\Reff^\infty$ predicted by kinetic theory for different memory rates $\lambda_\mem$. \textbf{(d)} Optimum $b^*$ of $\Reff^\infty(b)$ as determined analytically from kinetic theory $b^*=\mu_b^{\infty}$ (KT), Eq.~(\ref{eq:asymptotics:app1}) and shifted memory $V_\odot + \delta \mem^*$ as analytically derived in Eq.~(\ref{eq:delta_mem}). KT is compared to the long-time average of the system-averaged policy $\avAS{b}$ and actual mobility $\avAS{V}/F_x$  as extracted from agent-based simulations (AS). Parameters for all plots are $R_\odot = 0$, $F_x=1$, $V_\odot=2.0$, $\tilde \lambda_T \rho =0.01$, $\Dmut = 5\cdot 10^{-4}$,  $k_B T = 0.1$, $\mu_b(t=0)=3.0$ and $\sigma_b^2(t=0)=1.0$.}
    \label{fig:brownian:phys}
\end{figure}
When the teaching time scale, rescaled by the density, is much longer than the memory time scale, the policy $b$ can be considered as frozen when evaluating the dynamics of $\mem$. In the present case, $\mem$ then obeys the Ornstein Uhlenbeck process described by Eq.~(\ref{eq:mem:app1}), with  constant mobility $b$. Accordingly, its distribution is Gaussian at all times, with mean and variance given by,
\begin{align}
\mu_\mem(b,t) &= bF_x + \left(\mu_\mem(b,t=0) - bF_x\right)\, e^{-\lambda_\mem t},\\
\sigma_\mem^2(b,t) &= \lambda_\mem k_BT \,b \left(1-e^{-2\lambda_\mem t} \right) +\sigma_\mem^2(b,t=0)e^{-2\lambda_\mem}t,
\end{align}
which one also recovers from the integration of Eqs~(\ref{eq:mu_mem:app1}) and~(\ref{eq:sigma_mem:app1}), neglecting the terms associated with teaching.

From the knowledge of $\mu_\mem(b,t)$ and $\sigma_\mem^2(b,t)$, one immediately infers the knowledge of $\Reff(b,t) \simeq \reward(\mu_\mem(b,t)) - \sigma_\mem^2(b,t)$ [Eq.~(\ref{eq:Reff:approx})]. Yet, even in this simplified setting, one cannot solve analytically for the dynamics of the policy, using Eqs.~(\ref{eq:hatP_unif:gauss}) and (\ref{eq:sigmahatP_unif:gauss}).
Exploiting the separation of time scales, we can safely consider times that are large as compared to $\lambda_\mem^{-1}$, while remaining small as compared to the teaching times $(\tilde \lambda_T \rho)^{-1}$ and $\Dpol_b^{-1}$. In this regime, $\mu_\mem(b,t)$ and $\sigma_\mem^2(b,t)$ converge to their asymptotic values $\mu_\mem^{\infty} = b F_x$ and ${\sigma_\mem^2}^{\infty} = \lambda_\mem k_BT b$, and the asymptotic effective rewards reads:
\begin{equation}
\Reff^{\infty}(b) = R_\odot- (bF_x - V_\odot)^2 - \lambda_\mem k_BT\, b.
\label{eq:Reff:app1}
\end{equation}
The first and second derivatives with respect to $b$ follow: $\Reff' = -2 F_x^2 \left(b - \left(\frac{V_\odot}{F_x}-\frac{\lambda_M k_B T}{2F_x^2}\right)\right)$, $\Reff'' = -2 F_x^2$. Since the second derivative $\Reff''$ is constant, the general Ricatti solutions, Eqs.~(\ref{eq:ricatti:mup}) and~(\ref{eq:ricatti:Dp}) for the dynamics of $\mu_b$ and $\sigma^2_b(t)$ apply. In the long time limit we thus find
\begin{equation}
\mu_b^{\infty} = b_\odot - \frac{\lambda_\mem k_B T}{2F_x^2}
\quad \text{and} \quad
{\sigma_b^2}^{\infty}= \sqrt{\frac{\Dpol_b}{\tilde \lambda_T \rho F_x^2}}
\label{eq:asymptotics:app1}
\end{equation} 
that are reached on time scales larger than $\tau_2 = (4 \Dpol_b \tilde \lambda_T \rho F_x^2)^{-1/2}$. We thus predict that the mean velocity reached by the learning dynamics $\avAS{ V } = \mu_b^{\infty} F_x$ misses the target velocity $V_\odot$ from below because of the thermal noise, whose amplitude depends on the mobility.

Figure~\ref{fig:brownian:phys} summarizes our findings and compares them with direct agent-based simulations of the model. $N=2000$ agents are initiated with a policy drawn for each agent from a normal distribution with mean $\mu_b(t=0)=3.0$ and variance $\sigma^2_b(t=0)=1.0.$ 
Figure~\ref{fig:brownian:phys}a,b respectively display the dynamics of the mean policy $\mu_b(t)$ and policy diversity $\sigma^2_b(t)$, which are essentially following the Ricatti solutions, that is an hyperbolic tangent interpolation between their initial conditions and their asymptotic values. The asymptotic value for $\sigma_b^2$ is independent of $\lambda_\mem$ as predicted in Eq.~(\ref{eq:asymptotics:app1}). 
Conversely, the asymptotic value for $\mu_b$ decays with $\lambda_\mem$, again in agreement with Eq.~(\ref{eq:asymptotics:app1}). The reason, illustrated on Fig~\ref{fig:brownian:phys}c, is that the effective reward is shifted by the memory diffusion term, which depends on the policy $b$ (see Eq.~(\ref{eq:Reff:app1})). 
As a result the optimal value $b^*$, which is also $\mu_b^{\infty}$, decays according to Eq.~(\ref{eq:asymptotics:app1}), as shown on Fig~\ref{fig:brownian:phys}d. In all cases the agreement between the kinetic theory and the agent-based simulations is very good. 

This result is also interesting from a general perspective. It shows that fluctuations in the memory, and therefore in the determination of the effective reward can lead to significant shifts in the final policy. In the present example, the effect is caused by the scaling of the thermal fluctuations with the mobility $b$ which was not present in the first application. Effectively, the agents prefer to obtain slightly lower mobilities $b$ than the target one, but therefore fewer fluctuations and thus more consistent rewards. One could imagine similar scenarios in real-world optimization processes or even in evolutionary settings where agents settle for policies which lead to consistent reward instead of high risk policies which would promise higher rewards. Importantly, if the goal is to promote a certain collective behavior characterized by $V_\odot$ the above example shows that deviations $\delta \mem^* \neq 0$ can emerge from non-trivial evolutionary dynamics. In this case, a significant time scale separation needs to be maintained to minimize $\delta \mem^*$, as derived in Eq.~(\ref{eq:delta_mem}).

\subsection{Overlapping memory and teaching time scales}

The situation becomes more complex if the teaching and memory time scales overlap. In this case we solve the dynamical equations~(\ref{eq:mu_mem:app1}-\ref{eq:sigma_mem:app1}) for a representative set of policies $b$, governing the mean and variance of the memory, and thereby obtain the temporal evolution of $\Reff(b,t)$. The time-dependent effective reward can then be used as input to solve Eqs.~(\ref{eq:hatP_unif:gauss}-\ref{eq:sigmahatP_unif:gauss}) for the dynamics of the mean policy and diversity within the Gaussian approximation.

\begin{figure}
    \centering
    \includegraphics[width=0.95\linewidth]{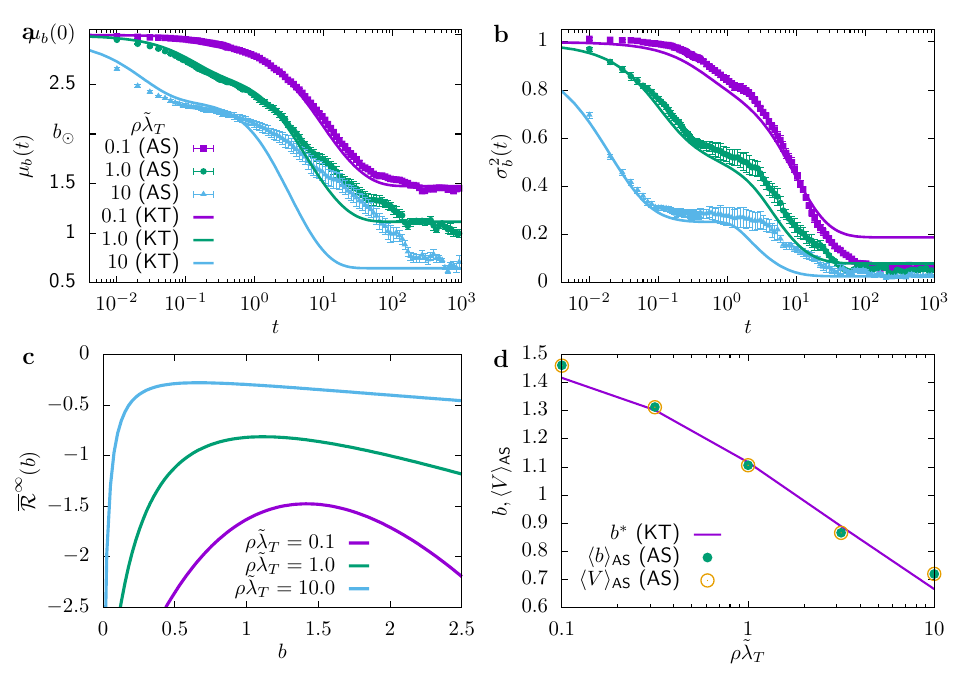}
    \caption{{\bf Learning  the mobility $\policy=b$ with overlapping memory and teaching time scales.} \textbf{(a)} Average policy $\mu_b(t)$ and \textbf{(b)} diversity $\sigma^2_b(t)$ for various effective teaching rates $\rho \tilde \lambda_T$. \textbf{(c)} Effective long-time reward $\Reff^\infty$ approximated from the kinetic theory for different effective teaching rates $\rho \tilde \lambda_T,$ leading to an overlap of the physical, memory, and teaching time scales. \textbf{(d)} Optimal policy $b^*$ estimated from the maximum of $\Reff(b)$ as compared to the long-time average of the system-averaged policy $\avAS{b}$ and the actual mobility $\avAS{V}/F_x$ as extracted from agent-based simulations (AS). The latter two quantities are perfectly overlapping. Parameters for all plots are $\reward_\odot = 0$, $F_x=1$, $V_\odot=2.0$, $k_B T = 1$, $\Dpol_b = 0.005$,  $\lambda_\mem = 1$.}
    \label{fig:brownian:teaching}
\end{figure}

Numerical results when the memory and teaching time scales overlap are shown in Fig.~\ref{fig:brownian:teaching}. In very good agreement between theory and simulations, we observe that a two-step relaxation is emerging in the dynamics of both the mean policy (Fig.~\ref{fig:brownian:teaching}a) and the diversity (Fig.~\ref{fig:brownian:teaching}b). The first step is connected to teaching on a time-scale on which the memory has little changed from its initial value $\mem_i(t=0) = b_i(t=0) F_x$.  Only on larger time scales, when the memory indeed had time to adjust, we find the final long-time steady-state solution. The most prominent deviation between theory and simulations is observed for $\rho \tilde \lambda_T=10$. Here, the kinetic theory features significantly faster learning and thus also a reduced diversity. This is likely because $\Reff$ is very flat for $b>0.5$ and thus small deviations between theory and simulations can make a huge impact (see Fig.~\ref{fig:brownian:teaching}c).

In the long-time limit one can actually find the solution of the memory dynamics analytically using Eqs.~(\ref{eq:mu_mem:app1}-\ref{eq:sigma_mem:app1}):\\
\begin{align}
\meanmem(b) &= \frac{b F_x + (\sqrt{1+\chi b }-1) V_\odot  }{\sqrt{1+\chi b  }} , \\
\sigma^2_\mem(b) &=  \frac{\lambda_\mem}{2 \rho \tilde{\lambda}_T}\left( \sqrt{1+\chi b } - 1 \right),
\end{align}
with $\chi=4 k_B T \rho \tilde{\lambda}_T$.
For the effective reward $\Reff(b)= \reward(\meanmem(b)) - \sigma^2_\mem(b)$ we thus find
\begin{equation}
\Reff^{\infty}(b) = \reward_\odot
-\frac{(bF_x - V_\odot)^2}{1+\chi b}
-\frac{\lambda_\mem}{2\rho \tilde{\lambda}_T}
\left(\sqrt{1+\chi b}-1\right).
\end{equation}
A plot of $\Reff^{\infty}(b)$ is displayed on Fig.~\ref{fig:brownian:teaching}c. 
As shown on Fig~\ref{fig:brownian:teaching}d, our analytical calculations match the results of agent-based simulations, indicating that despite the lack of time scale separation between the memory and reward dynamics, evaluating the asymptotics for the memory analytically remains a good proxy for evaluating the long-time policy.

Quite remarkably, we observe that for more frequent teaching, the optimal policy, maximizing the effective reward, shifts even further away from the desired value $b_\odot=2$. As discussed in the previous subsection, the existence of policy-dependent fluctuations favors a negative shift of $\mu_b$ away from $V_\odot.$ Since $\meanmem \approx \mu_b F_x$  the same observation therefore also holds for the average memory. Teaching, however, will always push $\meanmem$ towards $V_\odot$ since not only policies, but also memory is transferred during teaching events and the reward $\reward$ favors agents with an instantaneous memory $\mem =V_\odot$. This effect increases the reward towards $\reward \rightarrow \reward_\odot$ but it weakens the impact of the state-dependent contribution $\sim \lambda_\mem$ in the memory dynamics. In consequence, small mobilities $b$ with few fluctuations are even further promoted.

It is interesting to investigate whether the above effect depends on the nature of the teaching events. In App.~\ref{app:nomemory} we study the same system as above but for a decentralized learning scenario in which the memory is not adapted during teaching events and instead is kept constant. We find that the dependence of the long-time policy on $\tilde \lambda_T$ completely vanishes. Instead, the learning time, i.e., the time required to reach this long-time policy now depends sensitively on $\tilde \lambda_T$ (see Fig.~\ref{fig:brownian:teaching:nomemory}). We rationalize this finding by deriving an alternative kinetic theory without memory adaptation in App.~\ref{app:nomemory}.

Altogether this second application shows that even in the simplest settings the overlapping time scales can very significantly affect the final policies. Conversely, this short example highlights that it is very difficult to judge from a given learned policy whether this indeed is the ``ideal'' behavior or whether it just emerged as ``compromise'' in a highly complex and dynamical evolutionary process.


\section{ Discussion and Conclusion: Results, Limitations and Connections to other Fields}
\label{sec:discussion}

\subsection{Summary of important findings}

In this paper, we have provided a detailed theoretical framework to describe the evolution of the statistics of the collective learning process performed by a spatially homogeneous assembly of smart active agents, which model microrobots as used in experiments \cite{ben2023morphological}. Beside motility, such agents have the ability to communicate within some spatial interaction range. Communication between agents is a cornerstone of the learning process, as the agents progressively adapt their policy $\policy$ (i.e., the parameters determining their dynamical rules), by comparing the reward associated with their strategy to the ones of their neighbors, and copying the neighbors' policy (with a higher probability) when the corresponding reward is higher than the agent's reward. Evaluating the reward in turn requires the agent to keep track of a memory variable $\mem$, by performing a running average of a specific observable (related to the target collective state) averaged over a running time window. The reward $\reward(\mem)$ thus allows each agent to assess the optimality of its current policy with respect to the prescribed target average state, encoded in the reward function $\reward(\mem)$.

The present theoretical framework is based on a kinetic theory approach, which is well suited to deal with the pairwise teaching interactions as considered in the present model. A central quantity in the statistical characterization of the learning process is the policy density $\phi(\policy)$, namely the policy distribution normalized to the local, real-space agent density $\rho$. To characterize the learning process we have then focused on deriving evolution equations for the policy mean $\mu_\policy(t)$ and variance/diversity $\sigma_\policy^2(t)$.

The dynamics of $\phi(\policy)$ is dictated by the effective reward function $\Reff(\policy)$, which is an emerging quantity rating how well the policy $\policy$ manages to reach the target average state. The effective reward $\Reff(\policy)$ represents the connection between our microscopic theory in which $\Reff(\policy)$ emerges from the combined effect of physical and memory dynamics, and postulated macroscopic theories in which $\Reff(\policy)$ is a predefined input. Importantly, we find that $\Reff(\policy)$ often predicts non-trivial reward maxima which are not apparent from the microscopic reward $\reward(\mem).$  Evaluating $\Reff(\policy)$
 requires evolution equations for the moments of the memory $\mem$.
The latter can be closed by assuming a Gaussian statistics for the memory $\mem$, which is a natural assumption from the Central Limit Theorem, since $\mem$ is an average over many random contributions, assuming the memory time $\tau_\mem$ to be much larger than the physical characteristic time $\tau_{\rm phys}$.
Throughout the presentation of this theoretical framework, we have paid attention to remain general as long as possible, and to introduce additional restrictive assumptions only when needed. We also emphasized the role played by covariances of memory and/or policy observables in the dynamics of memory and policy means and variances.

We have applied our theoretical frameworks to two different explicit microscopic models to illustrate several important outcomes of our work, yielding a satisfactory and qualitative (or in some cases even quantitative) agreement between agent-based simulations and theoretical predictions. Starting from a simple example of the AOUP model of active particles coupled to an external field, we have first considered the basic case when the reward encodes the optimization of a single average observable, before turning to the two-dimensional case in both policy and memory space. Even in this simple example, we could show very non-trivial learning dynamics emerging from the impact of diversity and mutations, from overlapping times scales, and from reward fluctuations impacting the long-time steady-state policies. We also illustrated on an elementary model of a diffusive agent how policy-dependent fluctuations may bias the learning phase, and lead to a shifted optimal state as a result of an asymmetric noisy optimization process.

\subsection{Limitations of the present theoretical framework and potential future generalizations}

Although the present framework has proved successful in describing, at least qualitatively, the simple and illustrative models of Secs.~\ref{sec:AOU} and~\ref{sec:BROWNIAN},
the approach developed here suffers from several important limitations, as listed below, which were needed to make calculations tractable:
\begin{itemize}
\item Physical interactions between agents are neglected;
\item The statistical description is formulated in terms of the single-agent phase-space density, which does not include correlations between nearby agents;
\item Some parts of the derivation rely on a time scale separation between the physical dynamics and the learning (as well as memory) dynamics;
\item Evolution equations on the policy mean and variance require to be closed by assuming a given form (e.g., Gaussian or exponential-like) of the policy density $\phi(\policy)$, and expanding policy-dependent functions in terms of low-order moments;
\item The hyperbolic tangent in the expression (\ref{eq:def:PT}) of $p_T$ has been linearized by taking a small-$\alpha_T$ limit.
\end{itemize}

Lifting at least some of these assumptions or restrictions is certainly a desirable goal for future work. In particular, including physical interactions between agents will be a necessary step to account for experimental observations beyond the diluted regime, for instance when robots gather in a light spot \cite{ben2023morphological}.
Taking into account steric repulsive interactions can in principle be done by following the same steps as in usual active matter, like for the description of Motility-Induced Phase Separation (MIPS) \cite{Cates_MIPS_review2015}. However, this physical description of interactions will only be a starting point for the description of the learning process (including the memory variable), likely yielding overall rather complicated calculations.

Going beyond the time scale separation assumption may also be an interesting generalization to explore a potentially new phenomenology. If the teaching time scale is not well separated from the physical one, the behavior of the system may not purely consist in a learning process resulting in an optimized collective state, but one may rather observe the emergence of (possibly stationary) space-dependent policy currents, leading to a non-trivial spatial organization of the system where perpetual learning and exploration takes place.
In addition, it would in principle be possible to write down a kinetic theory in terms of the two-agent phase-space density to take into account pairwise correlations.
Yet, such an approach may face the usual difficulty of closing the hierarchy of multi-agent phase-space densities.

The remaining more technical assumptions or approximations may not be easily overcome. Various closures, like various forms of the policy distribution, may be used. However, most of them remain adhoc assumptions, although they may be supported by numerical simulation results (like the exponential-like policy distribution considered in Sec.~\ref{sec:AOU:degenerate}). A first-principle derivation of the shape of the policy distribution would be useful, but it presently remains out of reach. Also, going beyond the linearization of the hyperbolic tangent in the expression of $p_T$ a priori leads to untractable calculations in the ensuing steps due to non-linear terms.

More generally, it is also of interest to briefly comment on the differences with previously considered kinetic theories in other context, including e.g., gases~\cite{boltzmann1896KT}, granular gases~\cite{brilliantov2004kinetic}, and active particles with alignment interactions~\cite{peshkov2014boltzmann}.
A significant difference with our kinetic theory of learning, which actually lies in the definition of the microscopic dynamics, is that in previously considered kinetic theories, interactions were triggered by collisions, when two particles randomly meet due to their respective ballistic motion. The interaction frequencies thus emerge from the multiparticle dynamics itself.
Instead, in our kinetic theory for decentralized learning, the interaction frequency is a microscopically given parameter. The reason is that interactions between agents now take the form of actively triggered communications, which do not passively result from the physical dynamics like collisions between particles. In contrast to other kinetic theories, by comparing to agent-based simulations, we therefore observe that our theory of learning usually becomes more accurate for high densities since the assumption of replacing individual agents by phase-space densities is more accurate.

\subsection{Connection to swarm robotics}

The present theoretical framework provides a natural starting point to describe learning dynamics in swarms of robots. Indeed, the microscopic rules considered here are directly inspired by decentralized social learning protocols in which robots locally compare their performances and copy, with a higher probability, the policy of better performing neighbors~\cite{bredeche2022social, ben2023morphological}. From this point of view, the kinetic theory developed in the present work offers a statistical description of how a population-level distribution of policies evolves under the combined effects of social transmission, policy diversity and fluctuations in the evaluation of rewards. It therefore provides a useful bridge between microscopic robotic implementations of social learning and macroscopic descriptions in terms of effective reward landscapes and policy distributions. At the same time, applying this framework quantitatively to robotic swarms will require overcoming several of the limitations discussed above. 

A first important challenge is to include the physical interactions. In many robotic systems, and in particular in dense swarms, steric interactions, collisions, pushing, blocking, alignment through contact, or aggregation in confined regions are not perturbative corrections. They may instead become part of the computation performed by the swarm itself, a mechanism often referred to as morphological computation~\cite{pfeifer2009morphological,ben2023morphological}. In such situations, the policy does not act on an otherwise passive physical dynamics, but rather modifies a collective physical process which already performs part of the task. Understanding how social learning couples to this physical computation is therefore a central challenge for future theoretical developments.

A second limitation concerns the time scale separation assumed throughout most of the present approach. In robotic experiments, the physical motion, the memory update and the learning events may occur on comparable time scales. The reward of a given policy may then be evaluated before the corresponding physical response has reached a stationary regime, and policy changes may continuously reshape the physical state from which rewards are measured. In this regime, the effective reward function introduced here may no longer be a quasi-static function of the policy alone, but may depend on the history of the coupled physical and learning dynamics. This could lead to persistent exploration, oscillatory learning dynamics, or spatially organized policy currents rather than convergence to a fixed optimal policy. Such regimes are difficult to capture analytically, but they are also precisely where robotic experiments can be most informative.

A third point concerns the form of the teaching probability. For analytical tractability, the hyperbolic tangent entering the teaching rule has been linearized in the small-($\alpha_T$) limit. Many robotic implementations, however, operate closer to the opposite limit, in which the best-performing robot is almost always selected as the teacher. This large-($\alpha_T$) regime may amplify fluctuations in the reward evaluation, enhance fixation events, and make the learning dynamics more sensitive to finite-size effects. Extending the kinetic theory beyond the weak-selection approximation would therefore be important in order to compare more directly with experimental social-learning protocols.

Another important direction concerns the nature and dimensionality of the policy itself. In the examples considered in the present work, the policy is directly one or several physical parameters of the dynamics; it may also be a small set of parameters defining a simple controller that maps environmental cues onto physical parameters \cite{jung2026inhomogeneous}. In swarm robotics, one may instead wish to use more agnostic controllers, for instance neural networks whose parameters define a high-dimensional policy. The learning dynamics then takes place in a large-dimensional policy space, possibly with strong redundancy between parameters and many nearly equivalent controllers. The role of over-parametrization in this context is largely open. It may facilitate learning by providing many paths toward efficient behaviors, as in standard machine learning, but it may also slow down social learning by diluting meaningful policy variations in a large neutral space. Understanding how the structure and dimensionality of the controller affect the effective reward landscape, the maintenance of policy diversity and the convergence time is therefore an important question at the interface between statistical physics, machine learning and swarm robotics~\cite{ariosto2025replication}.

Finally, the present work has focused on purely social learning: agents collectively improve their behavior by comparing rewards and copying policies, while the individual policy of a robot does not improve through its own trial-and-error experience. A natural extension would be to couple this social learning dynamics to individual reinforcement learning performed by each robot. Such a hybrid setting would raise several questions. How should a robot balance exploration through individual reinforcement learning with exploitation of policies discovered by others? When is it more efficient to copy a successful neighbor, and when is it preferable to keep exploring individually? How should the relative rates of individual and social learning be tuned in order to maximize the speed, robustness and adaptability of the swarm? Addressing these questions would bring the present kinetic theory closer to realistic robotic learning architectures, while also opening a broader statistical-physics perspective on the collective dynamics of interacting learners.

\subsection{Connection to evolutionary biology}
\label{sec:evolution}

The collective learning process described in this manuscript bears a close resemblance to evolutionary dynamics. If we interpret the reward as a notion of fitness, the transfer of policies can be viewed as a death–birth process, in which a poorly performing parent is replaced by a better-adapted offspring. It is therefore not surprising that the reduced equations obtained through coarse-graining correspond to, or closely resemble, macroscopic equations commonly studied in population genetics \cite{CrowKimura1970} and evolutionary game theory \cite{hofbauer2003evolutionary}.

First, we note that Eq.~(\ref{eq:dt:phi:homog_v1}) is very similar to a replicator equation in continuous policy space \cite{taylor1978evolutionary,schuster1983replicator,hofbauer2003evolutionary}. In particular, the term $(\Reff(\policy) - \avMP{\reward})\phi(\policy)$ directly corresponds to the selection term in the replicator equation. Unlike standard formulations, which assume a predefined policy-dependent fitness function $\Reff(\policy)$, our framework provides a microscopic derivation of $\Reff(\policy)$ from the underlying learning dynamics. Importantly, this reveals that $\Reff(\policy)$ can itself depend on the learning process. Additionally, our microscopic derivation leads to a fitness function which does not depend on the relative abundances of the individual species. Understanding the implications of this would be an exciting path for future research.  The multiplicative mutations in the replicator-mutator equation \cite{komarova2004replicator} differ from the simple diffusive and additive mutations in our model. This suggests that realistic models of evolutionary dynamics may require more expressive mutation processes than currently used in our theory, potentially including reward-dependent diffusion coefficients $\Dmut = \Dmut(\reward)$. Such mechanisms may offer a route toward understanding phenomena such as stress-induced mutagenesis \cite{bjedov2003stress,foster2007stress}.

Another central macroscopic model in evolutionary game theory is the Moran process, which can be seen as a finite-population counterpart of the replicator equation \cite{moran1958random,lieberman2005evolutionary,cengio2026evolution}. The policy update rules in the Moran model closely resemble those of our agent-based formulation. However, our framework extends this setting by incorporating spatial dynamics in physical space and by inducing an emergent effective reward function $\Reff(\policy)$. Finally, the Lotka–Volterra (LV) equations constitute a widely used nonlinear model in biology, ecology, and statistical physics for describing interacting populations \cite{lotka1925elements,volterra1926variazioni,abrams2000evolution}. Since LV dynamics can be recovered as a special case of the replicator equation with two strategies and prescribed fitness functions, they can likewise emerge within our collective learning framework. In this interpretation, one considers two competing species with distinct reward functions: agents of species $x$ are rewarded for proximity to their own type and distance from species $y$, whereas agents of species $y$ are rewarded for proximity to species $x$ and separation from conspecifics.

Making these connections more explicit may help unify and rationalize macroscopic models used in evolutionary game theory. A natural direction for future work is therefore the systematic derivation of macroscopic equations from specific agent-based learning models with a given reward structure $\reward(\mem)$, including extensions that incorporate more complex mutation mechanisms or interacting species with competing reward functions
(in analogy to the Lotka–Volterra setting), and take into account the macroscopic space-dependence in the collective dynamics.

\subsection{Connection to machine learning}

Finally, it is also of interest to briefly discuss the similarities and differences between our decentralized learning protocol designed for groups of interacting simplified microrobots,
and more general decentralized learning protocols which have been attracting increasing attention in the machine-learning (ML) community. In a conventional machine-learning setting, a ML model is trained from a dataset collected on a central server, and the model parameters are optimized by a central computer. Although this centralized framework is often effective, it can raise important concerns regarding privacy, security, and robustness. For these reasons, decentralized federated learning has emerged as an alternative paradigm~\cite{yuan2024decentralized,gabrielli2023survey}. In such approaches, multiple agents or clients keep their own local datasets, train local ML models independently, and then communicate model parameters or updates with other agents to reach a consensus corresponding to a global model, without requiring a central server. Statistical mechanical descriptions of federated learning in this direction has only recently emerged~\cite{arola2024effective,catania2026solvable}.

Our decentralized learning framework for smart active matter is related to this general idea, in the sense that each agent possesses an internal policy that plays a role analogous to ML model parameters. However, there is also an important conceptual distinction. In standard machine learning, the local model is typically updated by minimizing an explicit loss function. In contrast, in our framework, agents do not perform gradient-based optimization. Instead, each agent evaluates an individual reward, compares this reward with that of neighboring agents, and copies the neighbor's policy with a probability based on the reward difference. In this sense, our learning rule is closer to evolutionary algorithms \cite{back1993overview,de2017evolutionary}, population-based training \cite{jaderberg2017population}, neuroevolution \cite{stanley2019designing} or imitation processes \cite{hussein2017imitation} than to a standard optimization-based machine-learning procedure. 

Another important point is the role of communication topology~\cite{beltran2023decentralized}. In decentralized federated learning, the network structure strongly affects convergence speed, robustness, and communication efficiency. Similar considerations arise in our case, but with a distinctive feature: because the agents move in space, the effective interaction network is a dynamic graph defined by spatial proximity rather than a fixed communication graph. This makes our framework particularly interesting from the viewpoint of statistical physics, since the learning process is intrinsically coupled to the physical motion of the agents. 

Discussions on the similarities and the discrepancies between decentralized federated learning and our decentralized learning framework for smart active agents may therefore be very fruitful. On the one hand, ideas from machine learning, such as consensus mechanisms, communication protocols, or topology-aware updates, may inspire new learning strategies for active matter. On the other hand, the statistical physics viewpoint developed here may further suggest new decentralized learning processes beyond the standard machine-learning setting.

\acknowledgments  
	
	This work has been supported by MIAI@Grenoble Alpes (under grant ANR-19-P3IA-0003) and by the grant ANR-24-CE33-7791 Spectral-Swarm-Robotics. This research was partly funded by the Austrian Science Fund (FWF) 10.55776/PAT1139125 (GJ).


\appendix

\section{Learning without memory exchange}
\label{app:nomemory}

In Sec.~\ref{sec:BROWNIAN} we found that if the effective teaching rate $\tilde{\lambda}_T \rho$ is not significantly smaller than the memory rate $\lambda_\mem$ this can have severe implications for the learning dynamics and, in particular, the long-time policy adopted by the agents. In this appendix we investigate how this observation is connected to memory adaptation during teaching events, as described after Eq.~(\ref{eq:def:PT}). As a second scenario we therefore study the case in which the memory is not adapted during teaching.

\begin{figure}
    \centering
    \includegraphics[width=0.99\linewidth]{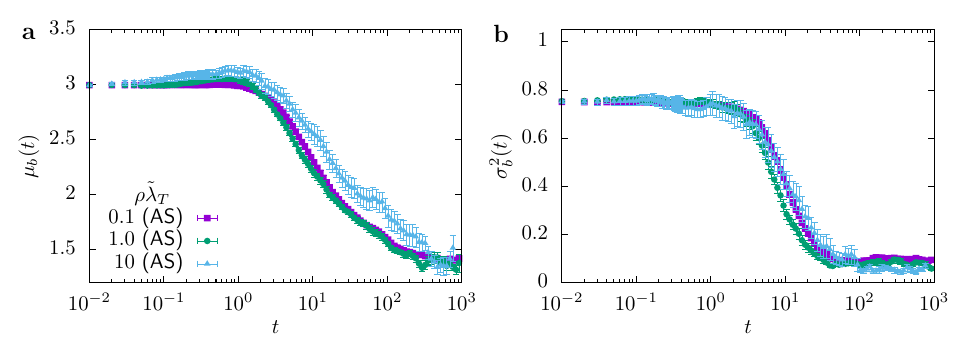}
    \caption{{\bf Learning the mobility $\policy=b$ with overlapping memory and teaching time scales without memory adaptation.} \textbf{(a)} Average policy $\mu_b(t)$ and \textbf{(b)} diversity $\sigma^2_b(t)$ for various effective teaching rates $\rho \tilde \lambda_T$. This figure shows data for the same parameters as Figs.~\ref{fig:brownian:teaching}a,b with the exception that the memory is not adapted during teaching events and instead is just kept at the same value as before.}
    \label{fig:brownian:teaching:nomemory}
\end{figure}

By performing agent-based simulations we find in Fig.~\ref{fig:brownian:teaching:nomemory} that the situation without memory adaptation differs significantly from the results presented in the main text in Fig.~\ref{fig:brownian:teaching}. In particular, the long-time policy seems to be independent of the effective teaching rate, as shown in Fig.~\ref{fig:brownian:teaching:nomemory}a. In contrast, now the learning time, i.e., the time scale at which the long-time policy is reached, depends sensitively on $\tilde \lambda_T.$ There is clearly an optimal value for $\tilde \lambda_T$ to minimize the learning time, which is reached for approximately $\tilde \lambda_T \rho= 1.$
In the following we want to rationalize this simulation result by adapting the kinetic theory to the case without memory adaptation.

We start by rewriting Eq.~(\ref{eq:def:WL}) without memory adaptation, 
	\begin{align}
	\hspace{-0.5cm} \nonumber 
	W_{f}(\mem',\policy'|\mem,\policy; \br,t) &= 2 \lambda_T \hspace{-0.1cm} \int \hspace{-0.1cm} d\state_2 \int \hspace{-0.1cm} d\mem_2 \hspace{-0.1cm}\int \hspace{-0.1cm} d \policy_2 \int \hspace{-0.1cm} d\br_2 K(\br_2,\br)\\ \nonumber
	& \qquad \times p_T(\reward(\mem_2),\reward(\mem)) f(\state_2,\mem_2,\policy_2,\br_2,t)  \delta(\policy^\prime-\policy_{2}) \delta(\mem^\prime-\mem).
	\label{eq:def:WL}
	\end{align}
Here, we have exchanged $\delta(\mem^\prime-\mem_2)$ by $\delta(\mem^\prime-\mem)$, i.e., the agent keeps its old memory instead of adopting the one of agent 2. We now immediately insert $p_T \approx \frac{1}{2}[1+\alpha_T(\reward(\mem_2)-\reward(\mem))]$, because otherwise the missing $\delta(\mem^\prime-\mem_2)$ makes it hard to simplify this expression. We find,
\begin{equation}
W_{f_0}(\mem',\policy'|\mem,\policy) =   \lambda_T \rho A_c \left[ 1 + \alpha_T (\avM{\reward}(\policy') - \reward(\mem))  \right] \phi(\policy') \delta(\mem^\prime-\mem) .
\label{eq:nomem:Wf0}
\end{equation}
Inserting this expression into the Master equation yields (see Eq.~(\ref{eq:Iteach})),
\begin{align}
I_{\rm teach}[f] =& \lambda_T \rho A_c \int d\policy' \Big[  \left[ 1 + \alpha_T (\avM{\reward}(\policy) - \reward(\mem))  \right] \phi(\policy) f(\state,\mem,\policy') -\left[ 1 + \alpha_T (\avM{\reward}(\policy')- \reward(\mem))  \right] \phi(\policy') f(\state,\mem,\policy) \Big] \nonumber\\
=&  \lambda_T \rho A_c \int d\policy'   \left[ 1 + \alpha_T (\avM{\reward}(\policy) - \reward(\mem))  \right] \phi(\policy) f(\state,\mem,\policy') - \lambda_T \rho A_c \left[ 1 + \alpha_T (\avMP{\reward}- \reward(\mem))  \right] f(\state,\mem,\policy)  .\label{eq:Iteach_nomem}
\end{align}
In the following, since we focus on the teaching term, we omit the other terms such as the memory term in dynamical equations for simplicity.
The teaching term  in the time-evolution of $f_0$ is therefore,
\begin{align}
\partial_t f_0(\mem,\policy) =  \lambda_T \rho A_c \int d\policy'   \left[ 1 + \alpha_T (\avM{\reward}(\policy) - \reward(\mem))  \right] \phi(\policy) f_0(\mem,\policy') - \lambda_T \rho A_c \left[ 1 + \alpha_T (\avMP{\reward}- \reward(\mem))  \right] f_0(\mem,\policy) . \label{eq:f0_nomem}
\end{align}
First, we derive the dynamics of $\phi(\policy) = \rho^{-1} \int d\mem f_0(\mem,\policy)$,
\begin{align}
\partial_t \phi(\policy) &=  \lambda_T \rho A_c \left[ 1 + \alpha_T (\avM{\reward}(\policy) - \avMP{\reward})  \right] \phi(\policy)  - \lambda_T \rho A_c \left[ 1 + \alpha_T (\avMP{\reward}- \avM{\reward}(\policy))  \right] \phi(\policy)  \\
&= \tilde{\lambda}_T \rho \phi(\policy) (\avM{\reward}(\policy) - \avMP{\reward}) ,
\label{eq:phi_nomem}
\end{align}
where we used Eq.~(\ref{eq:def:tildelambda}).
This is identical to the learning with memory exchange shown in Eq.~(\ref{eq:dt:phi:homog_v1}). 

Let us now introduce the moments $\mu_\mem(\policy) =[\rho \phi(\policy)]^{-1} \int d\mem \mem f_0(\mem,\policy)$ which we can derive from Eq.~(\ref{eq:dt:R:mem:v0}) using the updated time-evolution for $f_0,$
\begin{align}
[\rho \phi(\policy)]^{-1}\int d\mem \mem \partial_t f_0(\mem,\policy) &= \lambda_T \rho A_c  \left[ \avMP{\mu_\mem} + \alpha_T (\avM{\reward}(\policy) \avMP{\mu_\mem} - \avMP{\mem\reward})  \right]  - \lambda_T \rho A_c \left[ \mu_\mem(\policy) + \alpha_T (\avMP{\reward}\mu_\mem(\policy) - \avM{\mem \reward}(\policy))  \right] . \label{eq:mumem_nomem}
\end{align}
As a result, we find for the full time-evolution,
\begin{align}
\partial_t \mu_\mem(\policy) &=  \lambda_T A_c \rho  \left[ \avMP{\mu_\mem} - \mu_\mem(\policy) + \alpha_T \avM{\reward}(\policy) ( \avMP{\mu_\mem} - \mu_\mem(\policy)) + \alpha_T ( \avMP{\reward}\mu_\mem(\policy) - \avM{\reward}(\policy)\mu_\mem(\policy)   - \avMP{\mem\reward}   + \avM{\mem \reward}(\policy))  \right] . \label{eq:mumem_nomem}
\end{align}
In the same way we introduce $\sigma^2_\mem(\policy) = [\rho \phi(\policy)]^{-1} \int d\mem (\mem - \mu_\mem(\policy))^2 f_0(\mem,\policy)$. By applying Eq.~(\ref{eq:dt:R:mem:v0}) to $\mem^2 \; -2 \mu_\mem(\policy)\mem \; + \mu_\mem(\policy)^2$ we end up with
\begin{align}
[\rho \phi(\policy)]^{-1}\int d\mem (\mem - \mu_\mem(\policy))^2 \partial_t f_0(\mem,\policy) &=  \lambda_T A_c \rho  \left(1+\alpha_T \avM{\reward}(\policy)\right)\left( \avMP{\mem^2} - 2\mu_\mem(\policy) \avMP{\mu_\mem}+\mu_\mem(\policy)^2\right) \nonumber \\ & \quad - \lambda_T A_c \rho\alpha_T\left( \avMP{\mem^2\reward}  - 2 \mu_\mem(\policy) \avMP{\mem \reward} + \mu_\mem(\policy)^2 \avMP{\reward}\right) \nonumber \\ & \quad - \lambda_T A_c \rho \left(1+\alpha_T \avMP{\reward}\right) \sigma^2_\mem (\policy) \nonumber \\& \quad + \lambda_T A_c \rho \alpha_T\left(\avM{\mem^2\reward}(\policy) - 2 \mu_\mem(\policy)\avM{\mem\reward}(\policy) + \mu_\mem(\policy)^2 \avM{\reward}(\policy)\right) .
\end{align}
We then obtain the final evolution equation for $\sigma_\mem^2$,
\begin{align}
\partial_t \sigma_\mem^2(\policy) &= \lambda_T A_c \rho \left[\avMP{\mem^2} - \avM{\mem^2}(\policy) + 2 \mu_\mem(\policy)\left(\mu_\mem(\policy) - \avMP{\mu_\mem}\right) \right] \nonumber \\
& \quad + \lambda_T A_c \rho\alpha_T \left[ \left(\avMP{\reward} - \avM{\reward}(\policy)\right) \left(\sigma^2_\mem(\policy) - \mu_\mem(\policy)^2\right) + \avM{\reward}(\policy) \left( \avMP{\mem^2} - \avM{\mem^2}(\policy) + 2 \mu_\mem(\policy)\left(\mu_\mem(\policy) - \avMP{\mu_\mem}\right)\right) \right ] \nonumber\\
& \quad + \lambda_T A_c \rho\alpha_T \left[ \avM{\mem^2\reward}(\policy) - \avMP{\mem^2\reward}+ 2 \mu_\mem(\policy)\left( \avMP{\mem\reward} -  \avM{\mem\reward}(\policy)\right)\right].
\end{align}
We conclude two important points from this analytical result. First, the above equations of motion for the average memory and memory variance are much more complex than their counterparts in Eqs.~(\ref{eq:dm:dt}) and (\ref{eq:dsigma2:dt}) as far as the teaching terms are concerned. In particular, we find that the dynamics of the memory is strongly affected by the currently adapted policies through the averages $\avMP{X}.$ In contrast, Eqs.~(\ref{eq:dm:dt}) and (\ref{eq:dsigma2:dt}) were local in $\policy$ and did not depend on the policy dynamics. Second, if the policy variance is small $\sigma_\policy^2 \rightarrow 0$ we see immediately from the above equations that all terms vanish at $\policy \rightarrow \meanpol $ since $\avMP{X}  \rightarrow \avM{X}(\meanpol) $. This explains why we find the long-time policy to be independent of $\tilde \lambda_T.$
However, making more rigorous statements about the learning dynamics in the absence of memory adaptation using kinetic theory
goes beyond the goal the present paper, and is left for future work.

\section{Multidimensional policy and memory}
\label{app:multidimensional}

We provide in this appendix a generalization of our results when the policy is no longer a scalar but a vector 
$\policy=(\policy_1,\dots,\policy_q)$. 
In most cases, this also requires to consider a multidimensional memory $\mem=(\mem_1,\dots,\mem_n)$, where $n$ may differ from $q$, although the case $n=q$ is a natural one. To include all possible cases, we consider that $q,n \ge 1$. For the sake of simplicity, we generically call the policy and memory multidimensional, although the one-dimensional case is formally included in the calculation.
We briefly summarize below the main steps of the formalism in this multidimensional setting.

\subsection{Kinetic theory with multidimensional policy and memory}
\label{sec:multidim:policy}

A starting point is to consider a multidimensional observable $G(\state)=\big(G_1(\state),\dots,G_n(\state)\big)$ and memory $\mem=(\mem_1,\dots,\mem_n)$. Equation~(\ref{eq:mem}) then generalizes to
\begin{equation} \label{eq:mem:multidim}
\frac{d\mem_j}{d t} = \lambda_\mem \left(G_j(\state)-\mem_j\right) +  \xi_j(t) \qquad \qquad (j=1,\dots,n),
\end{equation}
where the Gaussian noises $\xi_j(t)$ have zero mean and correlation $\langle \xi_j(t) \xi_{j'}(t')\rangle_{\rm noise} = 2D_{\mem_j}(\policy)  \delta_{jj'}\delta(t-t')$,
with $D_{\mem_j}(\policy)$ the policy-dependent diffusion coefficient of memory component $\mem_j$.
%
In terms of the kinetic theory, the memory term $I_{\rm mem}[f]$ defined in Eq.~(\ref{eq:def:Imem}) generalizes for multidimensional observable and memory to
\begin{equation} \label{eq:def:Imem:multidim}
I_{\rm mem}[f] = \lambda_\mem \sum_{j=1}^n \frac{\p }{\p \mem_j}\left[\left(\mem_j-G_j(\state) + \lambda_\mem^{-1} D_{\mem_j}(\policy)  \frac{\p }{\p \mem_j}\right)f\right].
\end{equation}

For a multidimensional policy $\policy=(\policy_1,\dots,\policy_q)$, the learning term of the kinetic theory, as introduced in Eq.~(\ref{eq:def:Ilearn}), is modified into:
\begin{equation} \label{eq:def:Ilearn:multidim}
    I_{\rm learn}[f] = I_{\rm teach}[f] + \sum_{k=1}^q \Dpol_{\policy_k} \frac{\p^2 f}{\p \policy_k^2},
\end{equation}
where $\Dpol_{\policy_k}$ is the diffusive mutation rate associated with policy component $\policy_k$.
The teaching term $I_{\rm teach}[f]$ is also affected by the presence of a multidimensional policy, but it remains formally similar to the one-dimensional policy case,
and is still expressed in terms of the transition rate $W_f(\mem',\policy'|\mem,\policy)$ given in Eq.~(\ref{eq:def:WL}), where $\mem$ and $\policy$ are now multidimensional.
 
Integrating the evolution equation for $f(\state,\mem,\policy)$ over $\state$, taking into account Eqs.~(\ref{eq:mem:multidim}) and (\ref{eq:def:Ilearn:multidim}), 
one finds the evolution equation for $f_0(\mem,\policy)$,
\begin{equation}
\p_t f_0(\mem,\policy) = \lambda_\mem \sum_{j=1}^n \frac{\p}{\p \mem_j}\left[\left(\mem_j- \avS{G_j}(\mem,\policy)\right) f_0\right]
+ \sum_{j=1}^n D_{\mem_j}(\policy) \frac{\p^2 f_0}{\p \mem_j^2}
+ \tilde\lambda_T \rho f_0 \left(\reward(\mem) - \avMP{\reward} \right) + \sum_{k=1}^q \Dpol_{\policy_k} \frac{\p^2 f_0}{\p \policy_k^2},
\label{eq:dt:f0:multidim}
\end{equation}
which generalizes Eq.~(\ref{eq:dt:f0:v2}) to the multidimensional case.
The multidimensional memory $\mem$ is then used to define the reward function $\reward(\mem)=\reward(\mem_1,\dots,\mem_n)$.

\subsection{Dynamics of the multidimensional policy distribution and moments}

For a multidimensional policy $\policy=(\policy_1,\dots,\policy_q)$, the evolution equation for $\phi(\policy)$ generalizing Eq.~(\ref{eq:dt:phi:homog_v1}) 
is obtained by integrating Eq.~(\ref{eq:dt:f0:multidim}) over the multidimensional memory $\mem$, yielding
\beq
	\p_t \phi(\policy) = \tilde\lambda_T \,\rho\,  \left( \Reff (\policy)- \avMP{\reward}  \right)\phi(\policy) + 
    \sum_{k=1}^q \Dpol_{\policy_k} \frac{\p^2 \phi}{\p \policy_k^2}.
    \label{eq:dt:phi:homog:multidim}
\eeq
The effective reward $\Reff(\policy)$ remains formally defined by Eq.~(\ref{eq:def:Reff}), and can be evaluated along the same lines as for the one-dimensional policy case presented in the main text. In the multidimensional policy case, $\Reff(\policy)=\Reff(\policy_1,\dots,\policy_q)$ can be evaluated by generalizing Eq.~(\ref{eq:Reff:approx}) into
\begin{equation} \label{eq:Reff:approx:multidim}
    \Reff(\policy) \approx \reward\big(\meanmem(\policy)\big) +
    \frac{1}{2} \sum_{j=1}^n \sigma_{\mem_j}^2(\policy) \frac{\p^2 \reward}{\p \mem_j^2} \big(\meanmem(\policy)\big)
\end{equation}
with $\meanmem(\policy)=\big(\mu_{\mem_1}(\policy),\dots,\mu_{\mem_n}(\policy)\big)$.
Equation~(\ref{eq:Reff:approx:multidim}) becomes exact if $\reward(\mem_1,\dots,\mem_n)$ is at most quadratic in the variables $\mem_j$.
 In Eq.~(\ref{eq:Reff:approx:multidim}) we have neglected correlations between memory components $\mem_j$, which is a plausible assumption 
since $\mem_j$ and $\mem_{j'}$ have no direct coupling for $j\ne j'$, but only an indirect coupling through the policy dynamics, which is likely to be weak if memory and learning processes have well separated time scales.
The dynamics of the mean and variance $\mu_{\mem_j}$ and $\sigma_{\mem_j}^2$ of each memory component $\mem_j$ is then given by Eqs.~(\ref{eq:dm:dt:noteach}) and (\ref{eq:dsigma2:dt:noteach}), which hold independently for each $\mem_j$.

We now aim to generalize Eqs.~(\ref{eq:dP:dt}) and (\ref{eq:dsigma2P:dt}) for the dynamics of the mean policy and variance to the multidimensional case.
The general evolution equations for the averages $\avMP{Y_k}$ of a function $Y_k(\policy_k)$ and $\avMP{Z_{k\ell}}$ of a function $Z_{k\ell}(\policy_k, \policy_\ell)$, with $k,\ell \in \{1,\dots,q\}$, reads
\begin{align}
\frac{d}{dt} \avMP{Y_k} &= \tilde{\lambda}_T \rho\, \mathrm{Cov}_{\policy}\big(Y_k,\Reff(\policy, t)\big)
    + \sum_{m=1}^q \Dpol_{\policy_m} \avMP{\frac{Y_k}{\phi(\policy)}\,\frac{\p^2 \phi}{\p \policy_m^2}},\\
\frac{d}{dt} \avMP{Z_{k\ell}} &= \tilde{\lambda}_T \rho \mathrm{Cov}_\policy \left(Z_{k\ell}, \avM{\reward}(\policy,t)\right) +  \sum_{m=1}^q \Dpol_{\policy_m} \avMP{\frac{Z_{k\ell}}{\phi(\policy)}\,\frac{\p^2 \phi}{\p \policy_{m}^2}} .
\end{align}
\noindent Particular functions of interest are $Y_k=\policy_k$ and $Z_{k\ell}=\delta\policy_k \delta\policy_\ell$, with $\delta\policy_k=\policy_k-\avMP{\policy_k}$, which
define the mean vector and covariance matrix of the distribution: $\mu_{\policy_k} = \avMP{\policy_k}$ and
$\Sigma_{\policy_k \policy_\ell} = \avMP{Z_{k\ell}} = \avMP{\policy_k \policy_\ell} - \mu_{\policy_k}\mu_{\policy_\ell}$:
\begin{align}
    \partial_t \mu_{\policy_k} &= \tilde{\lambda}_T \rho \,\mathrm{Cov}_\policy \left(\policy_k, \avM{\reward}(\policy,t)\right) + \sum_{m=1}^q \Dpol_{\policy_m} \int d\policy \policy_k\,\frac{\p^2 \phi}{\p \policy_m^2}, \label{eq:dt:meanpol:multidimensional}\\
    \partial_t \Sigma_{\policy_k \policy_\ell} &= \tilde{\lambda}_T \rho \,\mathrm{Cov}_\policy \left(\delta \policy_k \delta \policy_\ell, \avM{\reward}(\policy,t)\right) + \sum_{m=1}^q \Dpol_{\policy_m} \int d\policy \left(\policy_k \policy_\ell - \mu_{\policy_k} \policy_\ell - \mu_{\policy_\ell} \policy_k \right) \frac{\partial^2 \phi }{\partial \policy_m^2} \label{eq:dt:cov:multidimensional}.
\end{align}
We discuss below how to deal with these equations in practical situations.

\subsection{Closures of the dynamics of the policy mean vector and covariance matrix}

Equations~(\ref{eq:dt:meanpol:multidimensional}) and (\ref{eq:dt:cov:multidimensional}) are in general not closed, and can be dealt with in at least two ways. A first procedure consists in assuming that the policy distribution takes the form of a multivariate Gaussian distribution, which is fully parameterized by the knowledge of
the policy means $\mu_{\policy_k}$ and covariance matrix $\Sigma_{\policy_k \policy_\ell}$.
Under this Gaussian assumption, the covariance terms on the right hand sides of Eqs.~(\ref{eq:dt:meanpol:multidimensional}) and (\ref{eq:dt:cov:multidimensional}) can be evaluated using
the multidimensional Stein identities, leading to
\begin{align}
    \partial_t \mu_{\policy_k} &= \tilde{\lambda}_T \rho \sum_{k'=1}^q \avMP{\frac{\p \Reff}{\p \policy_{k'}}} \Sigma_{\policy_k \policy_{k'}},\label{eq:dt:meanpol:multidim:gauss}\\
    \partial_t \Sigma_{\policy_k \policy_\ell} &= \tilde{\lambda}_T \rho \sum_{k'\!,\ell'=1}^q \avMP{\frac{\p^2 \Reff}{\p \policy_{k'} \p \policy_{\ell'}}} \Sigma_{\policy_k \policy_{k'}} \Sigma_{\policy_{\ell} \policy_{\ell'}} + 2 \delta_{k\ell} \Dpol_{\policy_k} \label{eq:dt:cov:multidim:gauss} \,,  
\end{align}
which generalizes the one-dimensional equations (\ref{eq:hatP_unif:gauss}) and (\ref{eq:sigmahatP_unif:gauss}).
The Gaussian assumption allows one to take correlations between policy components $\policy_k$ into account, at the price of imposing Gaussian marginal distributions for each policy component.

However, in some situations it is useful to relax the Gaussian assumption, for instance if a marginal distribution is strongly influenced by a bound of the policy definition interval, or has an asymmetric shape. We thus assume that the policy component $\policy_k$ is defined
over an interval $[\policy_{k,\min},\policy_{k,\max}]$, where the bounds $\policy_{k,\min}$ and $\policy_{k,\max}$ may be either finite or infinite.
In this case, an alternative way to simplify and close Eqs.~(\ref{eq:dt:meanpol:multidimensional}) and (\ref{eq:dt:cov:multidimensional}) is to neglect correlations by assuming that policy components $\policy_k$ are statistically independent (which is in most cases an approximation), 
\begin{equation} \label{eq:fact:dist:ansatz}
    \phi(\policy) = \prod_{k=1}^q \tilde{\phi}_{\policy_k}(\policy_k).
\end{equation}
Appropriate parameterizations, which may be non-Gaussian, are then chosen for each marginal distribution $\tilde{\phi}_{\policy_k}(\policy_k)$.
Note that in general, different policy components $\policy_k$ have different marginal distributions $\tilde{\phi}_{\policy_k}(\policy_k)$.
Under the factorization assumption (\ref{eq:fact:dist:ansatz}), the dynamics of the policy mean $\mu_{\policy_k}$ and variance
$\sigma^2_{\policy_k}=\Sigma_{\policy_k \policy_k}$ can be written as:
\begin{align}
\frac{d}{dt}  \mu_{\policy_k} &= \tilde{\lambda}_T \rho\, \mathrm{Cov}_{\policy}\big(\policy_k,\Reff(\policy)\big)
- \Dpol_{\policy_k} \left[ \tilde{\phi}_{\policy_k}(\policy_k)\right]^{\policy_{k,\max}}_{\policy_{k,\min}},
\label{eq:dP:dt:multidim:fact} \\
\frac{d}{dt}  \sigma^2_{\policy_k}  &= \tilde{\lambda}_T \rho\, \mathrm{Cov}_{\policy}\big( (\delta\policy_k)^2,\Reff(\policy)\big) + 2 \Dpol_{\policy_k}
- 2 \Dpol_{\policy_k} \left[  \delta\policy_k \, \tilde{\phi}_{\policy_k}(\policy_k)\right]^{\policy_{k,\max}}_{\policy_{k,\min}}.
\label{eq:dsigma2P:dt:multidim:fact}
\end{align}
Here, we always have $\p_{\policy_k}\tilde{\phi}_{\policy_k}(\policy_{k,\min})=\p_{\policy_k}\tilde{\phi}_{\policy_k}(\policy_{k,\max})=0$, either because $\tilde{\phi}_{\policy_k}(\policy_k)$ vanishes at infinity, or due to the no-flux condition at finite boundaries $\policy_{k,\min}$ or $\policy_{k,\max}$.
Due to the factorization assumption Eq.~(\ref{eq:fact:dist:ansatz}), the covariance $\mathrm{Cov}_{\policy}(\policy_{k},\policy_{k'})$ with $k\ne k'$ vanishes, so that only the variance $\sigma^2_{\policy_k}$ needs to be considered.

Equations~(\ref{eq:dP:dt:multidim:fact}) and (\ref{eq:dsigma2P:dt:multidim:fact}) will be useful to treat the AOU model with two-dimensional policy studied in Sec.~\ref{sec:AOU:degenerate}
and \ref{sec:2D:policy:memory} under the factorization assumption (\ref{eq:fact:dist:ansatz}).
In addition, the two-dimensional policy and memory model studied in Sec.~\ref{sec:2D:policy:memory} can also be treated using the bivariate Gaussian ansatz, leading to
Eqs.~(\ref{eq:dt:meanpol:multidim:gauss}) and (\ref{eq:dt:cov:multidim:gauss}) with $q=n=2$.
Calculations relative to the latter case are described in Appendix~\ref{app:Gauss:correl:AOU2d}.


\subsection{Bivariate Gaussian analysis of the two-dimensional policy dynamics of AOU agents}
\label{app:Gauss:correl:AOU2d}
We provide in this appendix a more detailed analysis of the two-dimensional policy dynamics $\policy=(b,D)$ of the AOU agents studied in Sec.~\ref{sec:2D:policy:memory}, now taking into account correlations between the policy components $b$ and $D$ through a bivariate Gaussian ansatz, $\phi(b,D)=\mathcal{N}(\bm{\mu},\bm{\Sigma})$, with $\bm{\mu}=(\mu_b,\mu_D)$ and $\bm{\Sigma}$
the covariance matrix of $b$ and $D$,
\begin{equation}
    \bm{\Sigma} =
    \begin{pmatrix}
    \Sigma_{bb} & \Sigma_{bD}\\
    \Sigma_{bD} & \Sigma_{DD}
    \end{pmatrix} ,
\end{equation}
with $\Sigma_{bb}=\sigma_b^2$, $\Sigma_{DD}=\sigma_D^2$, and $\Sigma_{bD}=\mathrm{Cov}_{b,D}(b,D)$.
Using formula Eqs.~(\ref{eq:dt:meanpol:multidim:gauss}) and (\ref{eq:dt:cov:multidim:gauss}) together with the first- and second-order multi-dimensional Stein identities we therefore finally find for the equations of motion,
\begin{align}
\frac{d}{dt} \mu_b &= \tilde\lambda_T \,\rho \left( \avMP{\frac{\partial \Reff}{\partial b}}\sigma_{b}^2 + \avMP{\frac{\partial \Reff}{\partial D}}\Sigma_{b D} \right)\nonumber\\
&=  \tilde\lambda_T \,\rho \left[ - 2 F_x (\mu_b F_x - V_\odot)\sigma_{b}^2 - 2F_x^2(\mu_D-\tau \sigma_\odot^2)\Sigma_{b D}  - \frac{\lambda_\mem}{1 + \lambda_\mem \tau}\Sigma_{b D} -  \frac{4F_x^2 \lambda_\mem \tau \mu_D }{ 2+\lambda_\mem \tau}\Sigma_{b D} \right]\label{eq:AOU:multi:b_dyanmics} , \\
\frac{d}{dt} \mu_D &=\tilde\lambda_T \,\rho  \left( \avMP{\frac{\partial \Reff}{\partial b}}\Sigma_{b D} + \avMP{\frac{\partial \Reff}{\partial D}}\sigma_{D}^2 \right)   \nonumber\\
&=  \tilde\lambda_T \,\rho \left[ - 2 F_x (\mu_b F_x - V_\odot)\Sigma_{b D} - 2F_x^2(\mu_D-\tau \sigma_\odot^2)\sigma_{D}^2  - \frac{\lambda_\mem}{1 + \lambda_\mem \tau}\sigma_{D}^2 - \frac{4F_x^2 \lambda_\mem \tau \mu_D }{ 2+\lambda_\mem \tau}\sigma_{D}^2 \right] \label{eq:AOU:multi:D_dyanmics} , \\
\frac{d}{dt}  \sigma_{b}^2 &= \tilde \lambda_T \rho \left( \avMP{\frac{\partial^2 \Reff}{\partial b^2}}\sigma_{b}^4 +2 \avMP{\frac{\partial^2 \Reff}{\partial b \partial D}} \Sigma_{b D} \sigma_{b}^2 + \avMP{\frac{\partial^2 \Reff}{ \partial D^2}} \Sigma_{b D}^2  \right)  + 2 \Dpol_b = - 2 \tilde \lambda_T \rho F_x^2 \sigma_{b}^4 - \tilde \lambda_T \rho F_x^2 \, \frac{4+6\lambda_\mem \tau}{ 2+\lambda_\mem \tau}   \Sigma_{b D}^2  + 2 \Dpol_b , \label{eq:AOU:multi:b_variance}\\
\frac{d}{dt}  \Sigma_{b D} &= \tilde \lambda_T \rho \left( \avMP{\frac{\partial^2 \Reff}{\partial b^2}} \Sigma_{b D} \sigma_{b}^2 + \avMP{\frac{\partial^2 \Reff}{\partial b \partial D}} \left(\sigma_{D}^2 \sigma_{b}^2 + \Sigma_{b D}^2 \right) + \avMP{\frac{\partial^2 \Reff}{ \partial D^2}} \Sigma_{b D} \sigma_{D}^2 \right)  \nonumber \\&=  - 2 \tilde \lambda_T \rho F_x^2 \Sigma_{b D} \sigma_{b}^2 - \tilde \lambda_T \rho F_x^2\, \frac{4+6 \lambda_\mem \tau  }{2+\lambda_\mem \tau}   \Sigma_{b D} \sigma_{D}^2 , \label{eq:AOU:multi:bD_variance}\\
\frac{d}{dt}  \sigma_{D}^2 &= \tilde \lambda_T \rho \left( \avMP{\frac{\partial^2 \Reff}{\partial b^2}} \Sigma_{b D}^2 + 2 \avMP{\frac{\partial^2 \Reff}{\partial b \partial D}}\sigma_{D}^2 \Sigma_{b D} + \avMP{\frac{\partial^2 \Reff}{ \partial D^2}} \sigma_{D}^4 \right) + 2 \Dpol_D =  - 2 \tilde \lambda_T \rho F_x^2 \Sigma_{b D}^2 - \tilde \lambda_T \rho F_x^2 \, \frac{4+6 \lambda_\mem \tau  }{ 2+\lambda_\mem \tau}  \sigma_{D}^4  + 2 \Dpol_D . \label{eq:AOU:multi:D_variance}
\end{align}
While these equations look quite complex they become much simpler when realizing that the covariance $\Sigma_{b D}$ between $b$ and $D$ rapidly vanishes
due to the absence of source term, so that $b$ and $D$ become uncorrelated on long time scales. 
The assumption of uncorrelated policy components made in Sec.~\ref{sec:2D:policy:memory} is thus well justified a posteriori.
This is related to the fact that Eq.~(\ref{eq:effective_reward_b_and_D}) has a separable form with additive contributions in $b$ and $D$, and hence the source term $\avMP{\frac{\partial^2 \Reff}{\partial b \partial D}}$ vanishes.

From a broader perspective, this is an interesting result since it suggests that learning high-dimensional policies may a priori be possible with the learning algorithm. Complexities might only arise when there are non-trivial correlation between the policies, for example because both policies combined affect the visited states and thus the memory in a non-trivial way.

    \FloatBarrier
	
\bibliography{biblio.bib}

@article{ariosto2025replication,
  title={Replication and Information Extraction in a Minimal Agent-Environment Model},
  author={Ariosto, Sebastiano and Garnier-Brun, Jerome and Saglietti, Luca and Straziota, Davide},
  journal={arXiv preprint arXiv:2509.23212},
  year={2025}
}

@incollection{pfeifer2009morphological,
  title={Morphological computation--connecting brain, body, and environment},
  author={Pfeifer, Rolf and G{\'o}mez, Gabriel},
  booktitle={Creating brain-like intelligence: From basic principles to complex intelligent systems},
  pages={66--83},
  year={2009},
  publisher={Springer}
}

@article{houchmandzadeh2011fixation,
	title={The fixation probability of a beneficial mutation in a geographically structured population},
	author={Houchmandzadeh, Bahram and Vallade, Marcel},
	journal={New Journal of Physics},
	volume={13},
	number={7},
	pages={073020},
	year={2011},
	publisher={IOP Publishing}
}

@article{chia2011statistical,
	title={Statistical mechanics of horizontal gene transfer in evolutionary ecology},
	author={Chia, Nicholas and Goldenfeld, Nigel},
	journal={Journal of Statistical Physics},
	volume={142},
	pages={1287--1301},
	year={2011},
	publisher={Springer}
}

@article{sella2005application,
	title={The application of statistical physics to evolutionary biology},
	author={Sella, Guy and Hirsh, Aaron E},
	journal={Proceedings of the National Academy of Sciences},
	volume={102},
	number={27},
	pages={9541--9546},
	year={2005},
	publisher={National Acad Sciences}
}

@article{drossel2001biological,
	title={Biological evolution and statistical physics},
	author={Drossel, Barbara},
	journal={Advances in physics},
	volume={50},
	number={2},
	pages={209--295},
	year={2001},
	publisher={Taylor \& Francis}
}

@article{castellano2009statistical,
	title = {Statistical physics of social dynamics},
	author = {Castellano, Claudio and Fortunato, Santo and Loreto, Vittorio},
	journal = {Rev. Mod. Phys.},
	volume = {81},
	issue = {2},
	pages = {591--646},
	numpages = {0},
	year = {2009},
	month = {May},
	publisher = {American Physical Society},
	doi = {10.1103/RevModPhys.81.591},
	url = {https://link.aps.org/doi/10.1103/RevModPhys.81.591}
}

@article{lorenz2007continuous,
	title={Continuous opinion dynamics under bounded confidence: A survey},
	author={Lorenz, Jan},
	journal={International Journal of Modern Physics C},
	volume={18},
	number={12},
	pages={1819--1838},
	year={2007},
	publisher={World Scientific}
}

@article{watson2002embodied,
	title = {Embodied Evolution: Distributing an evolutionary algorithm in a population of robots},
	journal = {Robotics and Autonomous Systems},
	volume = {39},
	number = {1},
	pages = {1-18},
	year = {2002},
	issn = {0921-8890},
	doi = {10.1016/S0921-8890(02)00170-7},
	url = {https://www.sciencedirect.com/science/article/pii/S0921889002001707},
	author = {Watson, R. A. and Ficici, S. G. and Pollack, J. B.}
}

@ARTICLE{bredeche2018embodied,
	AUTHOR={Bredeche, Nicolas and Haasdijk, Evert  and Prieto, Abraham },
	TITLE={Embodied Evolution in Collective Robotics: A Review},
	JOURNAL={Frontiers in Robotics and AI},
	VOLUME={5},
	YEAR={2018},
	URL={https://www.frontiersin.org/journals/robotics-and-ai/articles/10.3389/frobt.2018.00012},
	DOI={10.3389/frobt.2018.00012},
	ISSN={2296-9144}}

@inproceedings{long2018towards,
	title={Towards optimally decentralized multi-robot collision avoidance via deep reinforcement learning},
	author={Long, Pinxin and Fan, Tingxiang and Liao, Xinyi and Liu, Wenxi and Zhang, Hao and Pan, Jia},
	booktitle={2018 IEEE international conference on robotics and automation (ICRA)},
	pages={6252--6259},
	year={2018},
	organization={IEEE}
}

@article{ben2023morphological,
	title={Morphological computation and decentralized learning in a swarm of sterically interacting robots},
	author={Ben Zion, Matan Yah and Fersula, Jeremy and Bredeche, Nicolas and Dauchot, Olivier},
	journal={Science Robotics},
	volume={8},
	number={75},
	pages={eabo6140},
	year={2023},
	publisher={American Association for the Advancement of Science},
	doi = {10.1126/scirobotics.abo6140}
}

@article{bredeche2022social,
	author = {Bredeche, Nicolas and Fontbonne, Nicolas},
	title = {Social learning in swarm robotics},
	journal = {Philosophical Transactions of the Royal Society B: Biological Sciences},
	volume = {377},
	number = {1843},
	pages = {20200309},
	year = {2022},
	doi = {10.1098/rstb.2020.0309},
	URL = {https://royalsocietypublishing.org/doi/abs/10.1098/rstb.2020.0309}
}

@inproceedings{verdoucq2022bioinspired,
	author={Verdoucq, Matthieu and Theraulaz, Guy and Escobedo, Ram\'on and Sire, Cl\'ement and Hattenberger, Gautier},
	booktitle={2022 International Conference on Unmanned Aircraft Systems (ICUAS)}, 
	title={Bio-inspired control for collective motion in swarms of drones}, 
	year={2022},
	volume={},
	number={},
	pages={1626-1631},
	doi={10.1109/ICUAS54217.2022.9836112}
}

@misc{cazenille2024signaling,
	title={Signaling and Social Learning in Swarms of Robots}, 
	author={Leo Cazenille and Maxime Toquebiau and Nicolas Lobato-Dauzier and Alessia Loi and Loona Macabre and Nathanael Aubert-Kato and Anthony Genot and Nicolas Bredeche},
	year={2024},
	eprint={2411.11616},
	archivePrefix={arXiv},
	primaryClass={cs.RO},
	url={https://arxiv.org/abs/2411.11616}, 
}

@misc{fersula2026aggregating,
      title={Aggregating swarms through morphology handling design contingencies: from the sweet spot to a rich expressivity}, 
      author={Jeremy Fersula and Nicolas Bredeche and Olivier Dauchot},
      year={2026},
      eprint={2601.07610},
      archivePrefix={arXiv},
      primaryClass={cond-mat.soft},
      url={https://arxiv.org/abs/2601.07610}, 
}

@article{ramaswamy2010mechanics,
	author = "Ramaswamy, Sriram",
	title = "The Mechanics and Statistics of Active Matter", 
	journal= "Annual Review of Condensed Matter Physics",
	year = "2010",
	volume = "1",
	number = "Volume 1, 2010",
	pages = "323-345",
	doi = "10.1146/annurev-conmatphys-070909-104101",
	url = "https://www.annualreviews.org/content/journals/10.1146/annurev-conmatphys-070909-104101",
	publisher = "Annual Reviews",
	issn = "1947-5462",
	type = "Journal Article",
}

@article{ramaswamy2019active,
	title = {Active fluids},
	author = {Ramaswamy, S.},
	journal = {Nat. Rev. Phys.},
	volume = {1},
	pages = {640–642},
	year = {2019},
	doi = {10.1038/s42254-019-0120-9}
}

@article{marchetti2013hydrodynamics,
	title = {Hydrodynamics of soft active matter},
	author = {Marchetti, M. C. and Joanny, J. F. and Ramaswamy, S. and Liverpool, T. B. and Prost, J. and Rao, Madan and Simha, R. Aditi},
	journal = {Rev. Mod. Phys.},
	volume = {85},
	issue = {3},
	pages = {1143--1189},
	numpages = {0},
	year = {2013},
	month = {Jul},
	publisher = {American Physical Society},
	doi = {10.1103/RevModPhys.85.1143},
	url = {https://link.aps.org/doi/10.1103/RevModPhys.85.1143}
}

@article{zeravzic2017colloquium,
	title = {Colloquium: Toward living matter with colloidal particles},
	author = {Zeravcic, Zorana and Manoharan, Vinothan N. and Brenner, Michael P.},
	journal = {Rev. Mod. Phys.},
	volume = {89},
	issue = {3},
	pages = {031001},
	numpages = {14},
	year = {2017},
	month = {Sep},
	publisher = {American Physical Society},
	doi = {10.1103/RevModPhys.89.031001},
	url = {https://link.aps.org/doi/10.1103/RevModPhys.89.031001}
}

@article{zhou2022programmable,
	doi = {10.1088/1361-6439/ac85fc},
	url = {https://dx.doi.org/10.1088/1361-6439/ac85fc},
	year = {2022},
	month = {aug},
	publisher = {IOP Publishing},
	volume = {32},
	number = {10},
	pages = {103001},
	author = {Zhou, Y. and Zu, J. and Liu, J.},
	title = {Programmable intelligent liquid matter: material, science and technology},
	journal = {Journal of Micromechanics and Microengineering}
}

@article{Li2021programming,
	author = {Li, S. and Dutta, B. and Cannon, S. and Daymude, J. J. and Avinery, R. and Aydin, E. and Richa, A. W. and Goldman, D. I. and Randall, D.},
	title = {Programming active cohesive granular matter with mechanically induced phase changes},
	journal = {Science Advances},
	volume = {7},
	number = {17},
	pages = {eabe8494},
	year = {2021},
	doi = {10.1126/sciadv.abe8494},
	URL = {https://www.science.org/doi/abs/10.1126/sciadv.abe8494}}

@article{wang2024robo,
	author = {Wang, J. and Wang, G. and Chen, H. and Liu, Y. and Wang, P. and Yuan, D. and Ma, X. and Xu, X. and Cheng, Z. and Ji, B. and Yang, M. and Shuai, J. and Ye, F. and Wang, J. and Jiao, Y. and Liu, L.},
	title = {Robo-Matter towards reconfigurable multifunctional smart materials},
	journal = {Nat. Commun.},
	volume = {15},
	pages = {8853},
	year = {2024},
	doi = {10.1038/s41467-024-53123-6}
}

@InProceedings{ma2024smarticle,
	author="Ma, Danna
	and Chen, Jiahe
	and Cutler, Sadie
	and Petersen, Kirstin",
	editor="Bourgeois, Julien
	and Paik, Jamie
	and Piranda, Beno{\^i}t
	and Werfel, Justin
	and Hauert, Sabine
	and Pierson, Alyssa
	and Hamann, Heiko
	and Lam, Tin Lun
	and Matsuno, Fumitoshi
	and Mehr, Negar
	and Makhoul, Abdallah",
	title="Smarticle 2.0: Design of Scalable, Entangled Smart Matter",
	booktitle="Distributed Autonomous Robotic Systems",
	year="2024",
	publisher="Springer Nature Switzerland",
	address="Cham",
	pages="509--522",
	isbn="978-3-031-51497-5"
}

@article{kaspar2021rise,
	title={The rise of intelligent matter},
	author={Kaspar, Corinna and Ravoo, Bart Jan and van der Wiel, Wilfred G. and Wegner, Seraphine V. and Pernice, Wolfram H. P.},
	journal={Nature},
	volume={594},
	number={7863},
	pages={345--355},
	year={2021},
	publisher={Nature Publishing Group UK London},
	doi = {10.1038/s41586-021-03453-y}
}

@article{levine2023physics,
	author = {Levine, Herbert and Goldman, Daniel I.},
	title = {Physics of smart active matter: integrating active matter and control to gain insights into living systems},
	journal = {Soft Matter},
	year  = {2023},
	volume  = {19},
	issue  = {23},
	pages  = {4204-4207},
	publisher  = {The Royal Society of Chemistry},
	doi  = {10.1039/D3SM00171G},
	url  = {http://dx.doi.org/10.1039/D3SM00171G}
}

@article{cichos2020machine,
	title = {Machine learning for active matter},
	author = {Cichos, F. and Gustavsson, K. and Mehlig, B. and Volpe, G.},
	journal = {Nat. Mach. Intell.},
	volume = {2},
	pages = {94–103},
	year = {2020},
	doi = {10.1038/s42256-020-0146-9}
}

@article{vansaders2023informational,
  title={Measurement-induced phase transitions in informational active matter},
  author={VanSaders, Bryan and Fruchart, Michel and Vitelli, Vincenzo},
  journal={PNAS Nexus},
  volume={5},
  number={4},
  pages={pgag077},
  year={2026},
  publisher={Oxford University Press US}
}

@misc{cocconi2024dissipation,
	title={Dissipation-accuracy tradeoffs in autonomous control of smart active matter}, 
	author={Cocconi, L. and Mahault, B. and Piro, L.},
	year={2024},
	eprint={2409.12595},
	archivePrefix={arXiv},
	primaryClass={cond-mat.stat-mech},
	url={https://arxiv.org/abs/2409.12595}, 
}

@article{ziepke2022multi,
author = {Ziepke, A. and Maryshev, I. and Aranson, I.S. and Frey, E.},
title = {Multi-scale organization in communicating active matter},
journal = {Nat. Commun.},
volume = {13},
pages = {6727},
year = {2022},
doi = {https://doi.org/10.1038/s41467-022-34484-2},
}

@article{majidi2019soft,
	author = {Majidi, Carmel},
	title = {Soft-Matter Engineering for Soft Robotics},
	journal = {Advanced Materials Technologies},
	volume = {4},
	number = {2},
	pages = {1800477},
	doi = {10.1002/admt.201800477},
	url = {https://onlinelibrary.wiley.com/doi/abs/10.1002/admt.201800477},
	year = {2019}
}

@article{ozkan2021collective,
	title = {Collective dynamics in entangled worm and robot blobs},
	author = {Ozkan-Aydin, Yasemin and Goldman, Daniel I. and Bhamla, M. Saad},
	journal = {Proc. Natl. Acad. Sci. USA},
	volume = {118},
	pages = {e2010542118},
	year = {2021},
	doi = {10.1073/pnas.2010542118}
}

@article{goldman2024robot,
	author = {Goldman, D. I. and Rocklin, D. Z.},
	title = {Robot swarms meet soft matter physics},
	journal = {Science Robotics},
	volume = {9},
	number = {86},
	pages = {eadn6035},
	year = {2024},
	doi = {10.1126/scirobotics.adn6035},
	URL = {https://www.science.org/doi/abs/10.1126/scirobotics.adn6035}
}

@article{pishvar2020foundations,
	author = {Pishvar, Maya and Harne, Ryan L.},
	title = {Foundations for Soft, Smart Matter by Active Mechanical Metamaterials},
	journal = {Advanced Science},
	volume = {7},
	number = {18},
	pages = {2001384},
	doi = {10.1002/advs.202001384},
	url = {https://onlinelibrary.wiley.com/doi/abs/10.1002/advs.202001384},
	year = {2020}
}

@article{kotikian2019untethered,
	author = {Kotikian, Arda and McMahan, Connor and Davidson, Emily C. and Muhammad, Jalilah M. and Weeks, Robert D. and Daraio, Chiara and Lewis, Jennifer A.},
	title = {Untethered soft robotic matter with passive control of shape morphing and propulsion},
	journal = {Science Robotics},
	volume = {4},
	number = {33},
	pages = {eaax7044},
	year = {2019},
	doi = {10.1126/scirobotics.aax7044},
	URL = {https://www.science.org/doi/abs/10.1126/scirobotics.aax7044}
}

@article{liebchen2019optimal,
	doi = {10.1209/0295-5075/127/34003},
	url = {https://dx.doi.org/10.1209/0295-5075/127/34003},
	year = {2019},
	month = {sep},
	publisher = {EDP Sciences, IOP Publishing and Società Italiana di Fisica},
	volume = {127},
	number = {3},
	pages = {34003},
	author = {Liebchen, B. and L\"{o}wen, H.},
	title = {Optimal navigation strategies for active particles},
	journal = {Europhysics Letters}
}

@article{nasiri2023optimal,
	doi = {10.1209/0295-5075/acc270},
	url = {https://dx.doi.org/10.1209/0295-5075/acc270},
	year = {2023},
	month = {mar},
	publisher = {EDP Sciences, IOP Publishing and Società Italiana di Fisica},
	volume = {142},
	number = {1},
	pages = {17001},
	author = {Nasiri, M. and L\"{o}wen, H. and Liebchen, B.},
	title = {Optimal active particle navigation meets machine learning},
	journal = {Europhysics Letters}
}

@article{piro2021optimal,
	title = {Optimal navigation strategies for microswimmers on curved manifolds},
	author = {Piro, Lorenzo and Tang, Evelyn and Golestanian, Ramin},
	journal = {Phys. Rev. Res.},
	volume = {3},
	issue = {2},
	pages = {023125},
	numpages = {9},
	year = {2021},
	month = {May},
	publisher = {American Physical Society},
	doi = {10.1103/PhysRevResearch.3.023125},
	url = {https://link.aps.org/doi/10.1103/PhysRevResearch.3.023125}
}

@article{piro2022optimal,
	doi = {10.1088/1367-2630/ac9079},
	url = {https://dx.doi.org/10.1088/1367-2630/ac9079},
	year = {2022},
	month = {sep},
	publisher = {IOP Publishing},
	volume = {24},
	number = {9},
	pages = {093037},
	author = {Piro, L. and Mahault, B. and Golestanian, R.},
	title = {Optimal navigation of microswimmers in complex and noisy environments},
	journal = {New Journal of Physics}
}

@article{piro2022efficiency,
	title = {Efficiency of navigation strategies for active particles in rugged landscapes},
	author = {Piro, L. and Golestanian, R. and Mahault, B.},
	journal = {Front. Phys.},
	volume = {10},
	pages = {1034267},
	year = {2022},
	doi = {10.3389/fphy.2022.1034267}
}

@article{borra2021optimal,
	doi = {10.1088/1742-5468/ac12c6},
	url = {https://dx.doi.org/10.1088/1742-5468/ac12c6},
	year = {2021},
	month = {aug},
	publisher = {IOP Publishing and SISSA},
	volume = {2021},
	number = {8},
	pages = {083401},
	author = {Borra, F. and Cencini, M. and Celani, A.},
	title = {Optimal collision avoidance in swarms of active Brownian particles},
	journal = {Journal of Statistical Mechanics: Theory and Experiment}
}

@article{yang2022autonomous,
	title={Autonomous environment-adaptive microrobot swarm navigation enabled by deep learning-based real-time distribution planning},
	author={Yang, Lidong and Jiang, Jialin and Gao, Xiaojie and Wang, Qinglong and Dou, Qi and Zhang, Li},
	journal={Nature Machine Intelligence},
	volume={4},
	number={5},
	pages={480--493},
	year={2022},
	publisher={Nature Publishing Group UK London},
	doi = {10.1038/s42256-022-00482-8}
}

@article{bonnemain2023pedestrians,
	title = {Pedestrians in static crowds are not grains, but game players},
	author = {Bonnemain, Thibault and Butano, Matteo and Bonnet, Th\'eophile and Echeverr\'{\i}a-Huarte, I\~naki and Seguin, Antoine and Nicolas, Alexandre and Appert-Rolland, C\'ecile and Ullmo, Denis},
	journal = {Phys. Rev. E},
	volume = {107},
	issue = {2},
	pages = {024612},
	numpages = {12},
	year = {2023},
	month = {Feb},
	publisher = {American Physical Society},
	doi = {10.1103/PhysRevE.107.024612},
	url = {https://link.aps.org/doi/10.1103/PhysRevE.107.024612}
}

@article{echevarria2023body,
	title = {Body and mind: Decoding the dynamics of pedestrians and the effect of smartphone distraction by coupling mechanical and decisional processes},
	journal = {Transportation Research Part C: Emerging Technologies},
	volume = {157},
	pages = {104365},
	year = {2023},
	issn = {0968-090X},
	doi = {10.1016/j.trc.2023.104365},
	url = {https://www.sciencedirect.com/science/article/pii/S0968090X23003558},
	author = {Echeverr\'{\i}a-Huarte, I. and Nicolas, A.}
}

@article{mo2023challenges,
	title = {Challenges and attempts to make intelligent microswimmers},
	author = {Mo C., Li G. and Bian X.},
	journal = {Front. Phys.},
	volume = {11},
	pages = {1279883},
	year = {2023},
	doi = {10.3389/fphy.2023.1279883}
}

@book{boltzmann1896KT,
	title={Vorlesungen über Gastheorie. Bd. 1},
	author={Ludwig Boltzmann},
	year={1896},
	publisher={Barth}
}

@book{brilliantov2004kinetic,
	title={Kinetic theory of granular gases},
	author={Brilliantov, Nikolai V. and P{\"o}schel, Thorsten},
	year={2004},
	publisher={Oxford University Press, USA}
}

@article{ihle2011kinetic,
	title={Kinetic theory of flocking: Derivation of hydrodynamic equations},
	author={Ihle, Thomas},
	journal={Physical Review E—Statistical, Nonlinear, and Soft Matter Physics},
	volume={83},
	number={3},
	pages={030901},
	year={2011},
	publisher={APS},
	doi = {10.1103/PhysRevE.83.030901}
}

@article{ihle2014towards,
	title={Towards a quantitative kinetic theory of polar active matter},
	author={Ihle, Thomas},
	journal={The European Physical Journal Special Topics},
	volume={223},
	number={7},
	pages={1293--1314},
	year={2014},
	publisher={Springer},
	doi = {10.1140/epjst/e2014-02192-0}
}

@article{bertin2013mesoscopic,
	title={Mesoscopic theory for fluctuating active nematics},
	author={Bertin, Eric and Chat{\'e}, Hugues and Ginelli, Francesco and Mishra, Shradha and Peshkov, Anton and Ramaswamy, Sriram},
	journal={New journal of physics},
	volume={15},
	number={8},
	pages={085032},
	year={2013},
	publisher={IOP Publishing},
	doi = {10.1088/1367-2630/15/8/085032}
}

@article{peshkov2014boltzmann,
	title={Boltzmann-{G}inzburg-{L}andau approach for continuous descriptions of generic {V}icsek-like models},
	author={Peshkov, Anton and Bertin, Eric and Ginelli, Francesco and Chat{\'e}, Hugues},
	journal={The European Physical Journal Special Topics},
	volume={223},
	number={7},
	pages={1315--1344},
	year={2014},
	publisher={Springer}
}

@article{bertin2006boltzmann,
	title = {Boltzmann and hydrodynamic description for self-propelled particles},
	author = {Bertin, Eric and Droz, Michel and Gr\'egoire, Guillaume},
	journal = {Phys. Rev. E},
	volume = {74},
	issue = {2},
	pages = {022101},
	numpages = {4},
	year = {2006},
	month = {Aug},
	publisher = {American Physical Society},
	doi = {10.1103/PhysRevE.74.022101},
	url = {https://link.aps.org/doi/10.1103/PhysRevE.74.022101}
}

@article{bertin2009hydrodynamic,
	doi = {10.1088/1751-8113/42/44/445001},
	url = {https://dx.doi.org/10.1088/1751-8113/42/44/445001},
	year = {2009},
	month = {oct},
	publisher = {},
	volume = {42},
	number = {44},
	pages = {445001},
	author = {Bertin, Eric and Droz, Michel and Grégoire, Guillaume},
	title = {Hydrodynamic equations for self-propelled particles: microscopic derivation and stability analysis},
	journal = {Journal of Physics A: Mathematical and Theoretical}
}

@article{ewens1989interpretation,
	title = {An interpretation and proof of the fundamental theorem of natural selection},
	journal = {Theoretical Population Biology},
	volume = {36},
	number = {2},
	pages = {167-180},
	year = {1989},
	issn = {0040-5809},
	doi = {https://doi.org/10.1016/0040-5809(89)90028-2},
	url = {https://www.sciencedirect.com/science/article/pii/0040580989900282},
	author = {W.J. Ewens}
}

@article{PhysRevE.97.042604,
  title = {Markovian robots: Minimal navigation strategies for active particles},
  author = {Nava, Luis G\'omez and Gro\ss{}mann, Robert and Peruani, Fernando},
  journal = {Phys. Rev. E},
  volume = {97},
  issue = {4},
  pages = {042604},
  numpages = {17},
  year = {2018},
  month = {Apr},
  publisher = {American Physical Society},
  doi = {10.1103/PhysRevE.97.042604},
  url = {https://link.aps.org/doi/10.1103/PhysRevE.97.042604}
}

@article{nasiri2024smart,
  title={Smart active particles learn and transcend bacterial foraging strategies},
  author={Nasiri, Mahdi and Loran, Edwin and Liebchen, Benno},
  journal={Proceedings of the National Academy of Sciences},
  volume={121},
  number={15},
  pages={e2317618121},
  year={2024},
  publisher={National Acad Sciences}
}

@article{doi:10.1177/1059712320930418,
author = {Yating Zheng and Cristián Huepe and Zhangang Han},
title ={Experimental capabilities and limitations of a position-based control algorithm for swarm robotics},
journal = {Adaptive Behavior},
volume = {30},
number = {1},
pages = {19-35},
year = {2022},
doi = {10.1177/1059712320930418},
URL = { 
        https://doi.org/10.1177/1059712320930418
},
eprint = { 
        https://doi.org/10.1177/1059712320930418
}
}

@article{jung2025kinetic,
  title = {Kinetic Theory of Decentralized Learning for Smart Active Matter},
  author = {Jung, Gerhard and Ozawa, Misaki and Bertin, Eric},
  journal = {Phys. Rev. Lett.},
  volume = {134},
  issue = {24},
  pages = {248302},
  numpages = {10},
  year = {2025},
  month = {Jun},
  publisher = {American Physical Society},
  doi = {10.1103/5m44-kwhv},
  url = {https://link.aps.org/doi/10.1103/5m44-kwhv}
}

@article{cates2013when,
title = {When are active Brownian particles and run-and-tumble particles equivalent? Consequences for motility-induced phase separation},
author = {M. E. Cates and J. Tailleur},
journal = {Europhys. Lett.},
volume = {101},
pages = {20010},
year = {2013}
}

@article{beltran2023decentralized,
  title={Decentralized federated learning: Fundamentals, state of the art, frameworks, trends, and challenges},
  author={Beltr{\'a}n, Enrique Tom{\'a}s Mart{\'\i}nez and P{\'e}rez, Mario Quiles and S{\'a}nchez, Pedro Miguel S{\'a}nchez and Bernal, Sergio L{\'o}pez and Bovet, G{\'e}r{\^o}me and P{\'e}rez, Manuel Gil and P{\'e}rez, Gregorio Mart{\'\i}nez and Celdr{\'a}n, Alberto Huertas},
  journal={IEEE Communications Surveys \& Tutorials},
  volume={25},
  number={4},
  pages={2983--3013},
  year={2023},
  publisher={IEEE}
}

@article{yuan2024decentralized,
  title={Decentralized federated learning: A survey and perspective},
  author={Yuan, Liangqi and Wang, Ziran and Sun, Lichao and Yu, Philip S and Brinton, Christopher G},
  journal={IEEE Internet of Things Journal},
  volume={11},
  number={21},
  pages={34617--34638},
  year={2024},
  publisher={IEEE}
}

@article{gabrielli2023survey,
  title={A survey on decentralized federated learning},
  author={Gabrielli, Edoardo and Pica, Giovanni and Tolomei, Gabriele},
  journal={arXiv preprint arXiv:2308.04604},
  year={2023}
}

@article{saxe2013exact,
  title={Exact solutions to the nonlinear dynamics of learning in deep linear neural networks},
  author={Saxe, Andrew M and McClelland, James L and Ganguli, Surya},
  journal={arXiv preprint arXiv:1312.6120},
  year={2013}
}

@article{taylor1978evolutionary,
  title={Evolutionary stable strategies and game dynamics},
  author={Taylor, Peter D and Jonker, Leo B},
  journal={Mathematical biosciences},
  volume={40},
  number={1-2},
  pages={145--156},
  year={1978},
  publisher={Elsevier}
}

@article{schuster1983replicator,
  title={Replicator dynamics},
  author={Schuster, Peter and Sigmund, Karl},
  journal={Journal of theoretical biology},
  volume={100},
  number={3},
  pages={533--538},
  year={1983},
  publisher={Elsevier}
}

@article{hofbauer2003evolutionary,
  title={Evolutionary game dynamics},
  author={Hofbauer, Josef and Sigmund, Karl},
  journal={Bulletin of the American mathematical society},
  volume={40},
  number={4},
  pages={479--519},
  year={2003}
}

@article{komarova2004replicator,
  title={Replicator--mutator equation, universality property and population dynamics of learning},
  author={Komarova, Natalia L},
  journal={Journal of theoretical biology},
  volume={230},
  number={2},
  pages={227--239},
  year={2004},
  publisher={Elsevier}
}

@article{bjedov2003stress,
  title={Stress-induced mutagenesis in bacteria},
  author={Bjedov, Ivana and Tenaillon, Olivier and Gerard, Benedicte and Souza, Valeria and Denamur, Erick and Radman, Miroslav and Taddei, Fran{\c{c}}ois and Matic, Ivan},
  journal={Science},
  volume={300},
  number={5624},
  pages={1404--1409},
  year={2003},
  publisher={American Association for the Advancement of Science}
}

@article{foster2007stress,
  title={Stress-induced mutagenesis in bacteria},
  author={Foster, Patricia L},
  journal={Critical reviews in biochemistry and molecular biology},
  volume={42},
  number={5},
  pages={373--397},
  year={2007},
  publisher={Taylor \& Francis}
}

@inproceedings{moran1958random,
  title={Random processes in genetics},
  author={Moran, Patrick Alfred Pierce},
  booktitle={Mathematical proceedings of the cambridge philosophical society},
  volume={54},
  number={1},
  pages={60--71},
  year={1958},
  organization={Cambridge University Press}
}

@article{lieberman2005evolutionary,
  title={Evolutionary dynamics on graphs},
  author={Lieberman, Erez and Hauert, Christoph and Nowak, Martin A},
  journal={Nature},
  volume={433},
  number={7023},
  pages={312--316},
  year={2005},
  publisher={Nature Publishing Group UK London}
}

@book{lotka1925elements,
  title={Elements of physical biology},
  author={Lotka, Alfred James},
  year={1925},
  publisher={Williams \& Wilkins}
}

@book{volterra1926variazioni,
  title={Variazioni e fluttuazioni del numero d'individui in specie animali conviventi},
  author={Volterra, Vito},
  year={1926},
  publisher={Societ{\`a} anonima tipografica" Leonardo da Vinci"}
}

@article{abrams2000evolution,
  title={The evolution of predator-prey interactions: theory and evidence},
  author={Abrams, Peter A},
  journal={Annual Review of Ecology and Systematics},
  volume={31},
  number={1},
  pages={79--105},
  year={2000},
  publisher={Annual Reviews 4139 El Camino Way, PO Box 10139, Palo Alto, CA 94303-0139, USA}
}

@article{cengio2026evolution,
  title={When evolution realizes large deviations of fitness: from speciation to dynamical phase transitions},
  author={Dal Cengio, Sara and Laurenceau, Quentin and Lecomte, Vivien and Smadi, Charline and Tailleur, Julien},
  journal={arXiv preprint arXiv:2601.04325},
  year={2026}
}

@book{CrowKimura1970,
  title = {An Introduction to Population Genetics Theory},
  author = {Crow, James F. and Kimura, Motoo},
  year = {1970},
  publisher = {Harper \& Row},
  address = {New York},
  isbn = {978-0872910153}
}

@article{back1993overview,
  title={An overview of evolutionary algorithms for parameter optimization},
  author={B{\"a}ck, Thomas and Schwefel, Hans-Paul},
  journal={Evolutionary computation},
  volume={1},
  number={1},
  pages={1--23},
  year={1993},
  publisher={mit Press}
}

@inproceedings{de2017evolutionary,
  title={Evolutionary computation: a unified approach},
  author={De Jong, Kenneth},
  booktitle={Proceedings of the Genetic and Evolutionary Computation Conference Companion},
  pages={373--388},
  year={2017}
}

@article{hussein2017imitation,
  title={Imitation learning: A survey of learning methods},
  author={Hussein, Ahmed and Gaber, Mohamed Medhat and Elyan, Eyad and Jayne, Chrisina},
  journal={ACM Computing Surveys (CSUR)},
  volume={50},
  number={2},
  pages={1--35},
  year={2017},
  publisher={ACM New York, NY, USA}
}

@article{jaderberg2017population,
  title={Population based training of neural networks},
  author={Jaderberg, Max and Dalibard, Valentin and Osindero, Simon and Czarnecki, Wojciech M and Donahue, Jeff and Razavi, Ali and Vinyals, Oriol and Green, Tim and Dunning, Iain and Simonyan, Karen and others},
  journal={arXiv preprint arXiv:1711.09846},
  year={2017}
}

@article{stanley2019designing,
  title={Designing neural networks through neuroevolution},
  author={Stanley, Kenneth O and Clune, Jeff and Lehman, Joel and Miikkulainen, Risto},
  journal={Nature Machine Intelligence},
  volume={1},
  number={1},
  pages={24--35},
  year={2019},
  publisher={Nature Publishing Group UK London}
}

@misc{jung2026inhomogeneous,
title = {Hydrodynamic description of collective learning in smart active matter},
author = {Jung, G. and Asnacios, J. and Ozawa, M. and Dauchot, O. and Bertin, E.},
note = {in preparation},
year = {2026},
}

@incollection{lowen2026towards,
  title={Towards intelligent active particles},
  author={L{\"o}wen, Hartmut and Liebchen, Benno},
  booktitle={Artificial Intelligence and Intelligent Matter: Nanoscience, Soft Matter, Philosophy},
  pages={257--271},
  year={2026},
  publisher={Springer}
}

@article{garnier2025hydrodynamics,
  title={Hydrodynamics of cooperation and self-interest in a two-population occupation model},
  author={Garnier-Brun, J{\'e}r{\^o}me and Zakine, Ruben and Benzaquen, Michael},
  journal={Physical Review Letters},
  volume={135},
  number={10},
  pages={107402},
  year={2025},
  publisher={APS}
}

@article{novkoski2026graspion,
  title={GRASPion: An open-source, programmable brainbot for active matter research},
  author={Novkoski, F and M{\'e}lard, M and Delens, M and W{\'e}ry, F and Noirhomme, M and Pande, J and Maier, A and Smith, A-S and Vandewalle, N},
  journal={Review of Scientific Instruments},
  volume={97},
  pages = {014704},
  number={1},
  year={2026},
  publisher={AIP Publishing}
}

@article{han2025fluctuation,
  title={Fluctuation theorem and optimal control of an active tracking particle with information processing},
  author={Han, Tai and Meng, Fanlong},
  journal={arXiv preprint arXiv:2508.21487},
  year={2025}
}

@book{te2025artificial,
  title={Artificial intelligence and intelligent matter},
  author={te Vrugt, Michael},
  year={2025},
  publisher={Springer}
}

@article{floyd2024learning,
  title={Learning to control non-equilibrium dynamics using local imperfect gradients},
  author={Floyd, Carlos and Dinner, Aaron R and Vaikuntanathan, Suriyanarayanan},
  journal={arXiv preprint arXiv:2404.03798},
  year={2024}
}

@article{janzen2026active,
  title={Active matter as a framework for living systems-inspired Robophysics},
  author={Janzen, Giulia and Maselli, Gaia and Jimenez, Juan F and Garcia-Perez, Lia and Matoz Fernandez, DA and Valeriani, Chantal},
  journal={Europhysics Letters},
  volume={154},
  number={3},
  pages={37001},
  year={2026},
  publisher={EDP Sciences, IOP Publishing and Societ{\`a} Italiana di Fisica}
}

@article{Cates_MIPS_review2015,
   author = "Cates, Michael E. and Tailleur, Julien",
   title = "Motility-Induced Phase Separation", 
   journal= "Annual Review of Condensed Matter Physics",
   year = "2015",
   volume = "6",
   number = "Volume 6, 2015",
   pages = "219-244",
   doi = "https://doi.org/10.1146/annurev-conmatphys-031214-014710",
   url = "https://www.annualreviews.org/content/journals/10.1146/annurev-conmatphys-031214-014710",
   publisher = "Annual Reviews",
}

@article{szamel2014self,
  title={Self-propelled particle in an external potential: Existence of an effective temperature},
  author={Szamel, Grzegorz},
  journal={Physical Review E},
  volume={90},
  number={1},
  pages={012111},
  year={2014},
  publisher={APS}
}

@article{bonilla2019active,
  title={Active ornstein-uhlenbeck particles},
  author={Bonilla, Luis L},
  journal={Physical Review E},
  volume={100},
  number={2},
  pages={022601},
  year={2019},
  publisher={APS}
}

@article{catania2026solvable,
  title={A solvable model for unsupervised federated learning},
  author={Catania, Giovanni and Decelle, Aur{\'e}lien and Manzan, Gianluca and Seoane, Beatriz and Tantari, Daniele},
  journal={arXiv preprint arXiv:2606.13045},
  year={2026}
}

@article{arola2024effective,
  title={Effective theory of collective deep learning},
  author={Arola-Fern{\'a}ndez, Llu{\'\i}s and Lacasa, Lucas},
  journal={Physical Review Research},
  volume={6},
  number={4},
  pages={L042040},
  year={2024},
  publisher={APS}
}


\end{document}